\documentclass[aps,prd,twoside,nofootinbib]{revtex4}
\usepackage{graphicx}
\usepackage{latexsym}
\usepackage{hyperref}
\usepackage{fancyhdr}
\usepackage{color}
\usepackage{bm}
\newsavebox{\savepar}

    \newcommand{\ev}{\hbox{ eV}}
    \newcommand{\kev}{\hbox{ keV}}
    \newcommand{\mev}{\hbox{ MeV}}
    \newcommand{\gev}{\hbox{ GeV}}
    \newcommand{\tev}{\hbox{ TeV}}
    \newcommand{\cm}{\hbox{ cm}}
    \newcommand{\km}{\hbox{ km}}
    \newcommand{\s}{\hbox{ s}}
    \newcommand{\mpc}{\hbox{ Mpc}}
    
    \newcommand{\K}{\hbox{ K}}
    
    \newcommand{\cfrac}[2]{\textstyle{\frac{#1}{#2}}}

    \newcommand{\m}{\hbox{ m}}

    \newcommand{\lag}{\ensuremath{\mathcal{L}}}

  \def\ket#1{| #1\rangle}
\newcommand{\be}{$^8\mathrm{B}\;$}
\newcommand{\bes}{$^7\mathrm{Be}\;$}

\newcommand{\ewgg}{\ensuremath{\mathrm{SU(2)_L }\otimes \mathrm{U(1)}_Y}}
\newcommand{\sev}{\hbox{ s}/\hbox{eV}}
\newcommand{\abs}[1]{\left| #1\right|}
\def\vev#1{\langle #1\rangle_0}

\preprint{FNAL--CONF--07/417--T}

\begin{document}

\title{Cosmic Neutrinos}

%

\author{Chris Quigg}
\email{quigg@fnal.gov}
\affiliation{Fermi National Accelerator Laboratory,  P.O. Box 500, 
Batavia, Illinois 60510 USA \\ and \\ Theory Group, Physics Department, CERN, CH-1211 Geneva 23, Switzerland}

\begin{abstract}
I recall the place of neutrinos in the electroweak theory and summarize what we know about neutrino mass and flavor change.
I next review the essential characteristics expected for relic neutrinos and survey what we can say about the neutrino contribution to the dark matter of the Universe. Then I discuss the standard-model interactions of ultrahigh-energy neutrinos, paying attention to the consequences of neutrino oscillations, and illustrate a few topics of interest to neutrino observatories. I conclude with short comments on the remote possibility of detecting relic neutrinos through annihilations of ultrahigh-energy neutrinos at the $Z$ resonance.

\end{abstract}

\maketitle

\thispagestyle{plain}


\section{INTRODUCTION}
Neutrinos are abundant, and they infiltrate everything around us---even you and me! Each second, some $10^{14}$ neutrinos made in the Sun and about a thousand neutrinos made by cosmic rays in Earth's atmosphere pass through your body. Other neutrinos reach us from natural and artificial sources: decays of radioactive  elements inside the Earth~\cite{Krauss:1983zn,Mantovani:2003yd,Araki:2005qa}, and emanations from nuclear reactors and particle accelerators. 
Inside your body are more than $10^{7}$ neutrino relics from the early universe, provided that neutrinos are stable on cosmological time scales. The calculated relic density of neutrinos and antineutrinos in the current universe is $n_{\nu_{i}0} = n_{\bar{\nu}_{i}0} \approx 56\cm^{-3}$ for each flavor. The neutrino gas that we believe permeates the present Universe has never been 
detected directly. Imaginative schemes have been proposed to 
record the elastic scattering of the 1.95-K relic neutrinos, but all appear to require 
significant further technological development before they can 
approach the needed sensitivity~\cite{Duda:2001hd,Ringwald:2004np,Cocco:2007za}.  Because neutrinos have mass, relic neutrinos would constitute some of the dark matter of the universe. Neutrinos as dark matter will be one theme of this lecture. The second major topic will be the interactions of ultrahigh-energy neutrinos and a cursory look at the prospect that neutrino observatories may teach lessons about particle physics~\cite{Han:2004kq}. Then we will look briefly at the possibility of detecting the relic neutrinos by observing the resonant annihilation of extremely-high-energy 
cosmic neutrinos on the background neutrinos through the reaction 
$\nu\bar{\nu} \to Z^{0}$. 

The background in cosmology that underlies our discussion may be found in  \S19--23 of the \textit{Review of Particle Physics}~\cite{Yao:2006px} or in any modern textbook; Scott Dodelson reviewed the essentials in his course at this school~\cite{ScottSSI}. The monographs by Bahcall~\cite{Bahcall} and by Giunti \& Kim~\cite{GiuntiKim} treat many topics in neutrino astrophysics; Steen Hannestad's review article on primordial neutrinos~\cite{Hannestad:2006zg} and Gianpiero Mangano's lecture notes~\cite{Gianpiero} encompass cosmological constraints on neutrino properties. The 2007 Neutrino Physics Summer School at Fermilab~\cite{NUSS} provided a comprehensive survey of neutrino physics. For a guide to the Russian literature, consult Ref.~\cite{Khlopov:437816}.

According to the standard cosmology, neutrinos should be the most
abundant particles in the Universe, after the photons of the cosmic
microwave background, provided that they are stable over cosmological
times. Because they interact only
weakly, neutrinos fell out of equilibrium when the age of the Universe was $\approx
0.1\hbox{ s}$ (redshift $z \approx 10^{10}$) and the temperature of the Universe was $\hbox{a
few}\mev$.  Accordingly, relic neutrinos have been present---as witnesses
or participants---for landmark events in the history of the Universe: the
era of big-bang nucleosynthesis, a few minutes into the life of the
Universe~\cite{Steigman:2005uz,FieldsSarkarBBN}; the decoupling era around $379\,000\hbox{
y}$~\cite{Bennett:2003bz}, when the cosmic microwave background was
imprinted on the surface of last scattering; and the era of large-scale
structure formation, when the Universe was only a few percent  of its
current age~\cite{Bond:1980ha,Tegmark:2005cy}. For recent quantitative assessments
of evidence that neutrinos were 
present at these times, see Ref.~\cite{Barger:2003zg,Steigman:2006mv,deBernardis:2007bu}.

Some of the earliest cosmological bounds on neutrino masses 
followed from the requirement that massive relic neutrinos, present 
today in the expected numbers, do not saturate the critical density of the 
Universe~\cite{Gershtein:1966gg,Cowsik:1972gh}. Refined analyses, 
incorporating constraints from a suite of cosmological measurements, 
sharpen the bounds on the sum of light-neutrino masses~\cite{Fogli:2004as}. 
The discovery of neutrino 
oscillations~\cite{Fukuda:1998mi,Ahmad:2002jz,Eguchi:2002dm} implies 
that neutrinos have mass, but we cannot precisely compute the contribution 
of relic neutrinos to the dark matter of the Universe until we establish 
the absolute scale of neutrino masses. Current estimates for the 
neutrino fraction of the Universe's mass--energy density lie in the 
range $0.1\% \lesssim \Omega_{\nu} \lesssim \hbox{a few }\%$, under standard assumptions. The uncertainty reflects our incomplete knowledge of neutrino properties.

\section{Neutrino Interactions and Properties \label{sec:massmix}}
\subsection{Neutrinos in the Electroweak Theory}
\begin{figure}[tb]
\begin{center}
\includegraphics[width=8.0cm]{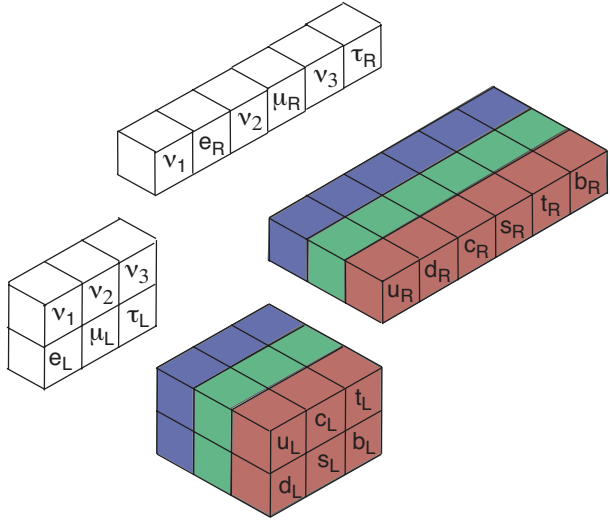}
\caption{Left-handed doublets and right-handed singlets of quarks 
and leptons that inspire 
the structure of the electroweak theory. 
\label{fig:SMsketch}}
\end{center}
\end{figure}

The electroweak theory emerged through trial and error, guided by experiment. The idealization that neutrinos are massless did not flow from any sound principle, but was inferred from kinematical evidence against measurably large masses. Since fermion mass normally requires linking left-handed and right-handed states, the presumed masslessness of the neutrinos could be captured by the omission of right-handed neutrinos from the theory, consistent with evidence that $\nu_e$~\cite{PhysRev.109.1015}, $\nu_{\mu}$~\cite{PhysRevLett.7.23,Possoz:1977jf}, and (much later) $\nu_{\tau}$\cite{PhysRevLett.78.4691} produced in charged-current interactions are left-handed.
On current evidence, the correct electroweak gauge symmetry melds the $\mathrm{SU(2)_{L}}$ family (weak-isospin) symmetry suggested by the left-handed doublets of Figure~\ref{fig:SMsketch} 
with a $\mathrm{U(1)}_{Y}$ weak-hypercharge phase symmetry.\footnote{See my lectures at the neutrino summer school~\cite{CQNUSS} for an informal but thorough development.}

We characterize the \ewgg\ theory by  the left-handed quarks
\begin{equation}
\mathsf{L}_q^{(1)} = 
\left(
		\begin{array}{c}
			u  \\
			d
		\end{array}
		 \right)_{\mathrm{L}} \quad
\mathsf{L}_q^{(2)} = 
		\left(
		\begin{array}{c}
			c  \\
			s
		\end{array}
		 \right)_{\mathrm{L}} \quad
		 \mathsf{L}_q^{(3)} = 
		\left(
		\begin{array}{c}
			t  \\
			b
		\end{array}
		 \right)_{\mathrm{L}}	\;,
		 \label{eq:lquarks}
	\end{equation}	
with weak isospin $I = \case{1}{2}$ and weak hypercharge $Y(\mathsf{L}_q) = \case{1}{3}$; their right-handed weak-isoscalar counterparts
\begin{equation}
\mathsf{R}_u^{(1,2,3)} = u_{\mathrm{R}}, c_{\mathrm{R}}, t_{\mathrm{R}}\hbox{ and }
\mathsf{R}_d^{(1,2,3)} = d_{\mathrm{R}}, s_{\mathrm{R}}, b_{\mathrm{R}}\;,
\label{eq:rightup}
\end{equation}
with weak hypercharges $Y(\mathsf{R}_u) = \case{4}{3}$ and $Y(\mathsf{R}_d) = -\case{2}{3}$;
the left-handed leptons
\begin{equation}
\mathsf{L}_e = 
\left(
		\begin{array}{c}
			\nu_{e}  \\
			e^{-}
		\end{array}
		 \right)_{\mathrm{L}} \;\;\;\;\;\;
\mathsf{L}_{\mu} = 
		\left(
		\begin{array}{c}
			\nu_{\mu}  \\
			\mu^{-}
		\end{array}
		 \right)_{\mathrm{L}} \;\;\;\;\;\;
\mathsf{L}_{\tau} = 
		\left(
		\begin{array}{c}
			\nu_{\tau}  \\
			\tau^{-}
		\end{array}
		\right)_{\mathrm{L}}	\;,
		\label{eq:lleptons}
	\end{equation}
with weak isospin $I = \case{1}{2}$ and weak hypercharge $Y(\mathsf{L}_{\ell}) = -1$; and the right-handed weak-isoscalar charged leptons
\begin{equation}
\mathsf{R}_{e,\mu,\tau} = e_{\mathrm{R}}, \mu_{\mathrm{R}}, \tau_{\mathrm{R}}\;,
\label{eq:rightlep}
\end{equation}
with weak hypercharge $Y(\mathsf{R}_{\ell}) = -2$. (Weak isospin and weak hypercharge are related to electric charge through $Q = I_{3} + \cfrac{1}{2}Y$.) Right-handed neutrinos are left out.

\begin{table}[bt]
    \centering
    \caption{Some properties of the leptons~\cite{Yao:2006px}.\label{tab:leptons}}
    \begin{tabular}{ccc}
        \hline
        Lepton & Mass & Lifetime  \\
        \hline\\[-6pt] 
        $\nu_{e}$ & $< 2\ev$ &   \\
        $e^{-}$ & $0.510\,998\,918 (44)\mev$ & 
	$>4.6 \times 10^{26}\hbox{ y}\; (90\%\ \mathrm{CL})$  \\[3pt]
        $\nu_{\mu}$ & $<0.19\mev\;(90\%\ \mathrm{CL})$ &   \\
        $\mu^{-}$ & $105.658\,369\,2 (94)\mev$ & $2.197\,03 
        (4)\times 10^{-6}\s$  \\[3pt]
        $\nu_{\tau}$ & $<18.2\mev\;(95\%\ \mathrm{CL})$ &   \\
        $\tau^{-}$ & $1776.90\pm0.20\mev$ & $290.6 \pm 1.0 
        \times 10^{-15}\s$  \\[4pt]
        \hline
    \end{tabular}
\end{table}

I do not think that we know enough to specify a new (``$\nu$'') standard model,\footnote{Some plausible definitions are explored in Refs.~\cite{Parke:2006mr,RabiNUSS}.} but the inference from neutrino oscillations that neutrinos have mass makes it tempting to suppose that right-handed neutrinos do exist, as indicated in Figure~\ref{fig:SMsketch}. These right-handed neutrinos are sterile---inert with respect to the known interactions with $\gamma$, left-handed $W$, and $Z$. As we shall see in more detail in \S~\ref{subsec:massmix}, neutrino masses can evade the usual requirement that a (Dirac) fermion mass link left-handed and right-handed states, provided that the neutrino is its own antiparticle. We cannot yet establish the existence of right-handed neutrinos, but I will take their existence as a working hypothesis. Given the absence of detectable right-handed charged-current interactions, it is not surprising that what we surmise about the right-handed neutrinos is of little consequence of most studies of neutrino interactions.

\subsection{First Look at Neutrino Properties \label{subsec:katrin}}
The leptons are all spin-$\cfrac{1}{2}$ particles that we idealize as pointlike, in light of experimental evidence that no internal structure can be discerned at a resolution $\lesssim \hbox{ few}\times 10^{-17}\cm$. What we know of their masses and lifetimes is gathered in Table~\ref{tab:leptons}. The \textit{kinematically determined} neutrino masses are consistent with zero; as we shall see in the following \S~\ref{subsec:osc}, the observation of neutrino oscillations imply that the neutrinos have nonzero, and unequal, masses. The preferred reaction for measuring the mass of the neutrino (mixture) associated with the electron is tritium $\beta$-decay,
\begin{equation}
^3\mathrm{H} \to \;^3\mathrm{He} \;\;e^- \;\bar{\nu}_e\;,
\label{eq:tritbeta}
\end{equation}
for which the endpoint energy is $Q \approx 18.57\kev$. Sources of the spectral distortions that limited the sensitivity of early experiments are absent in modern experiments using free tritium. Nevertheless, detecting a small neutrino mass is enormously challenging: the fraction of counts in the beta spectrum for a massless neutrino that lie beyond the endpoint associated with a 1-eV neutrino is but $2 \times 10^{-13}$ of the total decay rate. The KATRIN experiment~\cite{Drexlin}, which scales up the intensity of the tritium beta source as well as the size and precision of previous experiments by an order of magnitude, is designed to measure the mass of the electron neutrino directly with a sensitivity of $0.2\ev$. 

Massless neutrinos are stable, but massive neutrinos might decay. Over a distance $L$, decay would deplete the flux of extremely relativistic neutrinos of energy $E$, mass $m$, and lifetime $\tau$  by the factor $e^{-L/\gamma c\tau} = \exp{\left(-\frac{L}{Ec}\cdot\frac{m}{\tau}\right)}$, where $c$ is the speed of light and $\gamma$ is the Lorentz factor. A limit on depletion thus implies a bound on the reduced neutrino lifetime, $\tau/m$. The most stringent such bound, derived from solar $\gamma$- and x-ray fluxes, applies for radiative neutrino decay,  $\tau/m > 7 \times 10^9\sev$~\cite{PhysRevD.31.3002,Mirizzi:2007jd}. The present bound on \textit{nonradiative} decays, deduced from the survival of solar neutrinos, is far less constraining: $\tau/m \gtrsim 10^{-4}\sev$~\cite{Beacom:2002cb}. We shall have more to say about probing neutrino instability in \S~\ref{subsec:nudk}.

\subsection{Evidence for Neutrino Oscillations \label{subsec:osc}}
If neutrinos are massless, we have the freedom to identify the mass eigenstates with flavor eigenstates, so the leptonic weak interactions are flavor preserving: $W^- \to \ell^- \bar{\nu}_{\ell}$ and $Z \to \nu_{\ell}\bar{\nu}_{\ell}$, where $\ell = e, \mu, \tau$. A neutrino that moves at the speed of light cannot change character between production and subsequent interaction, so massless neutrinos do not mix.

Time passes for massive neutrinos, which do not move at the speed of light. If neutrinos of definite flavor ($\nu_e, \nu_{\mu}, \nu_{\tau}$) are superpositions of different mass eigenstates ($\nu_1, \nu_2, \nu_3$), the mass eigenstates evolve in time with different frequencies  and so the superposition changes in time: a beam created as flavor $\nu_{\alpha}$ evolves into a flavor mixture. The essential phenomenological framework is well known;\footnote{One convenient reference for this audience is Boris Kayser's course at the 2004 SLAC Summer Institute~\cite{Kayser:2005cd}. The Nobel lectures of Ray Davis~\cite{Davis:2003kh} and Masatoshi Koshiba~\cite{Koshiba:2003xy} are good sources for the history of neutrino oscillation studies. Strumia \& Vissani's protobook~\cite{Strumia:2006db} contains a wealth of experimental information and analysis.} we will review just enough to put the observations in context. We achieve an adequate orientation by simplifying to the case of two families.

Suppose that two flavor eigenstates $\nu_{\alpha}$ and $\nu_{\beta}$ are superpositions of the mass eigenstates $\nu_i$ and $\nu_j$, such that
\begin{equation}
\nu_{\alpha} = \nu_i\cos\theta + \nu_j\sin\theta;\;\;\nu_{\beta} = -\nu_i\sin\theta + \nu_j\cos\theta\;.
\label{eq:mix2}
\end{equation}
The mixing angle $\theta$ should be predicted by an eventual theory of fermion masses; for now, it is to be determined experimentally.

\begin{figure}[tb]
\begin{center}
\centerline{\includegraphics[height=0.275\textheight]{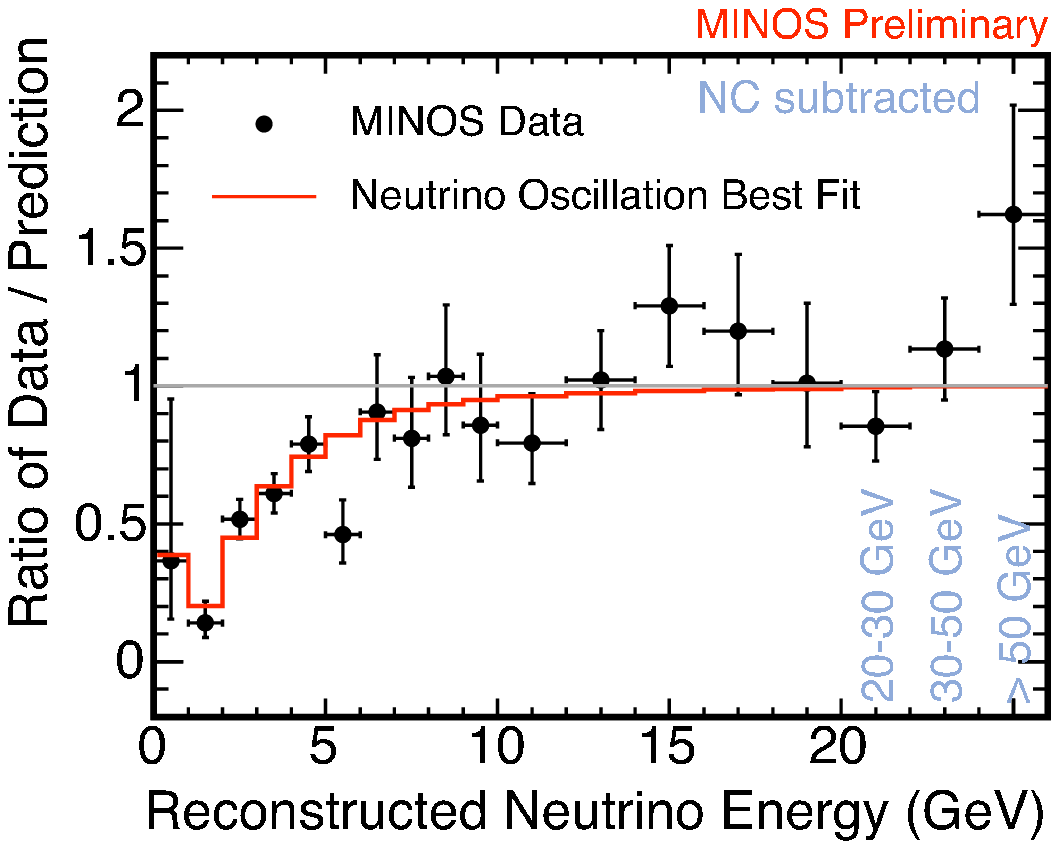} \quad \includegraphics[height=0.275\textheight]{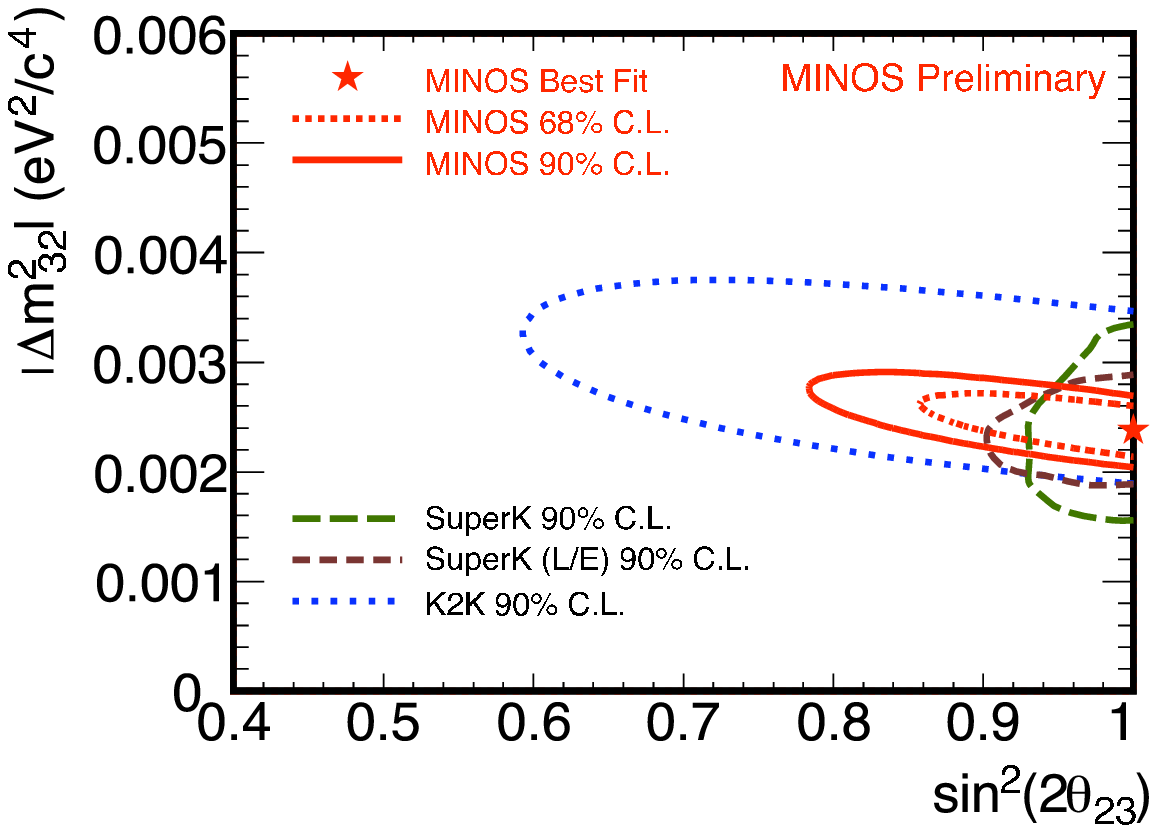}}
\caption{Left pane: Ratio of the spectrum of neutrino-induced events in the MINOS far detector, with neutral-current events subtracted, to the null-oscillation prediction (points). The best-fit oscillation expectation is overlaid as the solid red curve. Right pane: preliminary MINOS best fit point (star), 68\% and 90\% CL contours (red), compared with 90\% 
CL contours determined in the Super-Kamiokande zenith angle~\cite{Ashie:2005ik} and $L/E$ analyses~\cite{Ashie:2004mr}, as well as that 
from the K2K experiment~\cite{Ahn:2006zz} [Graphs from Ref.~\cite{MINOS:2007zz}]
\label{fig:MINOS}}
\end{center}
\end{figure}

After propagating over a distance $L$, a beam created as $\nu_{\alpha}$ with energy $E$ has a probability to mutate into $\nu_{\beta}$ given by
\begin{equation}
P_{\alpha \to \beta} = \sin^2 2\theta \sin^2\left(\Delta m^2 L/4E\right)\;,
\end{equation}
where $\Delta m^2 = m_j^2 - m_i^2$. Extending the observations of the KamiokaNDE experiment~\cite{Hirata:1992ku}, Super-K has produced very compelling evidence~\cite{Ashie:2005ik} that $\nu_\mu$ produced in the atmosphere disappear (into other flavors, dominantly $\nu_\tau$) during propagation over long distances. Their evidence, in the form of a zenith-angle distribution and the $L/E$ plot, has been confirmed and refined by the long-baseline accelerator experiments K2K~\cite{Ahn:2006zz} and MINOS~\cite{Michael:2006rx,MINOS:2007zz}. The most recent measurements come from MINOS, some $735\km$ distant from Fermilab at the Soudan mine in Minnesota. The left pane of Figure~\ref{fig:MINOS} shows how their yield of $\nu_\mu$ events, compared to the no-oscillation expectation, varies with beam energy at fixed baseline---just as anticipated in the oscillation scenario. The right pane of Figure~\ref{fig:MINOS} summarizes the constraints on the mixing angle and mass-squared difference from Super-K, K2K, and MINOS. Current evidence is consistent with maximal mixing and $|\Delta m^2|\approx 2.5 \times 10^{-3}\ev^2$. 

No appreciable oscillation of atmospheric $\nu_e$ has been observed, but flavor change has been observed in neutrinos created in the Sun. Analysis of the solar neutrino experiments is a bit involved, because neutrinos experience matter effects during their journey outward from the production region. The upshot is that the $E_\nu \approx 10\mev$ \be neutrinos emerge as a nearly pure $\nu_2$ mass eigenstate. Thus, they do not oscillate during the passage from Sun to Earth; the flavor change has already happened within the Sun. The lower energy $pp$ ($E_\nu \approx 200\kev$) and \bes ($E_\nu \approx 900\kev$) neutrinos are not strongly affected by matter in the Sun, and do undergo oscillations on their Sun--Earth trajectory. The KamLAND reactor experiment~\cite{Eguchi:2002dm,Araki:2004mb,KamLAND:2008ee} has seen a signal for vacuum neutrino oscillations of $\bar{\nu}_e$ that supports and refines the interpretation of the solar neutrino experiments. The Borexino experiment~\cite{Arpesella:2007xf} has for the first time detected the \bes neutrinos, again supporting the oscillation interpretation and parameters.

A deficit of solar neutrinos observed as $\nu_e$, compared with the expectations of the standard solar model, had been a feature of data for some time. Ruling in favor of the oscillation (neutrino-flavor-change) hypothesis, as opposed to a defective solar model, required the combination of several measurements to demonstrate that the missing $\nu_e$ are present as $\nu_\mu$ and $\nu_\tau$ arriving from the Sun. The solar neutrinos are not energetic enough to initiate $\nu_\mu \to \mu$ or $\nu_\tau \to \tau$ transitions, but indirect means were provided by the Sudbury Neutrino Observatory, SNO, a heavy-water $\mathrm{D}_2\mathrm{O}$ Cherenkov detector. The deuteron target makes it possible to distinguish three kinds of neutrino interactions sensitive to differently weighted mixtures of $\nu_e$, $\nu_\mu$. and $\nu_\tau$. The charged-current deuteron dissociation [CC] reaction,  
\begin{equation}
\nu_e \mathrm{d} \to e^-\; p\; p \;,
\label{eq:CC}
\end{equation}
proceeds by $W$-boson exchange, and is sensitive only to the $\nu_e$ flux. Neutral-current dissociation [NC],
\begin{equation}
\nu_{\ell} \mathrm{d} \to \nu_{\ell}\;p\;n\;,
\label{eq:NC}
\end{equation}
proceeds by $Z$ exchange, and is sensitive to the total flux of active neutrino species  $\nu_e + \nu_\mu+ \nu_\tau$. Elastic scattering from electrons in the target [ES] is sensitive to a weighted average  $\approx \nu_e + \cfrac{1}{7}(\nu_{\mu}+\nu_{\tau})$ of the active-neutrino fluxes, as we can see by inspecting the cross sections
\begin{eqnarray}
    \sigma(\nu_{\mu,\tau}e \to \nu_{\mu,\tau}e) & = &
    \displaystyle{\frac{G_{\mathrm{F}}^{2}m_{e}E_{\nu}}{2\pi}} \left[L_{e}^{2} + 
    R_{e}^{2}/3\right]  \quad(Z\hbox{-exchange only})
    \nonumber  \\
    \sigma(\nu_{e}e \to \nu_{e}e) & = &
    \displaystyle{\frac{G_{\mathrm{F}}^{2}m_{e}E_{\nu}}{2\pi}} \left[(L_{e}+2)^{2} + 
    R_{e}^{2}/3\right]   \;(W+Z\hbox{-exchange})
    \label{eq:ES}  
\end{eqnarray}
where the $Ze\bar{e}$ chiral couplings are 
{$L_{\ell} = 2 \sin^{2}\theta_{W} - 1 \approx -\cfrac{1}{2}$ and 
    $R_{\ell} = 2 \sin^{2}\theta_{W}  \approx \cfrac{1}{2}$}. This reaction can be studied as well in ordinary water-Cherenkov detectors.
    
\begin{figure}[tb]
\centerline{\includegraphics[width=4in]{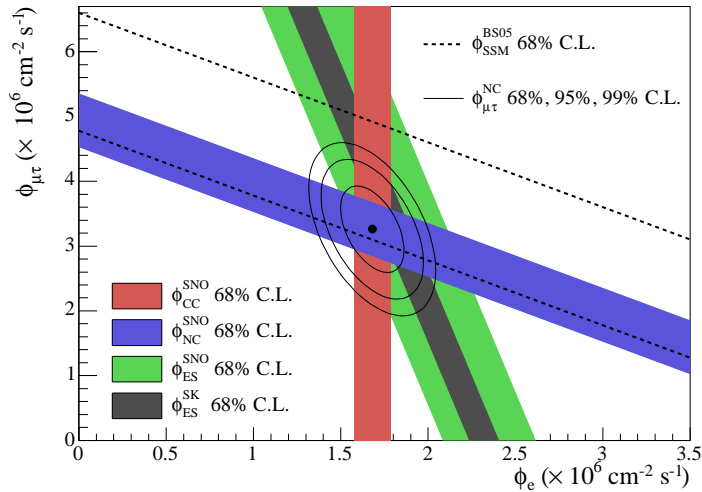}}
\caption{\label{fig:salt_flavor} 
Flux of $\mu+\tau$ neutrinos versus
 flux of electron neutrinos determined in solar neutrino experiments.  The filled bands indicate the CC, NC and ES flux measurements.  The Standard Solar Model~\cite{Bahcall:2004pz} predicts that the total \be solar neutrino flux lies between the dashed lines. The total flux measured with the NC channel is shown as the solid (blue) band parallel to the model prediction.  The Super-Kamiokande ES
 result~\cite{Fukuda:2002pe} is the dark narrow band within the (green) SNO ES band.  The intercepts of the experimental bands with the  axes represent $\pm 1\sigma$ uncertainties.   The point represents $\phi_e$ from the CC flux (red band) and
 $\phi_{\mu\tau}$ from the NC-CC difference; the contours  represent  68\%, 95\%, and 99\%
 C.L. [From Ref.~\cite{Aharmim:2005gt}]}
\end{figure}

Super-K and SNO observations~\cite{Fukuda:2001nj,Ahmad:2001an,Ahmad:2002jz} are summarized in Figure~\ref{fig:salt_flavor}.
Taken together, they indicate that the \textit{total} flux agrees with solar model, but only 30\% of neutrinos arrive from the Sun as $\nu_e$. The nonzero value
 of $\phi_{\mu\tau}$ provides strong evidence that neutrinos created as $\nu_e$ are transformed into other active flavors. All the evidence is consistent with the conclusion that $\nu_e$ is dominantly a mixture of two mass eigenstates, designated $\nu_1$ and $\nu_2$, with a ``solar'' mass-squared difference $\Delta m_\odot^2 = m_2^2 - m_1^2 \approx 7.9 \times 10^{-5}\ev^2$ and mixing angle $\sin^2\theta_{\odot} \approx 0.3$. No oscillation phenomena have yet been established beyond the ``atmospheric'' and ``solar'' sectors~\cite{AguilarArevalo:2007it}.
 
\begin{figure}[bt]
\begin{center}
\centerline{\includegraphics[width=0.45\textwidth]{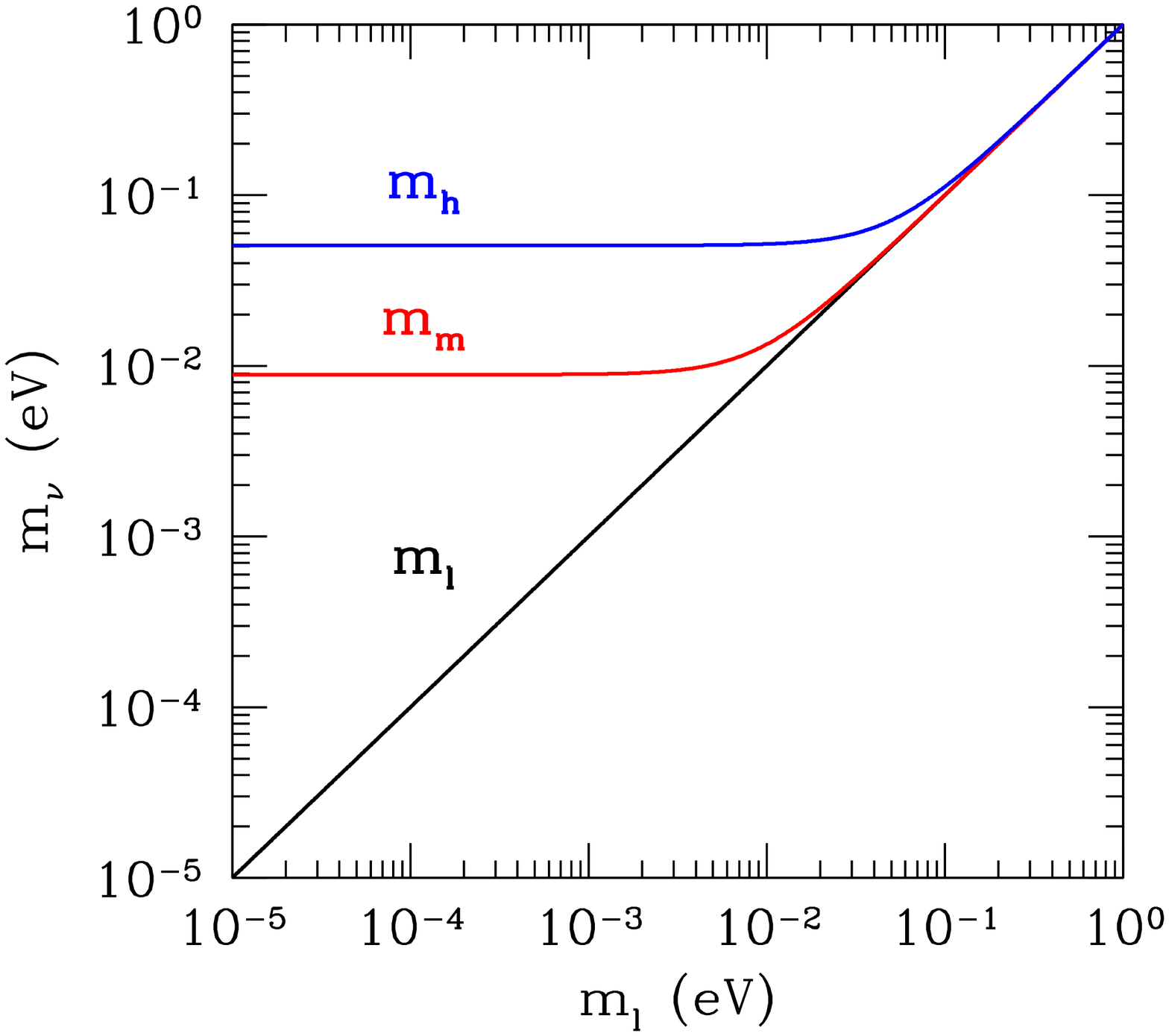} \quad \includegraphics[width=0.45\textwidth]{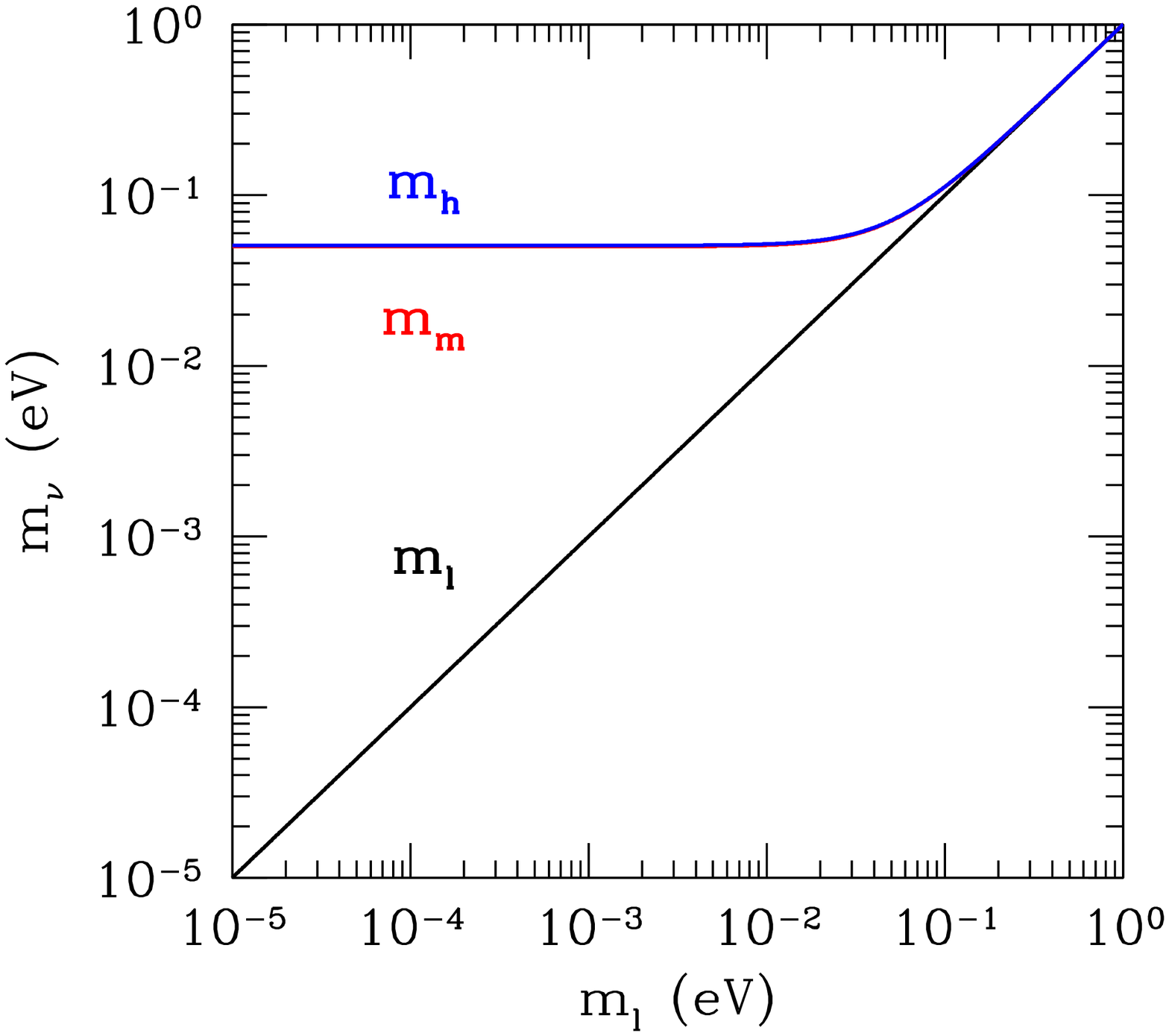}}
\caption{Favored values for the light, medium, and heavy neutrino masses $m_{\ell}$, $m_{\mathrm{m}}$, $m_{\mathrm{h}}$, as functions of the lightest neutrino mass  in the three-neutrino oscillation scenario for the normal (left pane) and inverted hierarchy (right pane). We take the solar mass-squared difference to be $\Delta m^2_{\odot} = m_2^2 - m_1^2 = 7.9\times 10^{-5}\ev^2$, and the atmospheric
$ \Delta m^2_{\mathrm{atm}} = |m_3^2 - m_1^2| = 2.5\times 10^{-3}\ev^2$.
\label{fig:neutrinomasses}}
\end{center}
\end{figure}

 The atmospheric and solar neutrino experiments, with their reactor and accelerator complements, have partially characterized the neutrino spectrum in terms of a closely spaced solar pair $\nu_1$ and $\nu_2$, where $\nu_1$ is taken by convention to be the lighter member of the pair, and a third neutrino, more widely separated in mass. We do not yet know whether $\nu_3$ lies above (``normal hierarchy'') or below (``inverted hierarchy'') the solar pair, and experiment has not yet set the absolute scale of neutrino masses. Figure~\ref{fig:neutrinomasses} shows the normal and inverted spectra as functions of assumed values for the mass of the lightest neutrino.

\subsection{Character of the Relic Neutrino Background 
\label{subsec:char}}
The cosmic microwave background is characterized by a Bose--Einstein blackbody 
distribution of photons (per unit volume)\footnote{The review article by Steigman~\cite{Steigman:1979kw} is a good introduction to this subject.  We adopt units such 
that $\hbar = 1 = c$, and we will measure temperature in kelvins or 
electron volts, as appropriate to the situation. The conversion factor 
is Boltzmann's constant, $k = 8.617343 \times 10^{-5}\ev\K^{-1}$.}
\begin{equation}
    \frac{dn_{\gamma}(T)}{d^{3}p} = \frac{1}{(2\pi)^{3}} \,
    \frac{1}{\exp{(p/T)} - 1}\; ,
    \label{eq:bbphotons}
\end{equation}
where $p$ is the relic momentum and $T$ is the 
temperature of the photon ensemble. The number density of photons throughout 
the Universe is
\begin{equation}
    n_{\gamma}(T) = \frac{1}{(2\pi)^{3}} \!\!\int \!\!d^{3}p \;
    \frac{1}{\exp{(p/T)} - 1} = \frac{2\zeta(3)}{\pi^{2}}\, 
    T^{3},
    \label{eq:bbgamnum}
\end{equation}
where $\zeta(3) \approx 1.20205$ is Riemann's zeta function. In the 
present Universe, with a photon temperature $T_{0} = 
(2.725 \pm 0.002)\K$~\cite{Bennett:2003bz}, the photon density is
\begin{equation}
    n_{\gamma0} \equiv n_{\gamma}(T_{0}) \approx 410\cm^{-3}\;.
    \label{eq:numgam}
\end{equation}

The present photon density provides a benchmark for other big-bang 
relics. The essential observation is that neutrinos decoupled when 
the cosmic soup cooled to around $1\mev$, so did not share in the 
energy released when electrons and positrons annihilated at $T \approx 
m_{e}$, the electron mass. Applying entropy conservation and counting relativistic degrees 
of freedom, it follows that the ratio of neutrino and photon 
temperatures (below $m_{e}$) is
\begin{equation}
    T_{\nu}/T = \left(\cfrac{4}{11}\right)^{\!1/3}\;,
    \label{eq:nutogam}
\end{equation}
so that the present neutrino temperature is 
\begin{equation}
    T_{\nu0} = \left(\cfrac{4}{11}\right)^{\!1/3}T_{0} = 1.945\K 
    \leadsto 1.697 \times 10^{-4}\ev\;.
    \label{eq:nutemp}
\end{equation}

The momentum distribution of relic neutrinos follows the Fermi--Dirac 
distribution (with zero chemical potential),
\begin{equation}
    \frac{dn_{\nu_{i}}(T_{\nu})}{d^{3}p} =
    \frac{dn_{\nu^{c}_{i}}(T_{\nu})}{d^{3}p} = \frac{1}{(2\pi)^{3}} \,
    \frac{1}{\exp{(p/T_{\nu})} + 1}\; .
    \label{eq:nuFD}
\end{equation}
The number distribution of relic neutrinos or antineutrinos is therefore
\begin{eqnarray}
    n_{\nu_{i}}(T_{\nu}) & =  &   \frac{1}{(2\pi)^{3}} \!\!\int \!\!d^{3}p \;
    \frac{1}{\exp{(p/T_{\nu})} + 1} \nonumber\\ & = &  \frac{3\zeta(3)}{4\pi^{2}}  \, 
    T_{\nu}^{3} = \cfrac{3}{22} n_{\gamma}(T)\;.
    \label{eq:numbnu} 
\end{eqnarray}
In the present Universe, the number density of each (active) neutrino or antineutrino
species is\footnote{Unconventional neutrino histories can alter this expectation. These include lepton asymmetries in the early universe, neutrino clustering, and a neutrinoless universe. For a brief survey, with references, see \S{V} of Ref.~\cite{Barenboim:2004di}.}
\begin{equation}
    n_{\nu_{i}0}  \equiv n_{\nu_{i}}(T_{\nu0}) 
    \approx 56\cm^{-3}\;, 
    \label{eq:numbnunow}
\end{equation}
plus a 1\% correction from reheating effects detailed in~\cite{Lopez:1998aq}.
The mean momentum of relic neutrinos today is
\begin{equation}
    \langle p_{\nu0} \rangle = \frac{7}{2} \, 
    \frac{\zeta(4)}{\zeta(3)} \cdot T_{\nu0} \approx 3.151 T_{\nu0}
    \approx 5.314 \times 10^{-4}\ev\;,
    \label{eq:numom}
\end{equation}
where we have used $\zeta(4) = \pi^{4}/90 = 1.08232$.  In the same way, 
the mean-squared neutrino momentum is given by
\begin{equation}
\label{eqn:p2mean}
\langle p^{2}_{\nu 0} \rangle = 15\,\frac{\zeta(5)}{\zeta(3)} \cdot 
T^{2}_{\nu 0} \approx
12.94\, T^{2}_{\nu 0}\;,
\end{equation}
so that
\begin{equation}
    \langle p_{\nu0}^{2} \rangle^{\cfrac{1}{2}} \approx 
    3.597\,T_{\nu0} \approx 6.044 \times 10^{-3}\ev\;.
    \label{eqn:prms}
\end{equation}

\begin{figure}[tb]
\begin{center}
\includegraphics[width=0.5\textwidth]{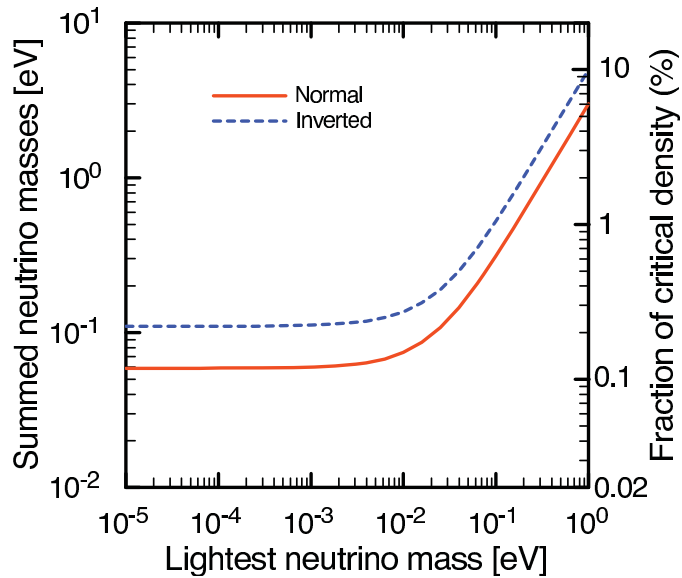}
\caption{Contributions of relic neutrinos to the mass density of the Universe, as functions of the mass of the lightest neutrino, for the normal (solid line) and inverted (dashed line) mass hierarchies.
\label{fig:neuDM}}
\end{center}
\end{figure}

With our (partial) knowledge of neutrino masses, we can estimate the contribution of neutrinos to the density of the current universe. The left-hand scale of Figure~\ref{fig:neuDM} shows the summed 
neutrino masses $m_1 + m_2 + m_3$ for the normal and inverted hierarchies, as functions of the lightest neutrino mass. The neutrino oscillation data imply that $\sum_i m_{\nu_i} \gtrsim 0.06\ev$ in the case of the normal hierarchy, and $\sum_i m_{\nu_i} \gtrsim 0.11\ev$ in the case of the inverted hierarchy.
Using the calculated number density (\ref{eq:numbnunow}), we can deduce the neutrino contribution to the mass density, expressed in units of the critical density
\begin{equation}
\rho_{\mathrm{c}} \equiv 3H_0^2/8\pi G_{\mathrm{N}} = 1.05 h^2 \times 10^4\ev\cm^{-3} = 5.6 \times 10^3\ev\cm^{-3}\;,
\label{eqn:rhocrit}
\end{equation}
where we have taken the reduced Hubble constant to be $h = 0.73$. This is measured by the right-hand scale in Figure~\ref{fig:neuDM}. We find that neutrinos contribute $\Omega_\nu \gtrsim (1.2, 2.2) \times 10^{-3}$ for the (normal, inverted) spectrum, and no more than 10\% of critical density, should the lightest neutrino mass approach $1\ev$. Combining (\ref{eqn:rhocrit}) and (\ref{eq:numbnunow}), we deduce that the condition for neutrinos not to overclose the Universe is\footnote{This line of argument may be traced to Refs.~~\cite{Gershtein:1966gg,Cowsik:1972gh,Cowsik:1973yj,Szalay:1976ef}.}
\begin{equation}
 \sum_i m_{\nu_i} \lesssim 50\ev \;,
\label{eqn:mnuconstraint}
\end{equation}
so long as neutrinos are stable on cosmological time scales and that the expected neutrino density is not erased by interactions beyond the standard electroweak theory~\cite{Beacom:2004yd}.\footnote{Lepton asymmetries in the early universe can lead to neutrino densities in the current Universe far from the standard expectation~\cite{Gelmini:2004hg}.}

Neutrinos influence fluctuations in the cosmic microwave background, affect the development of density perturbations that set the pattern of large-scale structure, and modulate the baryon acoustic oscillations. Analyses of the interplay between neutrinos and astronomical observables provide bounds on the sum of neutrino masses, many of which are compiled in Refs.~\cite{Fukugita:2005sb,Yao:2006px,NuinC,Lesgourgues:2006nd,ScottNUSS}. Depending on the richness of the data set considered and the specificity of the assumed cosmological scenario, recent inferences range from $\approx 1.5\ev$~\cite{Fogli:2004as,Tegmark:2003ud} to approximately $0.6\ev$~\cite{Spergel:2006hy, Fogli:2004as,Goobar:2006xz, Tegmark:2003ud} to $\lesssim 0.2\ev$~\cite{Goobar:2006xz,Melchiorri:2006nj,Seljak:2006bg, Cirelli:2006kt}. These provocative constraints are less secure than the bound (\ref{eqn:mnuconstraint}) derived from $\rho_{\nu} \lesssim \rho_{\mathrm{c}}$, but they are highly suggestive. It is worth noting that detection of late-time neutrino influence may imply an improved lower bound on the neutrino lifetime~\cite{Serpico:2007pt}. It will be very interesting to watch the evolving conversation between direct measurements and indirect inferences~\cite{Cooray:1999rv,Bilenky:2002aw,Tegmark:2005cy}.

\subsection{Fermion masses and mixings \label{subsec:massmix}}
In the standard electroweak theory, fermion mass arises  from gauge-invariant Yukawa interactions between the complex scalar fields $\phi$ introduced to hide the electroweak symmetry and the quarks or charged  leptons. For the leptons ($\ell$ runs over $e, \mu, \tau$), the Yukawa term is
\begin{equation}
      \lag_{{\rm Yukawa-}\ell} = -\zeta_{\ell}\left[(\overline{{\sf L}}_{\ell}\phi){\sf R}_{\ell} + 
      \overline{{\sf R}}_{\ell}(\phi^\dagger{\sf
L}_{\ell})\right]\;,
\label{eq:Yukterm}
\end{equation}
where the spinors ${{\sf L}}_{\ell}$ and ${\sf R}_{\ell}$ are defined in (\ref{eq:lleptons}) and (\ref{eq:rightlep});
a similar interaction appears for quarks.
Self-interactions among the scalars may contrive to hide the gauge symmetry.
The ``Higgs potential'' has the form
\begin{equation}
      V(\phi^\dagger \phi) = \mu^2(\phi^\dagger \phi) +
\abs{\lambda}(\phi^\dagger \phi)^2 .
\label{SSBpot}
\end{equation}
The electroweak symmetry is spontaneously broken if the parameter
$\mu^2$ is taken to be negative. In that event, gauge invariance gives us the freedom to choose the state of minimum energy---the vacuum state---to correspond to the vacuum expectation value
\begin{equation}
\vev{\phi} = \left(\begin{array}{c} 0 \\ v/\sqrt{2} \end{array}
\right),
\label{eq:vevis}
\end{equation}
where $v = \sqrt{-\mu^2/\abs{\lambda}}= (G_{\mathrm{F}}\sqrt{2})^{-1/2} \approx 246\gev$.

\begin{figure}[tb]
\begin{center}
\includegraphics[width=0.5\textwidth]{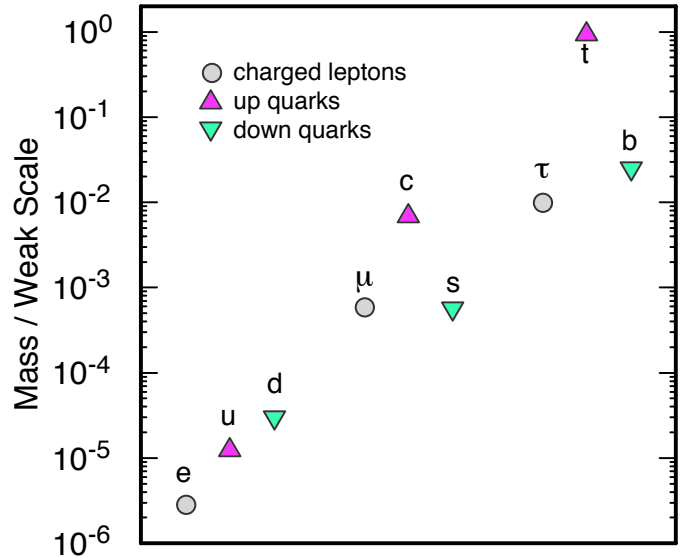}
\caption{Yukawa couplings $\zeta_i = m_i / (v/\sqrt{2})$ inferred from the masses of the quarks and charged leptons.
\label{fig:Yuk}}
\end{center}
\end{figure}

When the electroweak symmetry is spontaneously broken, the electron 
mass emerges as $m_{e} = \zeta_{e}v/\sqrt{2}$. Each fermion mass involves a distinct Yukawa coupling $\zeta$. The Yukawa couplings that reproduce the observed quark and charged lepton masses range over many orders of magnitude, as shown in Figure~\ref{fig:Yuk}.
The origin of the Yukawa couplings is obscure: they do not follow from a known symmetry principle, for example. In that sense, therefore, \textit{all fermion masses involve physics
beyond the standard model.} 

We may seek to accommodate neutrino mass\footnote{For surveys of attempts to understand neutrino masses, see Refs.~\cite{King:2003jb,Mohapatra:2004ht,Mohapatra:2006gs, RabiNUSS, King:2007nw}.} in the electroweak theory by adding to the spectrum a right-handed neutrino $N_{\mathrm{R}}$ and constructing the (gauge-invariant) Dirac mass term
\begin{equation}
\mathcal{L}_{\mathrm{D}}^{(\nu)} = -\zeta_{\nu}\left[(\bar{\mathsf{L}}_{\ell}\bar{\phi})N_{\mathrm{R}} +
\bar{N}_{\mathrm{R}}(\bar{\phi}^{\dagger}\mathsf{L}_{\ell})\right]
\to -m_{\mathrm{D}}\left[\bar{\nu}_{\mathrm{L}}N_{\mathrm{R}} +
\bar{N}_{\mathrm{R}}\nu_{\mathrm{L}}\right]\;,
\label{eqn:diracnu}
\end{equation}
where 
$\bar{\phi} = i\sigma_2 \phi^*$ is the complex conjugate of the Higgs doublet and $m_{\mathrm{D}} = \zeta_{\nu}v/\sqrt{2}$. A Dirac mass term conserves the additive lepton number $L$ that takes on the value $+1$ for neutrinos and negatively charged leptons, and $-1$ for antineutrinos and positively charged leptons. To account for the observed tinyness of the neutrino masses, $m_\nu \lesssim 1\ev$, the Yukawa couplings must be extraordinarily small, $\zeta_\nu \lesssim 10^{-11}$. Whether they are qualitatively more puzzling than the factor of $3 \times 10^5$ that separates the electron and top-quark Yukawa couplings is for now a question for intuition.

Matters of taste aside, we have another reason to consider alternatives to the Dirac mass term: unlike all the other particles that enter the electroweak theory, the right-handed neutrinos are standard-model singlets. Alone among the standard-model fermions, they might be their own antiparticles, so-called Majorana fermions. The charge conjugate of a right-handed field is left-handed,
$\psi_{\mathrm{L}}^{c} \equiv (\psi^{c})_{\mathrm{L}} =
(\psi_{\mathrm{R}})^{c}$. 
Majorana mass terms connect the left-handed and 
right-handed components of conjugate fields,
\begin{eqnarray}
-\mathcal{L}_{\mathrm{MA}} & = & A(\bar{\nu}_{\mathrm{R}}^{c}\nu_{\mathrm{L}} + 
\bar{\nu}_{\mathrm{L}}\nu_{\mathrm{R}}^{c}) = A \bar{\chi}\chi
\nonumber  \\
-\mathcal{L}_{\mathrm{MB}} & = & 
B(\bar{N}^{c}_{\mathrm{L}}N_{\mathrm{R}} + 
\bar{N}_{\mathrm{R}}N^{c}_{\mathrm{L}}) = B\bar{\omega}\omega\;.
\end{eqnarray}
The self-conjugate Majorana mass eigenstates are
\begin{eqnarray}
\chi & \equiv & \nu_{\mathrm{L}} + \nu^{c}_{\mathrm{R}} = \chi^{c}
\nonumber  \\
\omega & \equiv & N_{\mathrm{R}} + N^{c}_{\mathrm{L}} = \omega^{c} \;.
\end{eqnarray}
A Majorana fermion cannot carry any additive quantum number.
The mixing of particle and antiparticle fields means that the Majorana 
mass terms correspond to processes that violate lepton number by two 
units. Accordingly, the exchange of a Majorana neutrino can mediate neutrinoless 
double beta decay, $(Z,A) \rightarrow (Z+2,A) + e^{-} + e^{-}$. 
Detecting neutrinoless double beta decay~\cite{Elliott:2002xe,Avignone:2007fu,Piquemal,VogelPiepkeBB} would offer decisive 
evidence for the Majorana nature of the neutrino.

The mass of the active $\nu_{\mathrm{L}}$ may be generated by a Higgs triplet that acquires a vacuum expectation value~\cite{Gelmini:1980re}, or by an effective operator that involves two Higgs doublets combined to transform as a triplet~\cite{Weinberg:1979sa}.

It is interesting to consider both Dirac and Majorana terms, and specifically to examine the case in which Majorana masses corresponding to an active state $\chi$ and a sterile state $\omega$ arise from weak triplets and singlets, respectively, with masses $M_3$ and $M_1$. The neutrino mass matrix then has the form
\begin{equation}
(\begin{array}{lr}
\bar{\nu}_{\mathrm{L}} & \bar{N}^c_{\mathrm{L}}
\end{array})
\left(
\begin{array}{cc}
M_3 & m_{\mathrm{D}} \\
m_{\mathrm{D}} & M_1
\end{array}
\right)
\left(
\begin{array}{c}
\nu^c_{\mathrm{R}} \\ N_{\mathrm{R}}
\end{array}
\right)\;.
\label{eqn:mixedmass}
\end{equation}
In the highly popular seesaw limit~\cite{Minkowski:1977sc,Yanagida:1979as,Gell-Mann:1980vs,Mohapatra:1979ia,Schechter:1980gr}, with $M_3 = 0$ and $m_{\mathrm{D}} \ll M_1$, diagonalizing the mass matrix (\ref{eqn:mixedmass}) yields two Majorana neutrinos,
\begin{equation}
n_{1\mathrm{L}}  \approx  \nu_{\mathrm{L}} - \frac{m_{\mathrm{D}}}{M_1}N^c_{\mathrm{L}} \qquad
n_{2\mathrm{L}}  \approx  N^c_{\mathrm{L}} + \frac{m_{\mathrm{D}}}{M_1}\nu_{\mathrm{L}}\;,
\label{eqn:majopair}
\end{equation}
with masses
\begin{equation}
m_1 \approx  \frac{m_{\mathrm{D}}^2}{M_1} \ll m_{\mathrm{D}} \qquad m_2 \approx M_1\;.
\label{eqn:majomasses}
\end{equation}
The seesaw produces one very heavy ``neutrino'' and one neutrino much lighter than a typical quark or charged lepton. Many alternative explanations of the small neutrino masses have been explored in the literature~\cite{Smirnov:2004hs}, including some in which collider experiments exploring the Fermi scale could reveal the origin of neutrino masses~\cite{Chen:2006hn}.

The charged-current interactions among the left-handed leptonic mass eigenstates $\bm{\nu} = (\nu_1, \nu_2, \nu_3)$ and
$\bm{\ell}_{\mathsf{L}} = (e_{\mathsf{L}}, \mu_{\mathsf{L}}, \tau_{\mathsf{L}})$  are specified by 
\begin{equation}
\mathcal{L}_{\mathsf{CC}}^{(q)} = - \frac{g}{\sqrt{2}}\, \bar{\bm{\nu}}\,\gamma^\mu {\mathcal{V}^{\dagger}}\bm{\ell}_{\mathsf{L}}W_\mu^+ + \hbox{ h.c.}\;,
\label{eqn:leptoncc}
\end{equation}
where the neutrino mixing matrix~\cite{Lee:1977ti}, sometimes called the Pontecorvo~\cite{Pontecorvo:1957qd}--Maki-Nakagawa-Sakata~\cite{Maki:1962mu} (PMNS) matrix in tribute to neutrino-oscillation pioneers, is
\begin{equation}
\mathcal{V} = \left(
\begin{array}{ccc}
\mathcal{V}_{e1} & \mathcal{V}_{e2} & \mathcal{V}_{e3} \\
\mathcal{V}_{\mu1} & \mathcal{V}_{\mu2} & \mathcal{V}_{\mu3} \\
\mathcal{V}_{\tau1} & \mathcal{V}_{\tau2} & \mathcal{V}_{\tau3}
\end{array}
\right)\;.
\label{eqn:pmns}
\end{equation}

A recent global fit~\cite{Gonzalez-Garcia:2004jd} yields the following ranges for the magnitudes of the neutrino mixing matrix elements:
\begin{equation}
    |\mathcal{V}| =\left(
\begin{array}{ccc}
0.79-0.88&0.47-0.61&<0.20 \\
0.19-0.52& 0.42-0.73&0.58-0.82\\
0.20-0.53&0.44-0.74&0.56-0.81
\end{array}
\right)\; .
\end{equation}
It is conventional to factor the neutrino mixing matrix as 
\begin{equation}
\mathcal{V} = \left[\begin{array}{ccc}
1 & 0 & 0 \\
0 & c_{23} & s_{23} \\
0 & -s_{23} & c_{23}
\end{array} \right]
\left[\begin{array}{ccc}
c_{13} & 0 & s_{13}e^{-i\delta} \\
0 & 1 & 0 \\
s_{13}e^{i\delta} & 0 & c_{13}
\end{array} \right]
\left[\begin{array}{ccc}
c_{12} & s_{12} & 0 \\
-s_{12} & c_{12} & 0 \\
0 & 0 &1
\end{array} \right]
\end{equation}
where we abbreviate $s_{ij} = \sin\theta_{ij}$, $c_{ij} = \cos\theta_{ij}$, and $\delta$ is a \textsf{CP}-violating phase.

\begin{figure}[tb]
\begin{center}
\centerline{\includegraphics[width=0.45\textwidth]{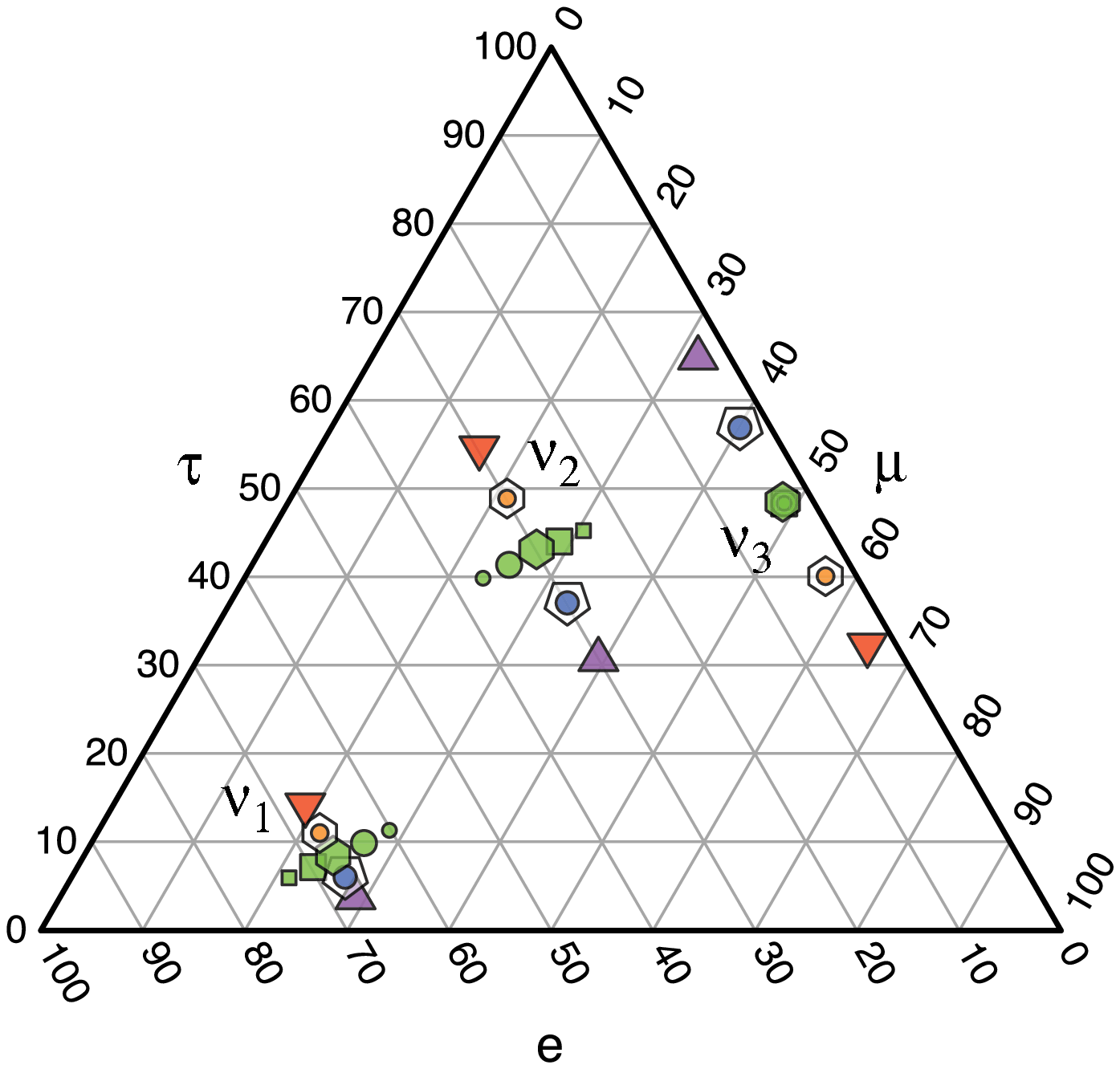} \quad\includegraphics[width=0.45\textwidth]{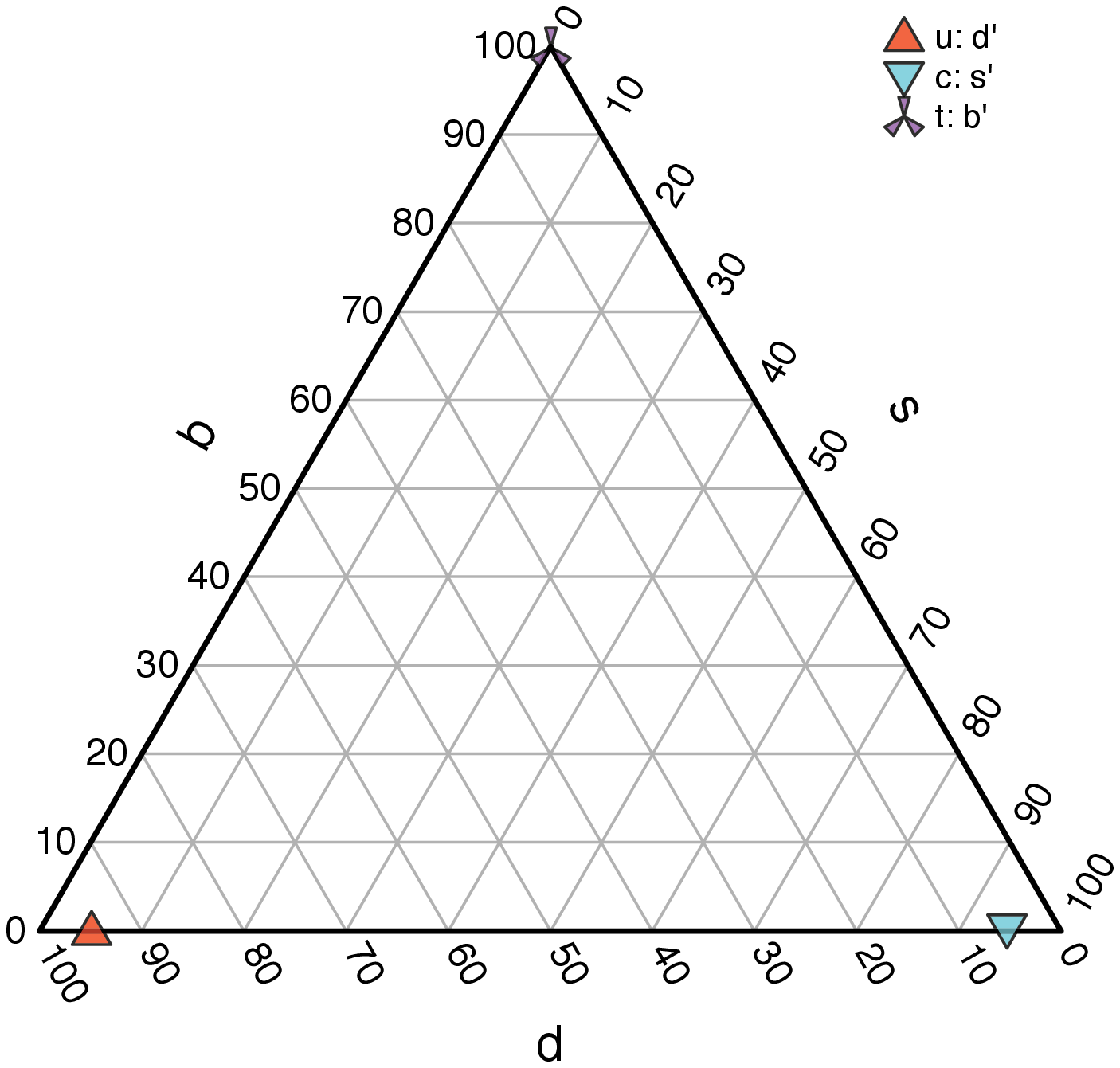}}
\caption{Left pane: $\nu_e, \nu_\mu, \nu_\tau$ flavor content of the neutrino mass eigenstates $\nu_1, \nu_2, \nu_3$. The green hexagons denote central values, with $\delta = 0$ and $\theta_{13} = 10^\circ$. Variations in the atmospheric angle $\theta_{23}$ are indicated by the points arrayed roughly parallel to the $\mu$ scale. Variations in the solar angle $\theta_{12}$ are depicted by the green symbols arrayed roughly perpendicular to the $\mu$ scale. Right pane: $d, s, b$ composition of the quark flavor eigenstates $d^\prime$ ($\bigtriangleup$), $s^\prime$ ($\bigtriangledown$), $b^\prime$ (tripod).
\label{fig:flavormixing}}
\end{center}
\end{figure}

Global fits~\cite{Gonzalez-Garcia:2004jd,GonzalezGarcia:2007ib,Maltoni:2004ei} restrict the mixing parameters to the ranges $30^\circ \lesssim \theta_{12} \lesssim 38^\circ$ for solar oscillations,\footnote{The latest KamLAND analysis~\cite{KamLAND:2008ee} tightens this constraint to $33^\circ \lesssim \theta_{12} \lesssim 36^\circ$.} $35^\circ \lesssim \theta_{23} \lesssim 55^\circ$ for atmospheric oscillations, 
$\theta_{13} \lesssim 10^\circ$, and leave $\delta$ unconstrained. These parameter ranges lead to the flavor content of the neutrino mass eigenstates depicted in the left pane of Figure~\ref{fig:flavormixing}, 
where central values (fixing $\delta = 0$ and $\theta_{13} = 10^\circ$) are indicated by the green hexagons. We observe that  $\nu_3$ consists of nearly equal parts of $\nu_{\mu}$ and $\nu_{\tau}$, perhaps with a trace of $\nu_e$, while $\nu_2$ contains similar amounts of $\nu_e$, $\nu_{\mu}$, and $\nu_{\tau}$, and $\nu_1$ is rich in $\nu_e$, with approximately equal minority parts of $\nu_{\mu}$ and $\nu_{\tau}$. The observed structure of the neutrino mixing matrix differs greatly from the pattern of the more familiar (Cabibbo--Kobayashi--Maskawa) quark mixing matrix, which is displayed graphically in the right pane of Figure~\ref{fig:flavormixing}.

\section{NEUTRINO OBSERVATORIES \label{sec:nuobs}}
\subsection{Generalities \label{subsec:gener}}

An early goal of the next generation of neutrino telescopes will be to
detect the flux of cosmic neutrinos that we believe will begin to show
itself above the atmospheric-neutrino background at energies of a few
TeV~\cite{Ackermann:2007km}. A short summary of the science program of these instruments is to
prospect for cosmic-neutrino sources, to characterize those sources, to
study neutrino properties, and to be sensitive to new phenomena in
particle physics~\cite{Han:2004kq,Albuquerque:2006fd,KaraSSI}.  The expected sources include active galactic nuclei
(AGN) at typical distances of roughly $100\mpc
\approx 3.1 \times 10^{24}\m$. If neutrinos are produced there in the 
decay of pions created in $pp$ or $p\gamma$ collisions, then we 
anticipate---at the source---equal numbers of neutrinos and 
antineutrinos, with a flavor mix $2 \gamma + 2 \nu_{\mu} + 
2\bar{\nu}_{\mu} + 1 \nu_e + 1 \bar{\nu}_e$, provided that all pions 
and their daughter muons decay. I  denote this standard flux at 
the source by
$\Phi^{0}_{\mathrm{std}} = \{\varphi_{e}^{0} = \cfrac{1}{3},
    \varphi_{\mu}^{0} = \cfrac{2}{3}, \varphi_{\tau}^{0} =
    0\}$.

We expect that a neutrino observatory with an instrumented volume of
$1\km^{3}$ will be able to survey the cosmic-neutrino flux over a broad
range of energies, principally by detecting the charged-current
interaction $(\nu_{\mu},\bar{\nu}_{\mu})N \rightarrow
(\mu^-,\mu^+)+\hbox{anything.}$ Important open questions are whether we
can achieve efficient, calibrated $(\nu_e, \bar{\nu}_e)$ and
$(\nu_{\tau},\bar{\nu}_{\tau})$ detection, and whether we can record
and determine the energy of neutral-current events.  Adding these capabilities will enhance
the scientific potential of neutrino observatories.
\subsection{Deeply inelastic scattering \label{subsec:dis}}
The cross section for deeply inelastic scattering on an isoscalar 
nucleon may be written in terms of the Bjorken scaling variables 
$x = Q^2/2M\nu$  and $y = \nu/E_\nu$ as
\begin{equation}
    \frac{d^2\sigma}{dxdy} = \frac{2 G_F^2 ME_\nu}{\pi} 
    {\left( 
    \frac{M_W^2}{Q^2 + M_W^2} \right)^{\!2}} \left[xq(x,Q^2) + x 
    \bar{q}(x,Q^2)(1-y)^2 \right]  \;,
    \label{eqn:sigsig}
\end{equation}
where $-Q^2$ is the invariant momentum transfer between
the incident
neutrino and outgoing muon, $\nu = E_\nu - E_\mu$ is the energy loss in
the lab (target) frame, $M$ and $M_W$ are the nucleon and
intermediate-boson masses, and $G_F = 1.16632 \times 10^{-5}\gev^{-2}$ is
the Fermi
constant. The parton densities are
\begin{eqnarray}
q(x,Q^2) & = & \frac{u_v(x,Q^2)+d_v(x,Q^2)}{2} +
\frac{u_s(x,Q^2)+d_s(x,Q^2)}{2} \nonumber\\ & & + s_s(x,Q^2) + b_s(x,Q^2)  
 \\[12pt]
	\bar{q}(x,Q^2) & = & \frac{u_s(x,Q^2)+d_s(x,Q^2)}{2} + c_s(x,Q^2) + 
	t_s(x,Q^2),\nonumber 
\end{eqnarray}
where the subscripts $v$ and $s$ label valence and sea contributions,
and $u$, $d$, $c$, $s$, $t$, $b$ denote the distributions for various
quark flavors in a {\em proton}.  The $W$-boson propagator, which has a
negligible effect at low energies, modulates the high-energy cross
section and has important consequences for the way the cross section is
composed.

I was drawn to this problem by the
observation~\cite{Andreev:1979cp} that the $W$-boson propagator squeezes
the significant contributions of the parton distributions toward
smaller values of $x$ with increasing energy.  There the QCD-induced
growth of the small-$x$ parton distribution enhances the high-energy
cross section.  This stands in contrast to the familiar effect of QCD
evolution at laboratory energies, which is to diminish the total cross
section as the valence distribution is degraded at high values of
$Q^{2}$. At that moment, my colleagues and I had developed for our 
study of supercollider physics~\cite{Eichten:1984eu} the first all-flavor set of parton 
distributions appropriate for applications at small $x$ and large 
$Q^{2}$, so I had in my hands everything needed for a modern 
calculation of the charged-current cross section at ultrahigh 
energies. In a sequence of works on the 
problem~\cite{Quigg:1986mb,Reno:1987zf,Gandhi:1995tf,Gandhi:1998ri}, we 
have tracked the evolving experimental understanding of parton distributions 
and investigated many facets of ultrahigh-energy neutrino interactions.

Let us recall some of the principal lessons.  The valence contribution, which dominates at
laboratory energies, becomes negligible above about $10^{16}\ev$,
whereas strange- and charm-quark contributions become significant.  Contributions from small values of $x$ become increasingly prominent as
the neutrino energy increases.  At $E_{\nu} = 10^{5}\gev$, nearly all of
the cross section comes from $x \gtrsim 10^{-3}$, but by $E_{\nu} =
10^{9}\gev$, nearly all of the cross section lies below $x = 10^{-5}$, 
where we lack direct experimental information. Reno has given 
a comprehensive review of small-$x$ uncertainties and the possible influence of 
new phenomena on the total cross section~\cite{Reno:2004cx}.\footnote{For a new calculation based on the ZEUS-S global fits and incorporating heavy-quark threshold corrections, see Ref.~\cite{CooperSarkar:2007cv}.}

\begin{figure}[tb]
\begin{center}
\centerline{\includegraphics[height=0.31\textheight]{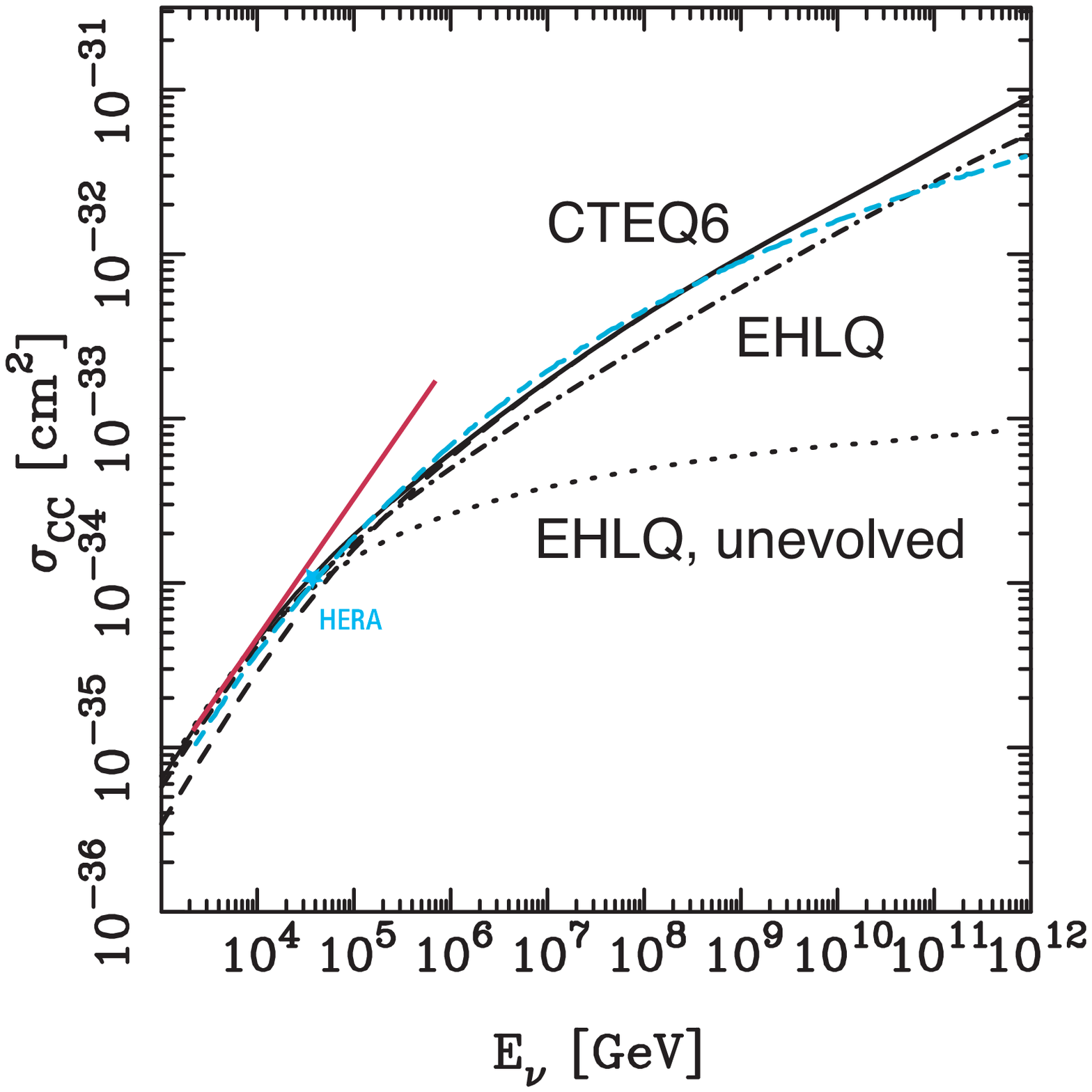}
\includegraphics[height=0.31\textheight]{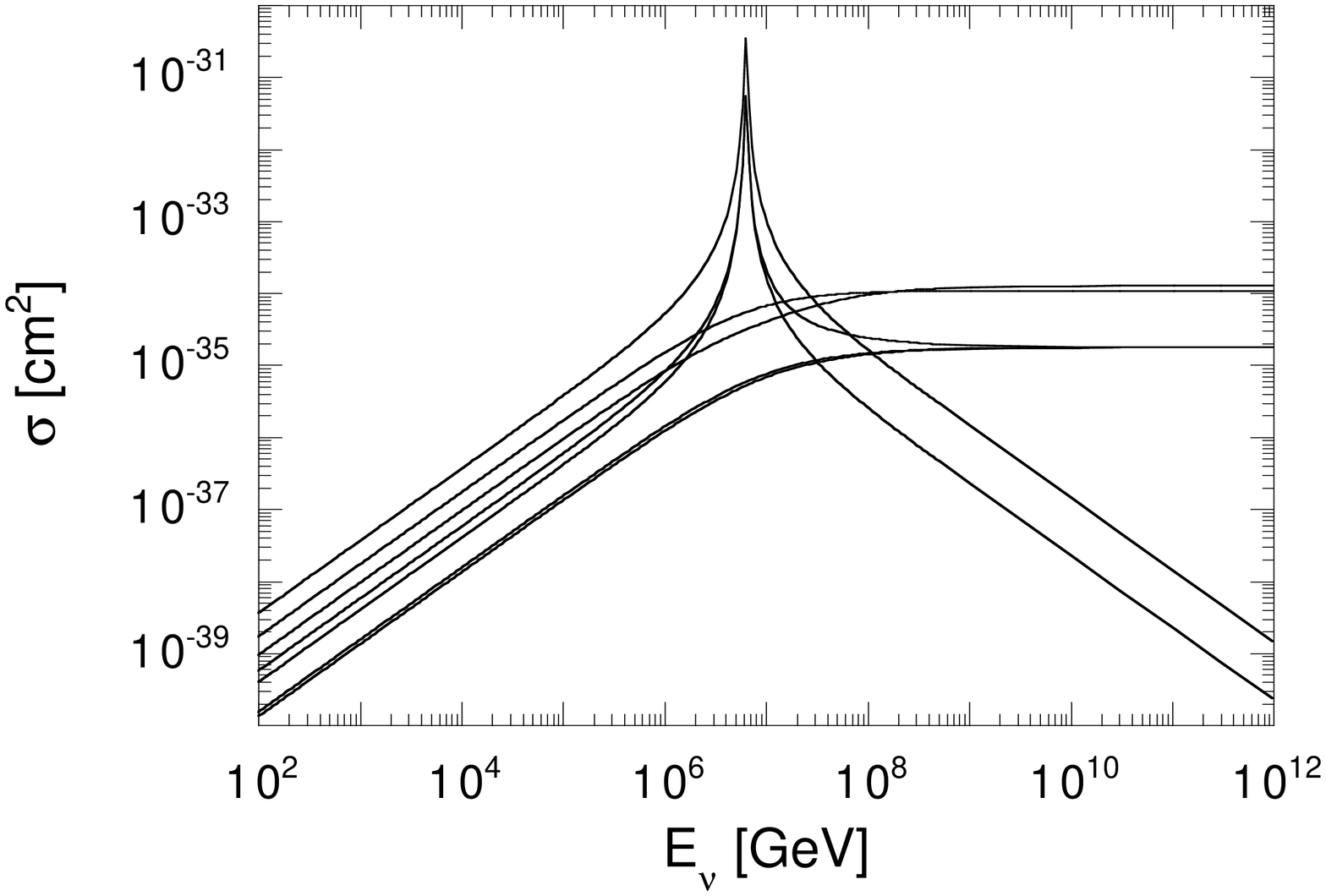}}
\caption{Left pane: The solid curve shows the charged-current $\nu N$ cross section 
    calculated using the CTEQ6 parton distributions~\protect\cite{Pumplin:2002vw};
    the dash-dotted line shows the situation in 1986, using Set~2 of 
    the EHLQ parton distributions~\protect\cite{Eichten:1984eu}. The dotted curve shows the energy 
    dependence of the cross section without QCD evolution, i.e., with 
    the EHLQ distributions frozen at $Q^{2} = 5\gev^{2}$~\protect\cite{Reno:2004cx}. The long-short dashed curve shows the prediction of a structure function that satisfies the Froissart bound for very low $x$ and very large $Q^2$~\cite{Berger:2007ic}. Right pane: Cross sections for
neutrino interactions on electron targets.  At low energies, from
largest to smallest cross section, the processes are (i)
$\bar{\nu}_{e}e \rightarrow \hbox{ hadrons}$, (ii) $\nu_{\mu}e
\rightarrow \mu\nu_{e}$, (iii) $\nu_{e}e \rightarrow \nu_{e}e$, (iv)
$\bar{\nu}_{e}e \rightarrow \bar{\nu}_{\mu}\mu$, (v) $\bar{\nu}_{e}e
\rightarrow \bar{\nu}_{e}e$, (vi) $\nu_{\mu}e \rightarrow \nu_{\mu}e$,
(vii) $\bar{\nu}_{\mu}e \rightarrow
\bar{\nu}_{\mu}e$~{\protect\cite{Gandhi:1995tf}.}
\label{fig:uhesigmas}}
\end{center}
\end{figure}

The left pane of Figure~\ref{fig:uhesigmas} compares our first calculation, using the 1984
EHLQ structure functions, with the modern CTEQ6 parton
distributions~\cite{Pumplin:2002vw}. At the highest energies plotted,
the cross section is about $1.8 \times$ our original estimates, because
today's parton distributions rise more steeply at small $x$ than did
those of two decades ago.  HERA measurements have provided the decisive
new information~\cite{Adloff:2003uh,Chekanov:2002pv,Chekanov:2006ff}. At $10^{12}\gev$, the QCD
enhancement of the small-$x$ parton density has increased the cross
section sixty-fold over the parton-model prediction without evolution.
An ongoing concern, addressed in recent studies~\cite{Berger:2007ic, Anchordoqui:2006ta}, is whether the density of ``wee'' partons becomes so large at relevant values of $Q^2$ that recombination or saturation effects suppress small-$x$ cross sections.
HERA measurements of the charged-current reaction $e p \to \nu +
\hbox{anything}$ at an equivalent lab energy near $40\tev$ observe the
damping due to the $W$-boson propagator and agree with standard-model.
 The 
right pane of Figure~\ref{fig:uhesigmas} shows the interaction cross 
section for neutrinos on electrons in the Earth, which is generally 
several orders of magnitude longer than the $\nu N$ interaction 
length. An important exception is the $\bar{\nu}_{e}e \to W^{-}$ 
resonance at $E_{\nu} \approx 6 \times 10^{15}\ev$.

\begin{figure}[tb]
\begin{center}
\centerline{\includegraphics[width=0.5\textwidth]{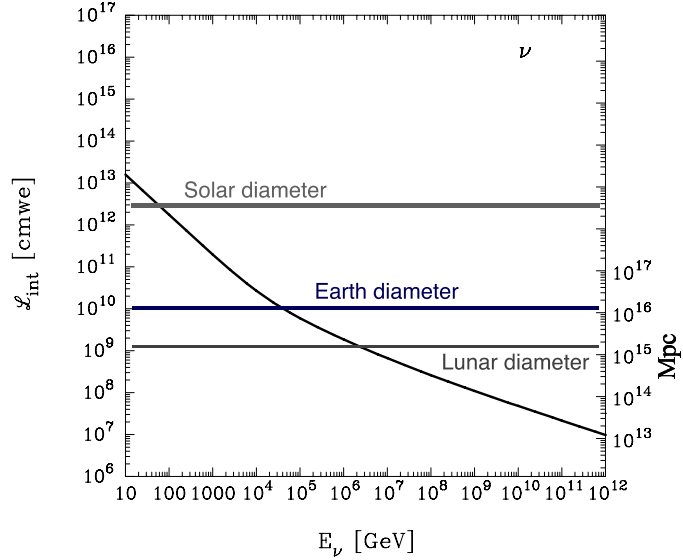}}
\caption{ Interaction length $\mathcal{L}_{\mathrm{int}}^{\nu 
N} = 1/\sigma_{\nu N}(E_{\nu})N_{\mathrm{A}}$, where $N_{\mathrm{A}}$ 
is Avogadro's number, for the reactions $\nu N \to
\hbox{anything}$ as a function of the incident neutrino energy.  The
left-hand scale, in cmwe, is appropriate for terrestrial applications;
the right-hand scale, in Mpc for the current Universe, is appropriate
for transport over cosmological distances~{\protect 
\cite{Gandhi:1998ri,Barenboim:2004di}}.  \label{fig:lint}}
\end{center}
\end{figure}

The rising cross sections have important implications for neutrino 
telescopes. Figure~\ref{fig:lint} shows that the 
Earth is opaque to neutrinos with energies above $40\tev$. This means 
that the strategy of looking down to distinguish charged-current 
interactions from the rain of cosmic-ray muons needs to be 
modified at high energies. On the other hand, the Universe at large is so
exceptionally poor in nucleons (only one in every $4\m^3$ on average), and so the $(\nu N)$ interaction length of 
ultrahigh-energy neutrinos in the cosmos is effectively infinite: a path of $8 \times 10^5\mpc$ in the current Universe has a depth of only $1\hbox{ cmwe}$.

\subsection{Influence of Neutrino Oscillations}
In the early days of planning for neutrino telescopes, people noticed
that observing $\tau$ production through the double-bang signature~\cite{Learned:1994wg}
might provide evidence for neutrino oscillations, since---to good
approximation---no $\nu_{\tau}$ are produced in conventional sources of
ultrahigh-energy neutrinos.  The discovery of neutrino oscillations is
of course already made; the phenomenon of neutrino oscillations means
that the flavor mixture at Earth, $\Phi = \{\varphi_{e},\varphi_{\mu},
\varphi_{\tau}\}$, will be different from the source mixture $\Phi^{0}
= \{\varphi_{e}^{0}, \varphi_{\mu}^{0}, \varphi_{\tau}^{0}\}$. The 
essential fact is that the vacuum oscillation length is very short, 
in cosmic terms. For  $|\Delta m^2| = 10^{-5}\ev^2$, the oscillation 
length
\begin{eqnarray}
\mathcal{L}_{\mathrm{osc}} & = & 4\pi E_{\nu}/|\Delta m^2| 
  \approx  2.5 \times 10^{-24}\mpc\cdot (E_{\nu}/1\ev)
  \label{eq:osclength}
\end{eqnarray}
is a fraction of a megaparsec, even for $E_{\nu}=10^{20}\ev$.
{Accordingly,} neutrinos oscillate many times  between cosmic source 
and terrestrial detector.

Neutrinos in the flavor basis $\ket{\nu_{\alpha}}$ are connected to the
mass eigenstates $\ket{\nu_{i}}$ by the unitary mixing matrix (\ref{eqn:pmns}), as
$\ket{\nu_{\alpha}} = \sum_i {\mathcal{V}_{\beta i}}\ket{\nu_i}$.  It is
convenient to idealize $\sin\theta_{13} = 0$, $\sin 2\theta_{23} = 
1$, and consider
\begin{equation}
\mathcal{V}_{\mathrm{ideal}} = \pmatrix{ c_{12} & s_{12} & 0 \cr
-s_{12}/\sqrt{2} & c_{12}/\sqrt{2} & 1/\sqrt{2} \cr
s_{12}/\sqrt{2} & -c_{12}/\sqrt{2} & 1/\sqrt{2}} \; .
\end{equation}
Then the transfer matrix $\mathcal{X}$ that maps the source flux 
$\Phi^{0}$ into the flux at Earth $\Phi$ takes the form
\renewcommand{\arraystretch}{1.5}
\begin{equation}
    \mathcal{X}_{\mathrm{ideal}} = \left(  
    \begin{array}{ccc}
    1-2x & x & x \\
    x & \textstyle{\frac{1}{2}}(1-x) & \textstyle{\frac{1}{2}}(1-x) \\
    x & \textstyle{\frac{1}{2}}(1-x) & \textstyle{\frac{1}{2}}(1-x)
    \end{array}\right)\;,
    \label{idealtrans}
\end{equation}
where $x = \sin^{2}\theta_{12}\cos^{2}\theta_{12}$.  Because the second
and third rows are identical, the $\nu_{\mu}$ and $\nu_{\tau}$ fluxes
that result from any source mixture $\Phi^{0}$ are equal:
$\varphi_{\mu} = \varphi_{\tau}$.  Independent of the value of $x$,
$\mathcal{X}_{\mathrm{ideal}}$ maps $\Phi^0_{\mathrm{std}} \rightarrow
\{\cfrac{1}{3},\cfrac{1}{3},\cfrac{1}{3}\}$.~\footnote{I owe this 
formulation to Stephen Parke.}

\begin{figure}[tb]
    \centerline{\includegraphics[width=7.5cm]{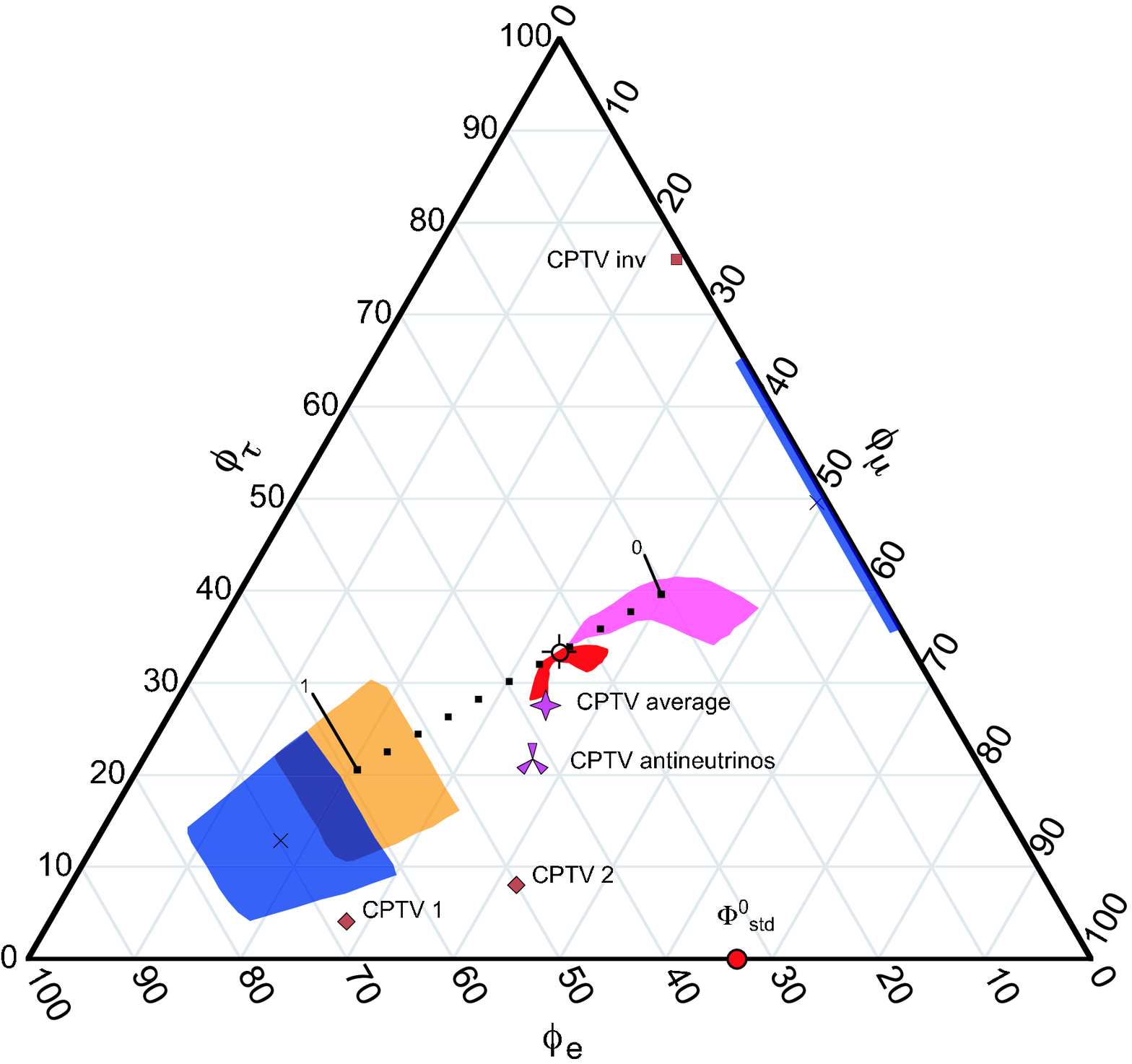}\quad
    \includegraphics[width=7.5cm]{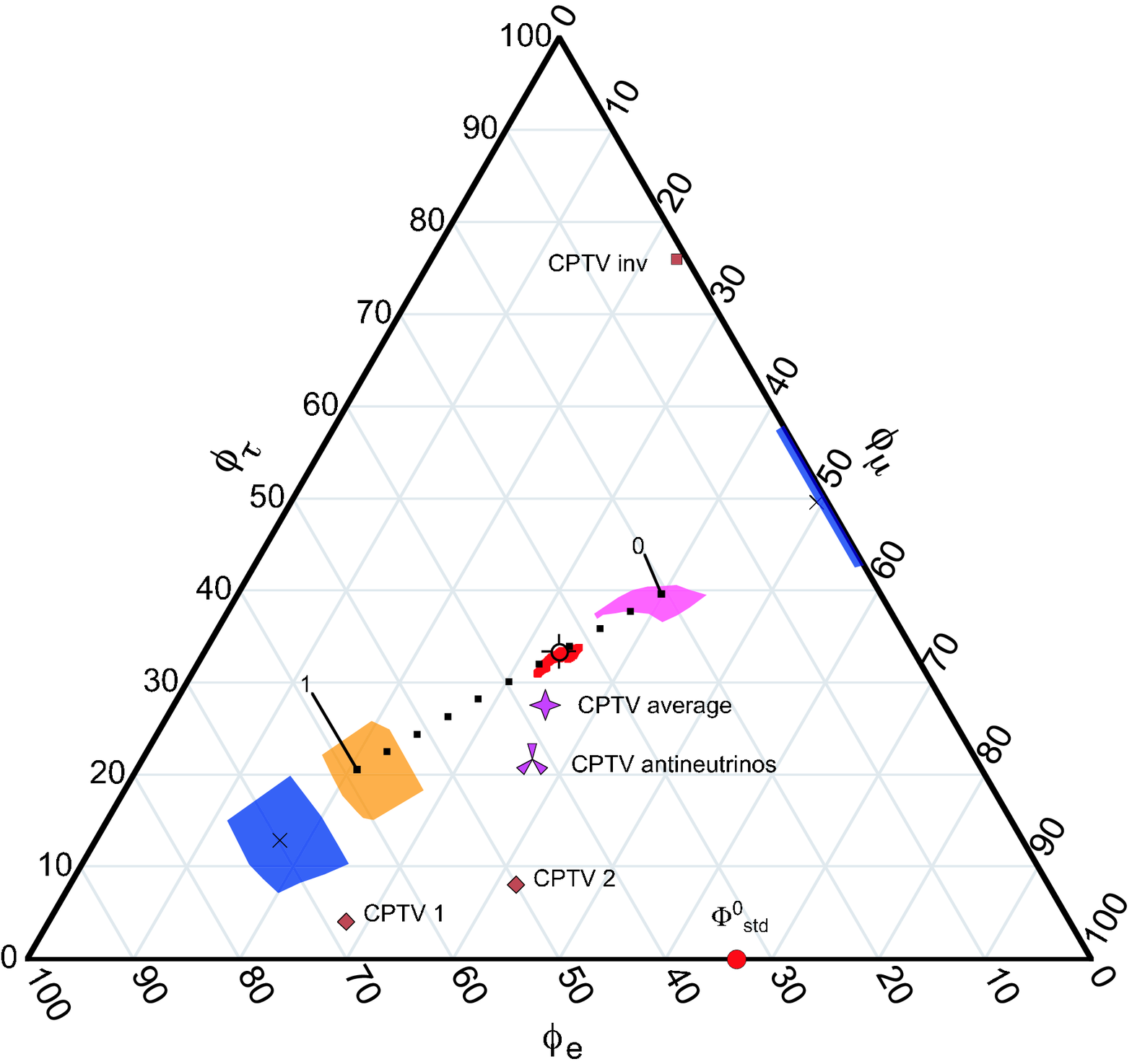}}
\caption{Ternary plots of the neutrino flux $\Phi$ at Earth, showing
the implications of current (left pane) and future (right pane)
knowledge of neutrino mixing.  The small black squares indicate the
$\nu_{e}$ fractions produced by the idealized transfer matrix
$\mathcal{X}_{\mathrm{ideal}}$ as $\varphi_{e}^{0}$ varies from 0 to 1
in steps of 0.1.  A crossed circle marks the standard mixed spectrum at
Earth, $\Phi_{\mathrm{std}} =
\{\cfrac{1}{3},\cfrac{1}{3},\cfrac{1}{3}\}$; for comparison, a red dot
marks the standard source spectrum, $\Phi_{\mathrm{std}}^{0} =
\{\cfrac{1}{3},\cfrac{2}{3},0\}$.  Colored swaths delimit the fluxes at
Earth produced by neutrino oscillations from the source mixtures (right to left)
$\Phi_0^0 = \{0,1,0\}$ (pink), $\Phi_{\mathrm{std}}^0$ (red), and
$\Phi_1^0 = \{1,0,0\}$ (orange), using 95\% CL ranges for the
oscillation parameters.  Black crosses ($\times$) show the mixtures at
Earth that follow from neutrino decay, assuming normal ($\varphi_e
\approx 0.7$) and inverted ($\varphi_e \approx 0$) mass hierarchies.
The blue bands at far left show how current and future uncertainties blur the
predictions for neutrino decays.  The violet tripod indicates how
\textsf{CPT}-violating oscillations shape the mix of antineutrinos that
originate in a standard source mixture, and the violet cross averages
that $\bar{\nu}$ mixture with the standard neutrino mixture.  The brown
squares denote consequences of \textsf{CPT} violation for antineutrino
decays~\protect\cite{Barenboim:2003jm}.}
\label{fig:oscnow}
\end{figure}

The variation of $\varphi_{e}$ with the $\nu_{e}$ source fraction
$\varphi_{e}^{0}$ is shown as a sequence of small black squares (for
$\varphi_e^0 = 0, 0.1, \ldots , 1$) in Figure~\ref{fig:oscnow} for the
value $x = 0.21$, which corresponds to $\theta_{12} = 0.57$, the
central value in a recent global analysis~\cite{deHolanda:2002iv}.  The $\nu_{e}$
fraction at Earth ranges from $0.21$, for $\varphi_{e}^{0} = 0$, to
$0.59$, for $\varphi_{e}^{0} = 1$.
\renewcommand{\arraystretch}{1}

The simple analysis based on $\mathcal{X}_{\mathrm{ideal}}$ is useful
for orientation, but it is important to explore the range of
expectations implied by global fits to neutrino-mixing parameters.  We
take~\cite{Barenboim:2003jm} $0.49 < \theta_{12} < 0.67$,
$\cfrac{\pi}{4}\times 0.8 < \theta_{23} < \cfrac{\pi}{4} \times 1.2$,
and $0 < \theta_{13} < 0.1$.  With current uncertainties in the
oscillation parameters, a standard source spectrum,
$\Phi_{\mathrm{std}}^{0} = \{\cfrac{1}{3},\cfrac{2}{3},0\}$, is mapped
by oscillations onto the red boomerang near $\Phi_{\mathrm{std}} =
\{\cfrac{1}{3},\cfrac{1}{3},\cfrac{1}{3}\}$ in the left pane of
Figure~\ref{fig:oscnow}.  Given that $\mathcal{X}_{\mathrm{ideal}}$
maps $\Phi_{\mathrm{std}}^{0} \rightarrow \Phi_{\mathrm{std}}$ for any
value of $\theta_{12}$, it does not come as a great surprise that the
target region is of limited extent. The variation of $\theta_{23}$ away
from $\cfrac{\pi}{4}$ breaks the identity
$\varphi_{\mu}\equiv\varphi_{\tau}$ of the idealized analysis.

One goal of neutrino observatories will be to characterize cosmic
sources by determining the source mix of neutrino flavors.  It is
therefore of interest to examine the outcome of different assumptions
about the source.  We show in the left pane of Figure~\ref{fig:oscnow}
the mixtures at Earth implied by current knowledge of the oscillation
parameters for source fluxes $\Phi_0^0 = \{0,1,0\}$ (the purple band
near $\varphi_e \approx 0.2$) and $\Phi_1^0 = \{1,0,0\}$ (the orange
band near $\varphi_e \approx 0.6$).
For the $\Phi^0_{\mathrm{std}}$ and $\Phi^0_1$ source spectra, the
uncertainty in $\theta_{12}$ is reflected mainly in the variation of
$\varphi_e$, whereas the uncertainty in $\theta_{23}$ is expressed in
the variation of $\varphi_{\mu}/\varphi_{\tau}$ For the $\Phi_0^0$
case, the influence of the two angles is not so orthogonal.  For all
the source spectra we consider, the uncertainty in $\theta_{13}$ has
little effect on the flux at Earth.  The extent of the three regions,
and the absence of a clean separation between the regions reached from
$\Phi_{\mathrm{std}}^{0}$ and $\Phi_0^0$ indicates that characterizing
the source flux will be challenging, in view of the current
uncertainties of the oscillation parameters.

We anticipated
improved information on $\theta_{12}$ and $\theta_{23}$ from KamLAND
and the long-baseline accelerator experiments at
Soudan  and Gran Sasso roughly on the time scale on which large-volume
neutrino telescopes will come into operation.  We based our
projections for the future on the ranges $0.54 < \theta_{12} < 0.63$\footnote{The latest KamLAND + solar-neutrino analysis does even better: $0.575 < \theta_{12} < 0.63$~\cite{KamLAND:2008ee}.}
and $\cfrac{\pi}{4} \times 0.9 < \theta_{23} < \cfrac{\pi}{4} \times
1.1$, still with $0 < \theta_{13} < 0.1$. The results are shown in the
right pane of Figure~\ref{fig:oscnow}.  The (purple) target region for
the source flux $\Phi_0^0$ shrinks appreciably and separates from the
(red) region populated by $\Phi_{\mathrm{std}}^0$, which is now tightly
confined around $\Phi_{\mathrm{std}}$.  The (orange) region mapped from
the source flux $\Phi_1^0$ by oscillations shrinks by about a factor of
two in the $\varphi_e$ and $\varphi_{\mu}-\varphi_{\tau}$ dimensions.

\subsection{Reconstructing the Neutrino Mixture at the Source}
\begin{figure}[tb]
    \centerline{\includegraphics[width=8cm]{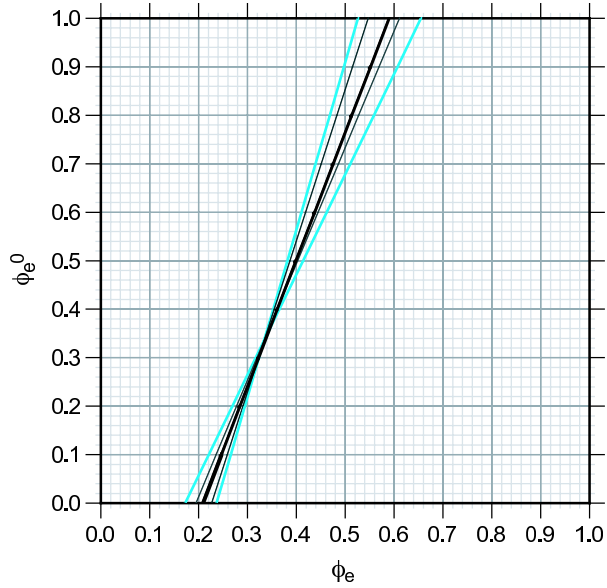}}
    \caption{The source flux $\varphi_e^0$ of electron neutrinos
    reconstructed from the $\nu_e$ flux $\varphi_e$ at Earth, using the
    ideal transfer matrix $\mathcal{X}_{\mathrm{ideal}}$ of Eqn.\
    (\ref{idealtrans}).  The heavy solid line refers to 
    $\theta_{12} = 0.57$.  The light blue lines refer to
    the current experimental constraints (at 95\% CL), and the thin
    solid lines refer to a projection of future experimental
    constraints~\protect\cite{Barenboim:2003jm}.}
\label{fig:invert}
\end{figure}

What can observations of the blend $\Phi$ of neutrinos arriving at
Earth tell us about the source~\cite{Barenboim:2003jm}?  Inferring the nature of the processes
that generate cosmic neutrinos is more complicated than it would be if 
neutrinos did not oscillate.  Because $\nu_{\mu}$ and
$\nu_{\tau}$ are fully mixed---and thus enter identically in
$\mathcal{X}_{\mathrm{ideal}}$---it is not possible fully to characterize
$\Phi^{0}$. We can, however, reconstruct the $\nu_{e}$ fraction at the 
source as $\varphi_{e}^{0} = (\varphi_{e} - x)/(1 - 3x)$, where $x = 
\sin^{2}\theta_{12}\cos^{2}\theta_{12}$. 
The reconstructed source flux $\varphi_e^0$ is shown in
Figure~\ref{fig:invert} as a function of the $\nu_e$ flux at Earth.  The
heavy solid line represents the best-fit value for $\theta_{12}$; the
light blue lines and thin solid lines indicate the current and future
95\% CL bounds on $\theta_{12}$.

A possible strategy for beginning to characterize a source of cosmic
neutrinos might proceed by measuring the $\nu_e/\nu_{\mu}$ ratio and
estimating $\varphi_e$ under the plausible assumption---later to be
checked---that $\varphi_{\mu} = \varphi_{\tau}$.  Very large
($\varphi_e \gtrsim 0.65$) or very small ($\varphi_e \lesssim 0.15$) $\nu_e$
fluxes cannot be accommodated in the standard neutrino-oscillation
picture.  Observing an extreme $\nu_e$ fraction would implicate
unconventional physics.  

Determining the energy dependence of $\varphi_{e}^{0}$ may also be of
astrophysical interest~\cite{Kashti:2005qa}.  In a thick source, the highest energy muons
may interact and lose energy before they can decay.  In the limit of
$\varphi_{e}^{0} = 0$, the arriving flux will be $\Phi =
\{x,\cfrac{1}{2}(1-x),\cfrac{1}{2}(1-x)\} \approx \{0.22, 0.39, 0.39\}$ (cf.\ Figure~\ref{fig:oscnow}).  More
generally, measured $\nu_e$ fractions that depart significantly from
the canonical $\varphi_e = \cfrac{1}{3}$ would suggest
neutrino sources that are in some way nonstandard.  An observed flux $\varphi_e = 0.5 \pm 0.1$ points to
a source flux $0.47 \lesssim \varphi_e^0 \lesssim 1$, with current
uncertainties, whereas $\varphi_e = 0.25 \pm 0.10$ indicates $0 \lesssim
\varphi_e^0 \lesssim 0.35$.

Constraining the source flux sufficiently to test the nature of the
neutrino production process will require rather precise determinations
of the neutrino flux at Earth.  Suppose we want to test the idea that
the source flux has the standard composition $\Phi^{0}_{\mathrm{std}}$.
With today's uncertainty on $\theta_{12}$, a 30\% measurement that
locates $\varphi_e = 0.33 \pm 0.10$ implies only that $0 \lesssim
\varphi_e^0 \lesssim 0.68$.  For a measured flux in the neighborhood of
$\cfrac{1}{3}$, the uncertainty in the solar mixing angle is of little
consequence: the constraint that arises if we assume the central value
of $\theta_{12}$ is not markedly better: $0.06 \lesssim \varphi_e^0 \lesssim
0.59$.  A 10\% measurement of the $\nu_e$ fraction, $\varphi_e = 0.33
\pm 0.033$, would make possible a rather restrictive constraint on the
nature of the source.  The central value for $\theta_{12}$ leads to
$0.26 \lesssim \varphi_e^0 \lesssim 0.43$, blurred to $0.22 \lesssim \varphi_e^0
\lesssim 0.45$ with current uncertainties.

\subsection{Influence of Neutrino Decays \label{subsec:nudk}}
To this point, we have considered the neutrinos to be stable particles. 
``Invisible'' decays, such as the decay of a heavy neutrino into a 
lighter neutrino plus a very light---or massless---(pseudo)scalar 
boson such as the majoron~\cite{Valle:1983ua,Gelmini:1983ea} are not 
very well constrained by observations~\cite{Beacom:2002cb,Beacom:2002vi}. [A majoron too massive 
to serve as a neutrino decay product can nevertheless have important 
consequences for cosmology, including deviations from the standard 
expectations for the radiation energy density and changes in the 
positions of peaks in the cosmic microwave background power 
spectrum~\cite{Chacko:2003dt}.]
If \textsf{CPT} invariance holds, SN1987a data set an upper limit on the lifetime of
the lightest neutrino of $\tau_{\ell}/m_{\ell} \gtrsim 10^{5}$ s/eV. 
Observations of solar neutrinos lead to $\tau_{2}/m_{2}\gtrsim 10^{-4}\s/\hbox{eV}$. Finally, if the neutrino
spectrum is normal, the data on  atmospheric neutrinos coming upward 
through the Earth, imply $\tau_{3}/m_{3}\gtrsim 10^{-10}$ s/eV. 

These rather modest limits open the possibility
that some neutrinos do not survive the journey from astrophysical
sources.  Decays of unstable neutrinos over cosmic distances can lead
to mixtures at Earth that are incompatible with the oscillations of
stable neutrinos~\cite{Beacom:2002cb,Beacom:2002vi,Barenboim:2003jm}.
The candidate decays are transitions between mass eigenstates by
emission of a very light particle, $\nu_i \rightarrow (\nu_j,
\bar{\nu}_j)+X$.  Dramatic effects occur when the decaying neutrinos
disappear, either by decay to invisible products or by decay into
active neutrino species so degraded in energy that they contribute
negligibly to the total flux at the lower energy.

If the lifetimes of the unstable mass eigenstates are short compared
with the flight time from source to Earth, all the unstable neutrinos 
will decay, and the (unnormalized) flavor
$\nu_{\alpha}$ flux at Earth will be $\widetilde{\varphi}_{\alpha} (E_{\nu})=  
\sum_{i = \mathrm{stable} } \sum_{\beta} \varphi^{0}_{\beta}(E_{\nu})
|U_{\beta i}|^2 |U_{\alpha i}|^2$,
with $\varphi_{\alpha} = \widetilde{\varphi}_{\alpha} / \sum_{\beta}
\widetilde{\varphi}_{\beta}$.
Should only the lightest neutrino survive, the flavor mix of neutrinos
arriving at Earth is determined by the flavor composition of the
lightest mass eigenstate, \textit{independent of the flavor mix at the
source.} 

For a normal mass hierarchy $m_1 < m_2 < m_3$, the $\nu_{\alpha}$ flux
at Earth is $\varphi_{\alpha} = |U_{\alpha 1}|^2$.  {Consequently,} the
neutrino flux at Earth is $\Phi_{\mathrm{normal}} = \{|U_{e1}|^2,
|U_{\mu1}|^2,|U_{\tau1}|^2\}$.  If the mass hierarchy is
inverted, $m_2 > m_1 > m_3$, the lightest (hence, stable) neutrino is
$\nu_3$, so the flavor mix at Earth is determined by $\varphi_{\alpha}
= |U_{\alpha 3}|^2$.  In this case, the neutrino flux at Earth is
$\Phi_{\mathrm{inverted}} = \{|U_{e3}|^2, |U_{\mu3}|^2,|U_{\tau3}|^2\}$.  Both $\Phi_{\mathrm{normal}}$ and
$\Phi_{\mathrm{inverted}}$, which are indicated by crosses ($\times$)
in Figure~\ref{fig:oscnow}, are very different from the standard flux
$\Phi_{\mathrm{std}} = \{\varphi_{e} = \cfrac{1}{3}, \varphi_{\mu} =
\cfrac{1}{3}, \varphi_{\tau} = \cfrac{1}{3}\}$ produced by the ideal
transfer matrix from a standard source.  Observing either mixture would
represent a departure from conventional expectations.

The fluxes that result from neutrino decays \textit{en route} from the
sources to Earth are subject to uncertainties in the neutrino-mixing
matrix.  The expectations for the two decay scenarios are indicated by
the blue regions in Figure~\ref{fig:oscnow}, where we indicate the
consequences of 95\% CL ranges of the mixing parameters now and in the
future.  With current uncertainties~\cite{GonzalezGarcia:2007ib}, the normal hierarchy populates
$0.60 \lesssim \varphi_e \lesssim 0.73$, and allows considerable departures
from $\varphi_{\mu} = \varphi_{\tau}$.  The normal-hierarchy decay
region based on current knowledge overlaps the flavor mixtures that
oscillations produce in a pure-$\nu_e$ source, shown in orange.  (It
is, however, far removed from the standard region that encompasses
$\Phi_{\mathrm{std}}$.)  With the projected tighter constraints on the
mixing angles, the range in $\varphi_e$ swept out by oscillation from a
pure-$\nu_e$ source or decay from a normal hierarchy shrinks by about a
factor of two, and is separated from the oscillations.  The degree of
separation between the region populated by normal-hierarchy decay and
the one populated by mixing from a pure-$\nu_e$ source depends on the
value of the solar mixing angle $\theta_{12}$.  For the apparently
excluded value $\theta_{12} = \cfrac{\pi}{4}$, both mechanisms would yield
$\Phi = \{\cfrac{1}{2}, \cfrac{1}{4}, \cfrac{1}{4}\}$.

The mixtures that result from the decay of the heavier members of an
inverted hierarchy entail $\varphi_e\approx 0$.  These mixtures are
well separated from any that would result from neutrino oscillations,
for any conceivable source at cosmic distances.

The energies of neutrinos that may be detected in the future from AGNs
and other cosmic sources range over several orders of magnitude,
whereas the distances to such sources vary over perhaps one order of
magnitude.  The neutrino energy sets the neutrino lifetime in the
laboratory frame; more energetic neutrinos survive over longer flight
paths than their lower-energy companions. [A similar phenomenon
is familiar for cosmic-ray muons.]  Under propitious circumstances of
reduced lifetime, path length, and neutrino energy, it might be
possible to observe the transition from more energetic survivor
neutrinos to less energetic decayed neutrinos.

\begin{figure}
    \centerline{\includegraphics[width=7.5cm]{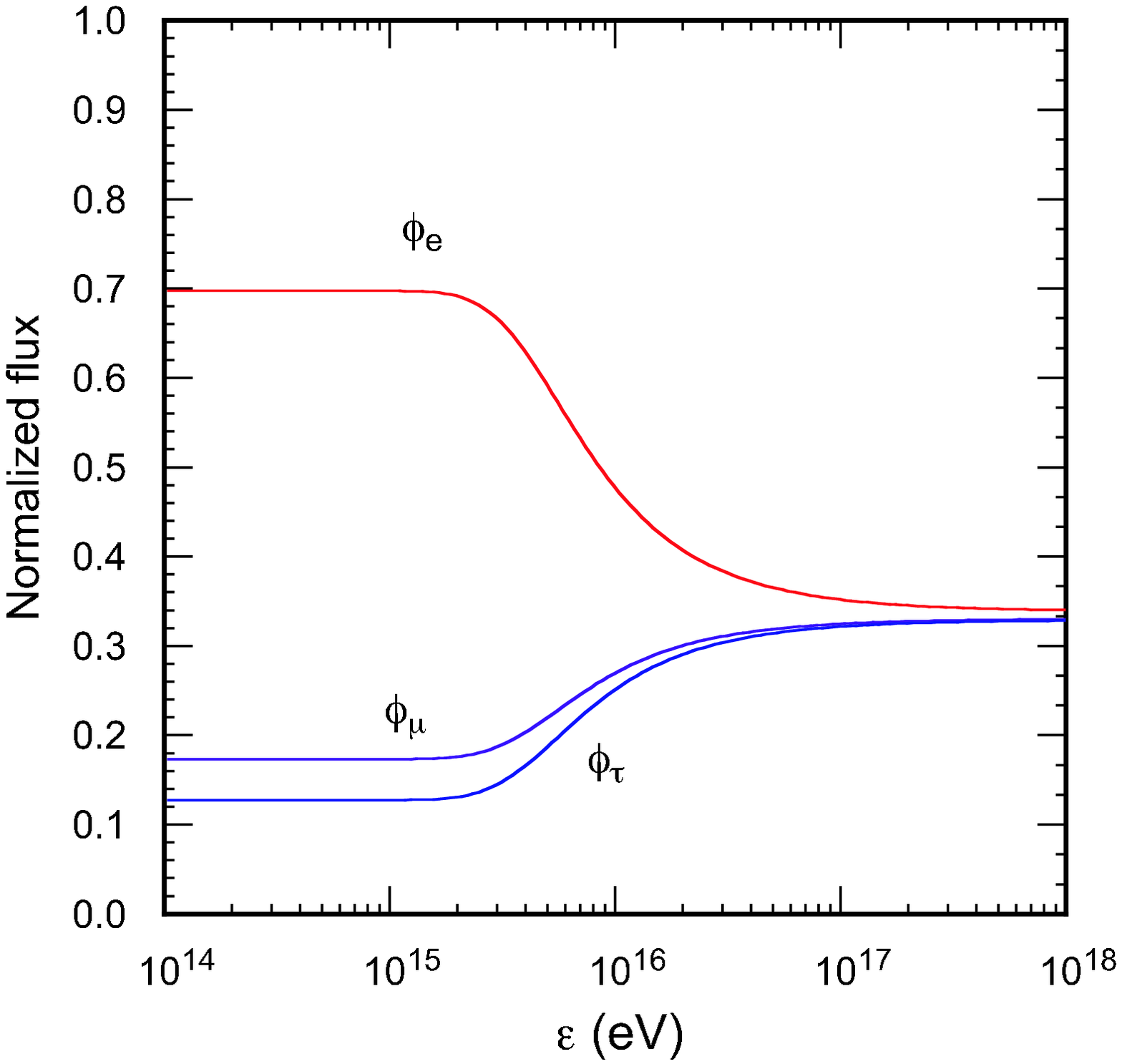}\quad
    \includegraphics[width=7.5cm]{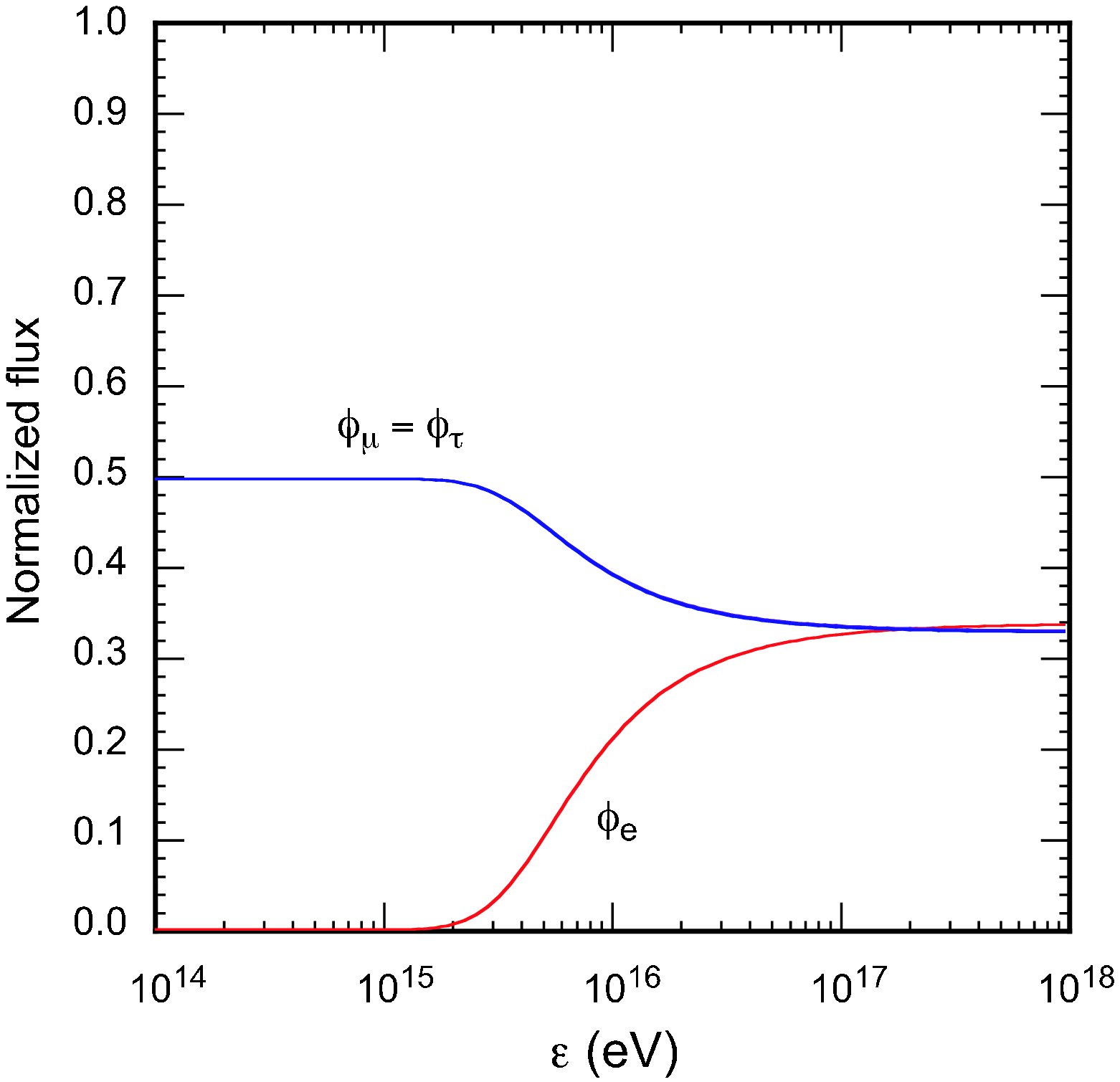}}
\caption{Energy dependence of normalized $\nu_e$, $\nu_{\mu}$, and
$\nu_{\tau}$ fluxes, for the two-body decay of the two upper mass
eigenstates, with the neutrino source at $L=100\hbox{ Mpc}$ from Earth
and $\tau/m = 1\hbox{ s/eV}$.  The left pane shows the result for a
normal mass hierarchy; the right pane shows the result for an inverted
mass hierarchy.  With suitable rescaling of the neutrino energy 
[$E_{\nu} = \varepsilon (1\s/\hbox{eV})/(\tau_{\nu}/m_{\nu})\cdot
L/(100\mpc)$], these plots apply for any combination of path length and
reduced lifetime~\protect\cite{Barenboim:2003jm}.}
\label{fig:decays}
\end{figure}

If decay is not complete, the (unnormalized) flavor $\nu_{\alpha}$ flux
arriving at Earth from a source at distance $L$ is 
$\widetilde{\varphi}_{\alpha} (E_{\nu})=  
\sum_{i } \sum_{\beta} \varphi^{0}_{\beta}(E_{\nu})
|U_{\beta i}|^2 |U_{\alpha i}|^2 e^{-(L/E_{\nu})(m_i/\tau_i)}$,
with the normalized flux $\varphi_{\alpha}(E_{\nu}) =
\widetilde{\varphi}_{\alpha}(E_{\nu}) / \sum_{\beta}
\widetilde{\varphi}_{\beta}(E_{\nu})$.  An idealized case will
illustrate the possibilities for observing the onset of neutrino decay
and estimating the reduced lifetime.  Assume a normal mass hierarchy,
$m_1 < m_2 < m_3$, and consider the special case $\tau_3/m_3 = \tau_2/m_2 \equiv \tau/m$.
For a given path length $L$, the neutrino energy at which the
transition occurs from negligible decays to complete decays is
determined by $\tau/m$.  The left pane of
Figure~\ref{fig:decays} shows the energy evolution of the normalized neutrino
fluxes arriving from a standard source; the energy scale is appropriate
for the case $\tau/m = 1\hbox{ s/eV}$ and $L = 100\hbox{ Mpc}$.  

If we locate the transition from survivors to decays at neutrino energy
$E^{\star}$, then we can estimate the reduced lifetime in terms of the
distance to the source as
\begin{equation}
\tau / m \approx 100\hbox{ s/eV} \cdot \left(\frac{L}{\hbox{1
Mpc}}\right) \left( \frac{1\hbox{ TeV}}{E^{\star}}\right)\; .
\label{taumest}
\end{equation}
In practice, ultrahigh-energy neutrinos are likely to arrive from a
multitude of sources at different distances from Earth, so the
transition region will be blurred.
Nevertheless, it would be rewarding to observe the decay-to-survival
transition, and to use Eqn.\ (\ref{taumest}) to estimate---even within
one or two orders of magnitude---the reduced lifetime.  If no evidence
appears for a flavor mix characteristic of neutrino decay, then Eqn.\
(\ref{taumest}) provides a lower bound on the neutrino lifetime.  For
that purpose, the advantage falls to large values of $L/E^{\star}$, and
so to the lowest energies at which neutrinos from distant sources can
be observed.  Observing the standard flux, $\Phi_{\mathrm{std}} =
\{\cfrac{1}{3},\cfrac{1}{3},\cfrac{1}{3}\}$, which is incompatible with
neutrino decay, would strengthen the current bound on $\tau/m$ by some
seven orders of magnitude, for 10-TeV neutrinos from sources at
$100\mpc$.

\section{NEUTRINO ABSORPTION SPECTROSCOPY \label{sec:dips}}
\begin{figure}
    \centerline{\includegraphics[width=10cm]{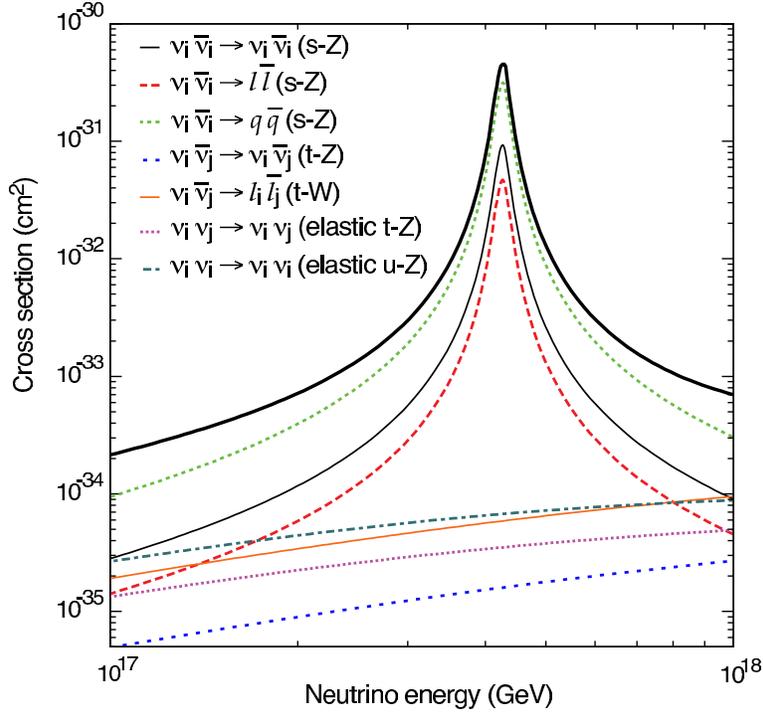}}
    \caption{Total neutrino annihilation cross section and the different
contributing channels as a function of the neutrino energy
assuming a relic neutrino mass of $m_\nu =10^{-5}\ev$ and zero
redshift~\protect\cite{Barenboim:2004di}.}
\label{fig:zed}
\end{figure}

The neutrino gas that we believe permeates the present Universe has
never been detected directly.  The example of the $\bar{\nu}_e e^- \to W^-$ ``Glashow resonance''~\cite{PhysRev.118.316} has motivated studies of the  resonant annihilation
of extremely-high-energy cosmic neutrinos on background neutrinos
through the reaction $\nu\bar{\nu} \to
Z^{0}$~\cite{Weiler:1982qy,Weiler:1983xx,Roulet:1992pz,Gondolo:1991rn,Yoshida:1996ie,Fargion:1997ft,Weiler:1997sh,Eberle:2004ua}. The components of the neutrino-(anti)neutrino cross sections are shown
in Figure~\ref{fig:zed}. 
The feature that matters is the $Z^{0}$-formation line that occurs 
near the resonant energy $E_{\nu}^{Z\mathrm{res}} = 
M_{Z}^{2}/2m_{\nu_{i}}$. The smallness of the neutrino masses (cf.\ Figure~\ref{fig:neutrinomasses}) means that the resonant energies are extremely high.
If it were possible to observe $Z$-bursts or absorption lines, one could hope to confirm the presence of relic neutrinos and learn the absolute neutrino masses and the flavor composition of the neutrino
mass eigenstates. To give an overview of the 
prospects for cosmic-neutrino annihilation spectroscopy, I summarize some of 
the main findings from a detailed study  that Gabriela 
Barenboim, Olga Mena, and I carried out recently~\cite{Barenboim:2004di}. 

Imagine first a toy experiment, in which an
extremely high-energy neutrino beam traverses a very long column with
the relic-neutrino properties of the current Universe.  Neglect for
now the expansion of the Universe and the thermal motion of the relic
neutrinos.  The ``cosmic neutrino attenuator'' is thus a column of
length $L$ with uniform neutrino density $n_{\nu0} = 56\cm^{-3}$ of
each neutrino species, $\nu_{e}, \bar{\nu}_{e}, \nu_{\mu},
\bar{\nu}_{\mu}, \nu_{\tau}, \bar{\nu}_{\tau}$.  If the column of relic
neutrinos is thick enough to attenuate neutrinos appreciably through
resonant absorption at the $Z^{0}$ gauge boson, the energies that
display absorption dips point to the neutrino masses through the
resonant-energy condition.  The relative depletion of
$\nu_{e},\nu_{\mu},\nu_{\tau}$ in each of the lines measures the flavor
composition of the corresponding neutrino mass eigenstate.

Even if we had at our disposal an adequate neutrino beam (with 
energies extending beyond $10^{26}\ev$), the time required
to traverse one interaction length for $\nu\bar{\nu} \to Z^{0}$
annihilation on the relic background in the current Universe is $1.2
\times 10^{4}\mpc = 39\hbox{ Gly}$. This exceeds the age of the
Universe, not to mention the human attention span!  If we are ever to
detect the attenuation of neutrinos on the relic-neutrino background,
we shall have to make use of astrophysical or cosmological neutrinos
sources traversing the Universe over cosmic time scales.  The expansion
of the Universe over the propagation time of the neutrinos entails 
three important effects: the evolution of relic-neutrino density, the
redshift of the incident neutrino energy, and the redshift of the 
relic-neutrino temperature.

\begin{figure}[tb]
    \centerline{\includegraphics[width=8cm]{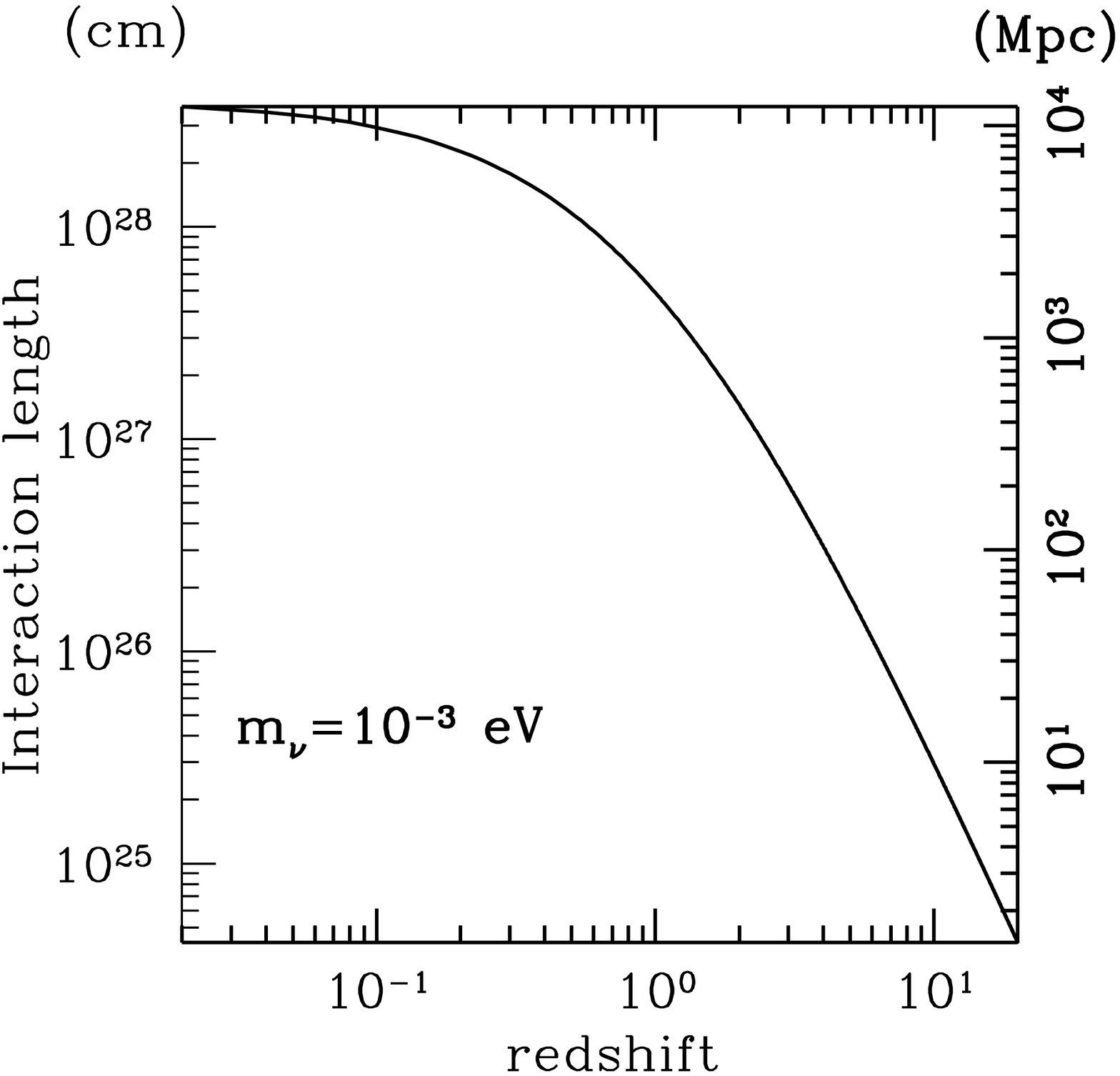}\quad
    \includegraphics[width=8cm]{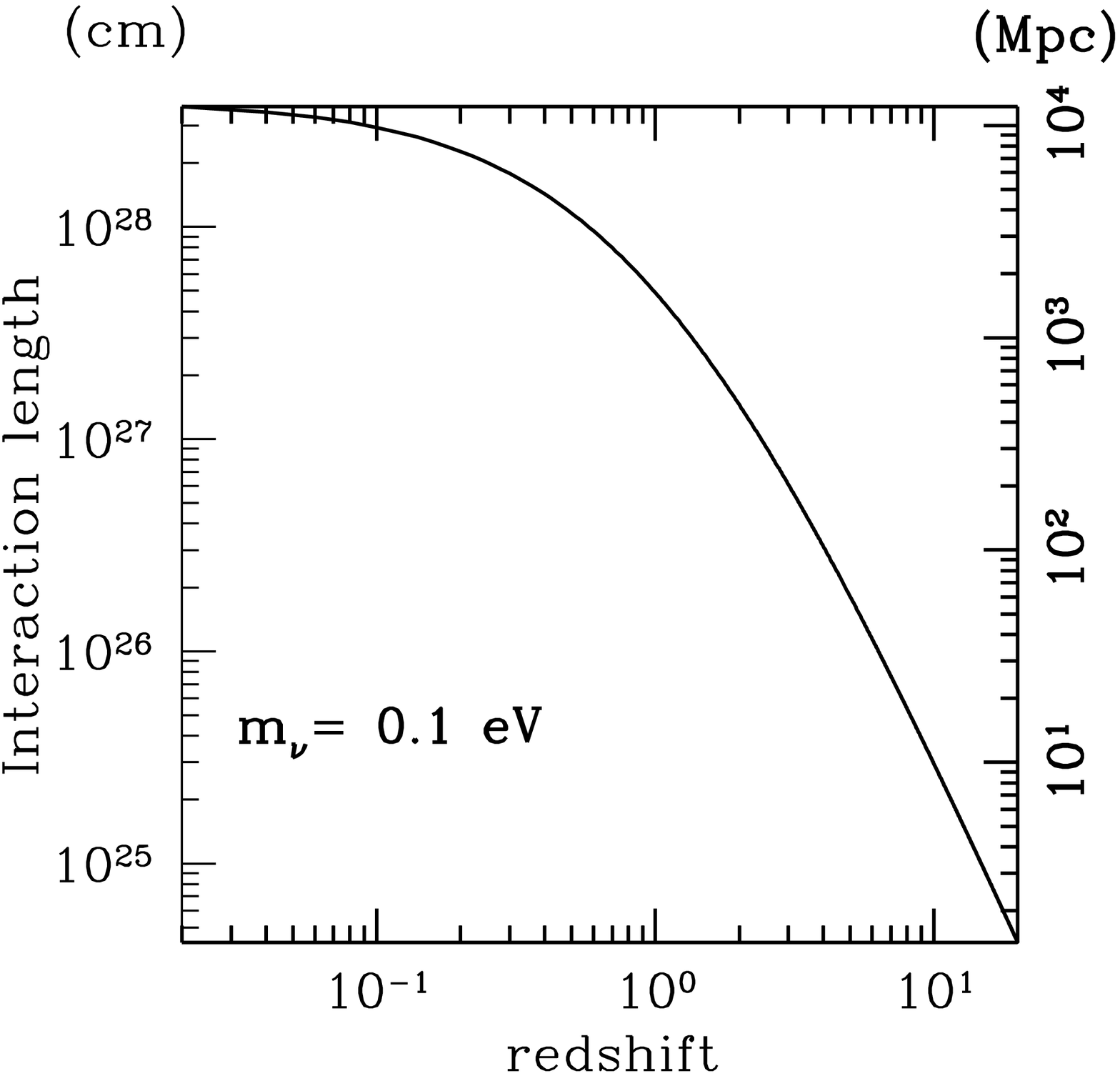}}
    \caption{Interaction lengths (for Dirac relic neutrinos)
versus redshift at the $Z^{0}$ resonance 
for neutrino masses $m_\nu =10^{-3}\ev$ (left pane) and  $10^{-1}\ev$ (right pane). The left-hand scales are in centimeters, the right-hand scales in megaparsecs~\protect\cite{Barenboim:2004di}.}
\label{fig:lintZ}
\end{figure}

In an evolving Universe, the column density of relic neutrinos is proportional to $(1 + z)^2/H(z)$, where $z$ is the redshift and $H(z)$ is the Hubble parameter. 
The resulting decrease of interaction lengths with increasing redshift shown in 
Figure~\ref{fig:lintZ} reveals that for $ 1 \lesssim z \lesssim 10$, the interaction length matches the distance to the AGNs we 
consider as plausible UHE neutrino sources \ldots  though not 
perhaps with the energies required for this application.
To compute the absorption lines, we must propagate ultrahigh-energy neutrinos through an evolving Universe.

The tiny neutrino mass makes for another complication: 
relic neutrinos are moving targets, with their momentum distribution 
characterized in the present Universe by Eq.~(\ref{eq:nuFD}). 
The thermal motion of the neutrinos gives rise to a Fermi (momentum) 
smearing of the UHE-$\nu$--relic-$\nu$ cross section. The resonant 
incident-neutrino energy for a relic neutrino in motion is given 
 by
\begin{equation}
    {E}_{\nu}^{Z \mathrm{res}} = 
    \frac{M_{Z}^{2}}{2(\varepsilon_{\nu} - p_{\nu}\cos\theta)}\;,
    \label{eq:smres}
\end{equation}
where $p_{\nu}$ and $\varepsilon_{\nu}$ are the relic-neutrino 
momentum and energy. The angle $\theta$ characterizes the direction of 
the relic neutrino with respect to the line of flight of the incident 
UHE neutrino. Accordingly, the resonant energy will be displaced 
downward from $M_{Z}^{2}/2m_{\nu}$ to approximately
\begin{equation}
    \widetilde{E}_{\nu}^{Z \mathrm{res}} = 
    \frac{M_{Z}^{2}}{2\langle\varepsilon_{\nu}\rangle}\;,
    \label{eq:smresavg}
\end{equation}
where $\langle\varepsilon_{\nu}\rangle = [\langle p_{\nu}^{2}\rangle + 
 m_{\nu}^{2}]^{1/2}$ plays the role of an \textit{effective relic-neutrino 
 mass.} 
The root-mean-square relic-neutrino momentum, which ranges from $6 
\times 10^{-4}\ev$ in the present Universe to $2.5 \times 10^{-2}\ev$ 
at redshift $z = 20$, thus serves as a rough 
lower bound on the effective neutrino mass. At a given redshift, the 
resonance peak for scattering 
from any neutrino with $m_{\nu} \lesssim \langle\varepsilon_{\nu}\rangle$ 
will be changed significantly.

\begin{figure}
    \centerline{\includegraphics[width=8cm]{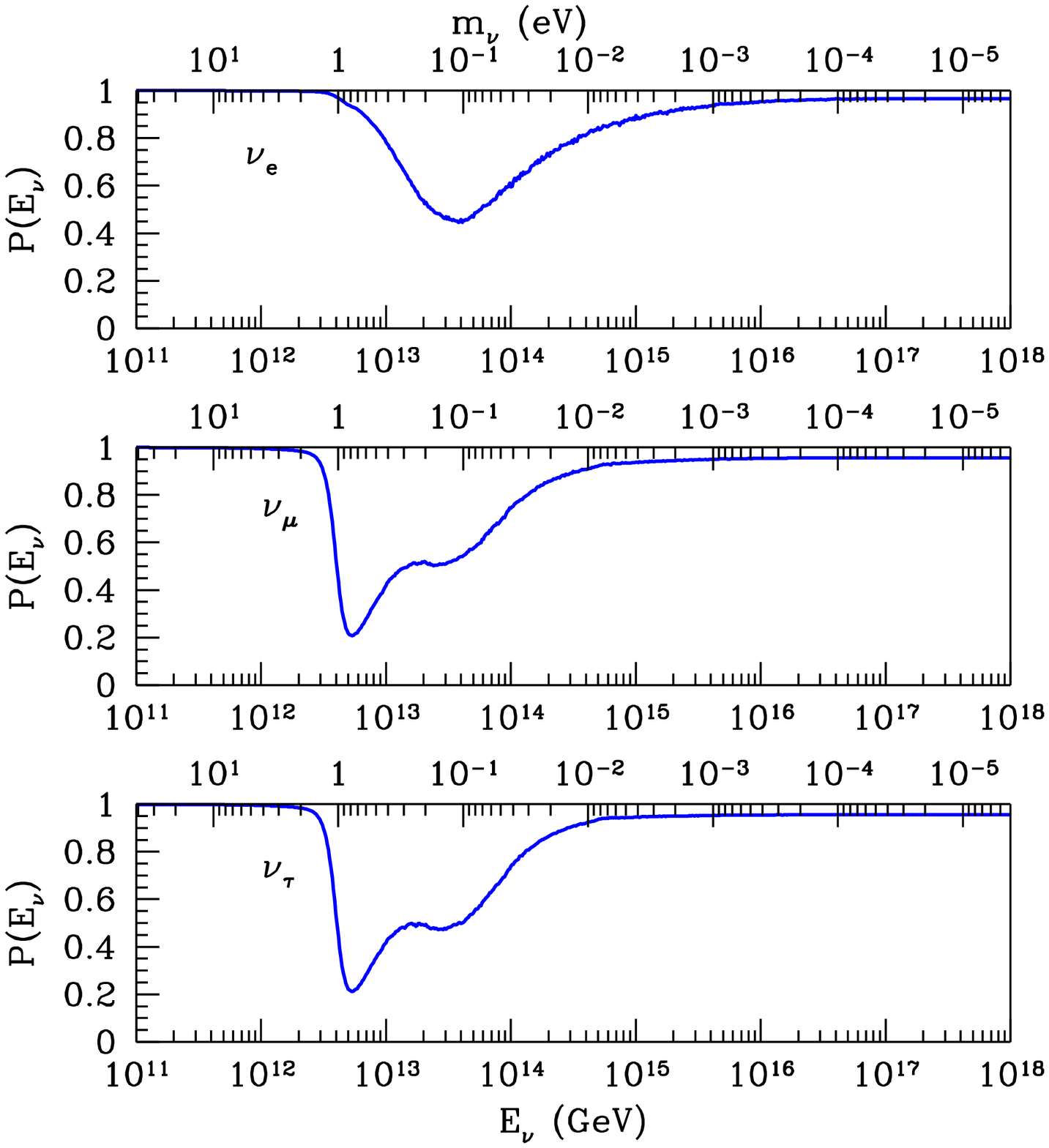}\quad
    \includegraphics[width=8cm]{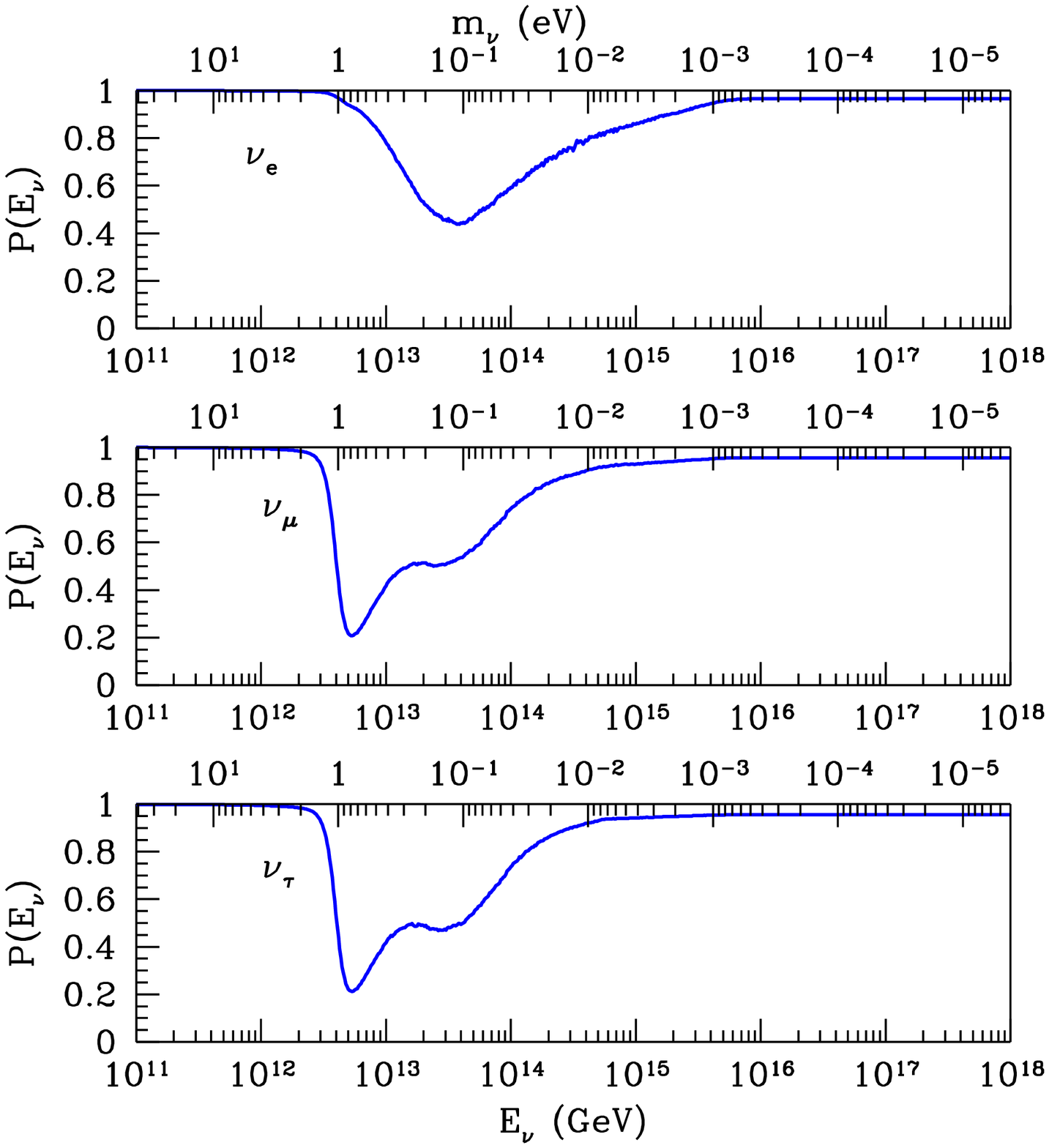}}
    \caption{Survival probabilities for $\nu_e$, $\nu_\mu$, and 
 $\nu_\tau$  as a function of the neutrino energy,
 after integration back to redshift $z = 20$, taking into account the 
 Fermi smearing induced by the thermal motion of the relic neutrinos.
The results apply for a normal hierarchy with lightest neutrino mass
 $m_{\ell}=10^{-5}\ev$ (left pane) or $m_{\ell}=10^{-3}\ev$ (right pane)~\protect\cite{Barenboim:2004di}.}
\label{fig:fermimo}
\end{figure}

\begin{figure}
    \centerline{\includegraphics[width=8cm]{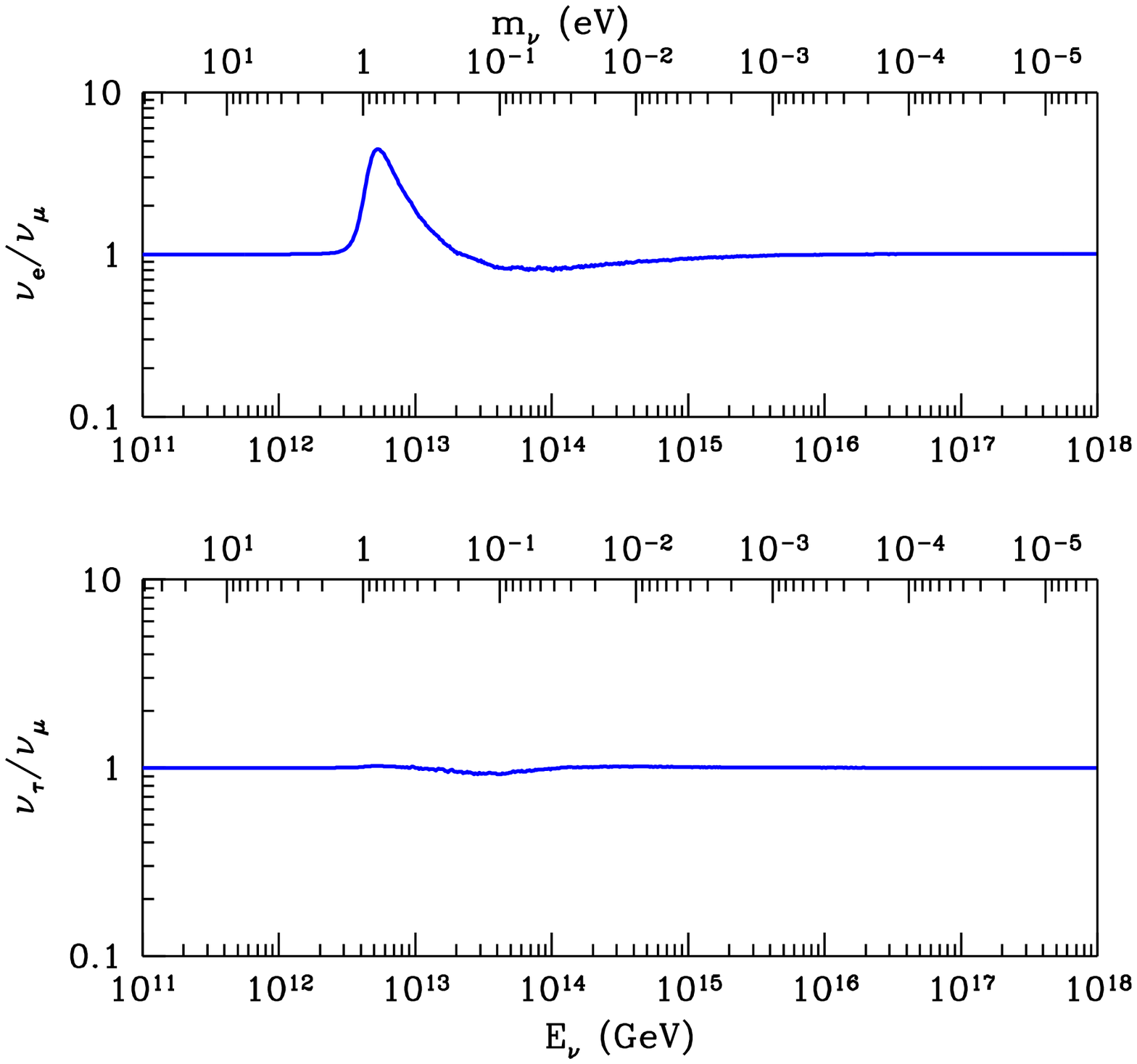}\quad
    \includegraphics[width=8cm]{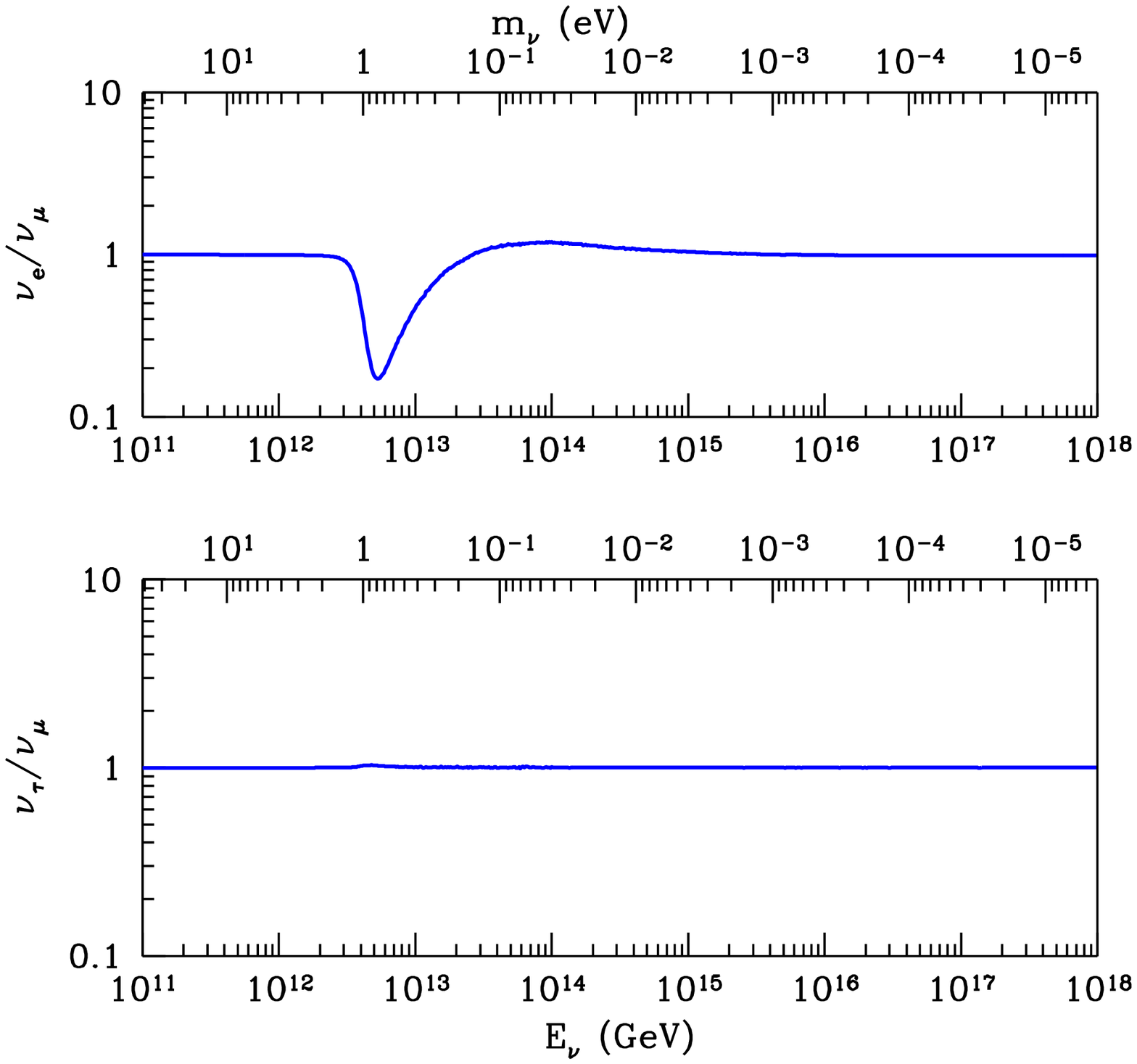}}
    \caption{Flux ratios $\nu_{e}/\nu_{\mu}$ and 
    $\nu_{\tau}/\nu_{\mu}$ at Earth, for normal (left pane) and 
    inverted (right pane) mass hierarchies with $m_{\ell} = 
    10^{-5}\ev$, after integration back to redshift $z=20$ and a 
    thermal averaging over the relic-neutrino momentum distribution.
    The scale at the top shows the neutrino mass 
    defined as $m_{\nu} = M_{Z}^{2}/2E_{\nu}$ that would be inferred if 
    neutrino energies were not redshifted~\protect\cite{Barenboim:2004di}.}
\label{fig:fermimorats}
\end{figure}

The absorption lines that result from a full calculation, including 
the effects of the relics' Fermi motion and the evolution of the 
Universe back to redshift $z = 20$, are shown in  Figure~\ref{fig:fermimo} 
for two values of the lightest neutrino mass, $m_{\ell} = 10^{-5}$ 
and $10^{-3}\ev$. Although the lines are distorted and displaced from 
their natural shapes and positions by redshifting and Fermi motion, 
they would nevertheless confirm our expectations for the relic 
neutrino background and give important information about the neutrino 
spectrum. In particular, the $\nu_{e}/\nu_{\mu}$ ratio, shown in 
Figure~\ref{fig:fermimorats}, is a marker for the normal or inverted mass 
hierarchy.


At least in principle, the observation of cosmic-neutrino absorption lines would open the
way to new insights about neutrino properties
and the thermal history of the universe.  How
the tale unfolds will depend on factors we cannot foresee.  The earlier
in redshift the relevant cosmic-neutrino sources appear, the lower the
present-day energy of the absorption lines and the denser the column of
relics the super-high-energy neutrinos must traverse.  In particular,
the appearance of dips at energies much lower that we expect points to
early---presumably nonacceleration---sources, that could give us insight
into fundamental physics at early times and high energy scales.  On the
other hand, integration over a longer range in redshift means more
smearing and distortion of the absorption lines.

If it can be achieved at all, the detection of neutrino absorption
lines will not be done very soon, and the interpretation is likely to
require many waves of observation and analysis.  Nevertheless,
observations of cosmic-neutrino absorption lines offer the possibility
to establish the existence of another relic from the big bang and,
conceivably, they may open a window on periods of the thermal history
of the universe not readily accessible by other means.

\section{Concluding Remarks \label{sec:sendoff}}
Just as new revelations about neutrino properties are enriching our understanding of particle physics, neutrino astronomy, astrophysics, and cosmology are blossoming. In this short tour, I have had to omit many topics of interest, so I close with a short guide to the literature on some of the neglected topics.

The r\^{o}le of light neutrinos in cosmology was explored early on~\cite{Tremaine:1979we}, and at a certain moment the possibility of a neutrino-dominated Universe had some currency~\cite{Schramm:1980xw}. Joel Primack reviewed the progression of dark-matter candidates, neutrinos included, in his SSI lecture~\cite{JoelPSSI}.

We now expect neutrinos to contribute a dash of hot dark matter to the composition of the Universe. Neutrinos may signal dark-matter annihilations at the center of our galaxy~\cite{Bertone:2004pz, Gondolo:2004sc,Cirelli:2005gh,Lehnert:2007fv,Barger:2007xf}, but rates may be unobservably small without a significant boost from dark-matter clumping~\cite{Berezinsky:2003vn}. The connection between dark matter and collider physics is emphasized in~\cite{Carena:2006nv,Peskin:2007nk}.
Coannihilation of cosmic neutrinos on dark-matter particles appears to be a negligible source of indirect dark-matter signals~\cite{Barenboim:2006dj}.

Neutrinos may be relevant to two outstanding challenges in cosmology. Under the right circumstances, a lepton--antilepton asymmetry in the early Universe drive the matter excess we observe today. The leptogenesis scenario is reviewed in~\cite{Buchmuller:2005eh,Fukugita:2005sb,Chen:2007fv}. Mass varying neutrinos can behave as a negative-pressure fluid that could be the origin of the accelerated expansion of the Universe~\cite{Hung:2000yg,Fardon:2003eh,Kaplan:2004dq}. This approach aims to correlate the range of neutrino masses suggested by neutrino-oscillation experiments with the scale of the cosmological constant employed to parametrize the cosmic acceleration, $\Lambda \approx (\hbox{a few meV})^4$.

\begin{acknowledgments}
Fermilab is operated by Fermi Research Alliance, LLC  under Contract No.~DE-AC02-07CH11359 with the United States Department of Energy.
 My enthusiastic thanks go to the organizers and 
participants in the XXXV SLAC Summer Institute for a very enjoyable 
and educational program. It is a pleasure to thank Luis \'{A}lvarez-Gaum\'{e} and other members of the CERN Theory Group for warm hospitality in Geneva. I thank Olga Mena for sharing her understanding of neutrino flavor change and for helpful comments on the manuscript.
\end{acknowledgments}

\bibliography{SSINeutrinos}

\begin{thebibliography}{153}
\expandafter\ifx\csname natexlab\endcsname\relax\def\natexlab#1{#1}\fi
\expandafter\ifx\csname bibnamefont\endcsname\relax
  \def\bibnamefont#1{#1}\fi
\expandafter\ifx\csname bibfnamefont\endcsname\relax
  \def\bibfnamefont#1{#1}\fi
\expandafter\ifx\csname citenamefont\endcsname\relax
  \def\citenamefont#1{#1}\fi
\expandafter\ifx\csname url\endcsname\relax
  \def\url#1{\texttt{#1}}\fi
\expandafter\ifx\csname urlprefix\endcsname\relax\def\urlprefix{URL }\fi
\providecommand{\bibinfo}[2]{#2}
\providecommand{\eprint}[2][]{\url{#2}}

\bibitem[{\citenamefont{Krauss et~al.}(1984)\citenamefont{Krauss, Glashow, and
  Schramm}}]{Krauss:1983zn}
\bibinfo{author}{\bibfnamefont{L.~M.} \bibnamefont{Krauss}},
  \bibinfo{author}{\bibfnamefont{S.~L.} \bibnamefont{Glashow}},
  \bibnamefont{and} \bibinfo{author}{\bibfnamefont{D.~N.}
  \bibnamefont{Schramm}}, \bibinfo{journal}{Nature}
  \textbf{\bibinfo{volume}{310}}, \bibinfo{pages}{191} (\bibinfo{year}{1984}).

\bibitem[{\citenamefont{Mantovani et~al.}(2004)\citenamefont{Mantovani,
  Carmignani, Fiorentini, and Lissia}}]{Mantovani:2003yd}
\bibinfo{author}{\bibfnamefont{F.}~\bibnamefont{Mantovani}},
  \bibinfo{author}{\bibfnamefont{L.}~\bibnamefont{Carmignani}},
  \bibinfo{author}{\bibfnamefont{G.}~\bibnamefont{Fiorentini}},
  \bibnamefont{and} \bibinfo{author}{\bibfnamefont{M.}~\bibnamefont{Lissia}},
  \bibinfo{journal}{Phys. Rev.} \textbf{\bibinfo{volume}{D69}},
  \bibinfo{pages}{013001} (\bibinfo{year}{2004}), \eprint{hep-ph/0309013}.

\bibitem[{\citenamefont{Araki et~al.}(2005{\natexlab{a}})}]{Araki:2005qa}
\bibinfo{author}{\bibfnamefont{T.}~\bibnamefont{Araki}} \bibnamefont{et~al.},
  \bibinfo{journal}{Nature} \textbf{\bibinfo{volume}{436}},
  \bibinfo{pages}{499} (\bibinfo{year}{2005}{\natexlab{a}}).

\bibitem[{\citenamefont{Duda et~al.}(2001)\citenamefont{Duda, Gelmini, and
  Nussinov}}]{Duda:2001hd}
\bibinfo{author}{\bibfnamefont{G.}~\bibnamefont{Duda}},
  \bibinfo{author}{\bibfnamefont{G.}~\bibnamefont{Gelmini}}, \bibnamefont{and}
  \bibinfo{author}{\bibfnamefont{S.}~\bibnamefont{Nussinov}},
  \bibinfo{journal}{Phys. Rev.} \textbf{\bibinfo{volume}{D64}},
  \bibinfo{pages}{122001} (\bibinfo{year}{2001}), \eprint{hep-ph/0107027}.

\bibitem[{\citenamefont{Ringwald and Wong}(2004)}]{Ringwald:2004np}
\bibinfo{author}{\bibfnamefont{A.}~\bibnamefont{Ringwald}} \bibnamefont{and}
  \bibinfo{author}{\bibfnamefont{Y.~Y.~Y.} \bibnamefont{Wong}},
  \bibinfo{journal}{JCAP} \textbf{\bibinfo{volume}{0412}}, \bibinfo{pages}{005}
  (\bibinfo{year}{2004}), \eprint{hep-ph/0408241}.

\bibitem[{\citenamefont{Cocco et~al.}(2007)\citenamefont{Cocco, Mangano, and
  Messina}}]{Cocco:2007za}
\bibinfo{author}{\bibfnamefont{A.~G.} \bibnamefont{Cocco}},
  \bibinfo{author}{\bibfnamefont{G.}~\bibnamefont{Mangano}}, \bibnamefont{and}
  \bibinfo{author}{\bibfnamefont{M.}~\bibnamefont{Messina}},
  \bibinfo{journal}{JCAP} \textbf{\bibinfo{volume}{0706}}, \bibinfo{pages}{015}
  (\bibinfo{year}{2007}), \eprint{hep-ph/0703075}.

\bibitem[{\citenamefont{Han and Hooper}(2004)}]{Han:2004kq}
\bibinfo{author}{\bibfnamefont{T.}~\bibnamefont{Han}} \bibnamefont{and}
  \bibinfo{author}{\bibfnamefont{D.}~\bibnamefont{Hooper}},
  \bibinfo{journal}{New J. Phys.} \textbf{\bibinfo{volume}{6}},
  \bibinfo{pages}{150} (\bibinfo{year}{2004}), \eprint{hep-ph/0408348}.

\bibitem[{\citenamefont{Yao et~al.}(2006)}]{Yao:2006px}
\bibinfo{author}{\bibfnamefont{W.-M.} \bibnamefont{Yao}} \bibnamefont{et~al.}
  (\bibinfo{collaboration}{Particle Data Group}), \bibinfo{journal}{J. Phys.}
  \textbf{\bibinfo{volume}{G33}}, \bibinfo{pages}{1} (\bibinfo{year}{2006}),
  \bibinfo{note}{and 2007 partial update for 2008},
  \urlprefix\url{pdg.lbl.gov}.

\bibitem[{\citenamefont{Dodelson}(2007{\natexlab{a}})}]{ScottSSI}
\bibinfo{author}{\bibfnamefont{S.}~\bibnamefont{Dodelson}}
  (\bibinfo{year}{2007}{\natexlab{a}}), \bibinfo{note}{\textit{Cosmology for
  Particle Physicists,} Lectures at the 2007 SLAC Summer Institute},
  \urlprefix\url{www-conf.slac.stanford.edu/ssi/2007/}.

\bibitem[{\citenamefont{Bahcall}(1989)}]{Bahcall}
\bibinfo{author}{\bibfnamefont{J.}~\bibnamefont{Bahcall}},
  \emph{\bibinfo{title}{Neutrino Astrophysics}} (\bibinfo{publisher}{Cambridge
  University Press}, \bibinfo{address}{Cambridge \& New York},
  \bibinfo{year}{1989}).

\bibitem[{\citenamefont{Giunti and Kim}(2007)}]{GiuntiKim}
\bibinfo{author}{\bibfnamefont{C.}~\bibnamefont{Giunti}} \bibnamefont{and}
  \bibinfo{author}{\bibfnamefont{C.~W.} \bibnamefont{Kim}},
  \emph{\bibinfo{title}{Fundamentals of Neutrino Physics and Astrophysics}}
  (\bibinfo{publisher}{Oxford University Press}, \bibinfo{address}{Oxford \&
  New York}, \bibinfo{year}{2007}).

\bibitem[{\citenamefont{Hannestad}(2006)}]{Hannestad:2006zg}
\bibinfo{author}{\bibfnamefont{S.}~\bibnamefont{Hannestad}},
  \bibinfo{journal}{Ann. Rev. Nucl. Part. Sci.} \textbf{\bibinfo{volume}{56}},
  \bibinfo{pages}{137} (\bibinfo{year}{2006}), \eprint{hep-ph/0602058}.

\bibitem[{\citenamefont{Mangano}(2007)}]{Gianpiero}
\bibinfo{author}{\bibfnamefont{G.}~\bibnamefont{Mangano}}
  (\bibinfo{year}{2007}), \bibinfo{note}{\textit{Neutrinos in Cosmology,}
  Lectures at the III International Pontecorvo Neutrino Physics School},
  \urlprefix\url{www.jinr.ru/pontecorvo07/}.

\bibitem[{\citenamefont{{Fermilab--KEK Neutrino School}}(2007)}]{NUSS}
\bibinfo{author}{\bibnamefont{{Fermilab--KEK Neutrino School}}}
  (\bibinfo{year}{2007}), \urlprefix\url{nuss.fnal.gov}.

\bibitem[{\citenamefont{Khlopov}(1999)}]{Khlopov:437816}
\bibinfo{author}{\bibfnamefont{M.~Yu.} \bibnamefont{Khlopov}},
  \emph{\bibinfo{title}{Cosmoparticle Physics}} (\bibinfo{publisher}{World
  Scientific}, \bibinfo{address}{Singapore}, \bibinfo{year}{1999}).

\bibitem[{\citenamefont{Steigman}(2006{\natexlab{a}})}]{Steigman:2005uz}
\bibinfo{author}{\bibfnamefont{G.}~\bibnamefont{Steigman}},
  \bibinfo{journal}{Int. J. Mod. Phys.} \textbf{\bibinfo{volume}{E15}},
  \bibinfo{pages}{1} (\bibinfo{year}{2006}{\natexlab{a}}),
  \eprint{astro-ph/0511534}.

\bibitem[{\citenamefont{Fields and Sarkar}(2006)}]{FieldsSarkarBBN}
\bibinfo{author}{\bibfnamefont{B.~D.} \bibnamefont{Fields}} \bibnamefont{and}
  \bibinfo{author}{\bibfnamefont{S.}~\bibnamefont{Sarkar}}
  (\bibinfo{year}{2006}), \bibinfo{note}{in~\cite{Yao:2006px}, \S~20},
  \urlprefix\url{pdg.lbl.gov/2007/reviews/bigbangnucrpp.pdf}.

\bibitem[{\citenamefont{Bennett et~al.}(2003)}]{Bennett:2003bz}
\bibinfo{author}{\bibfnamefont{C.~L.} \bibnamefont{Bennett}}
  \bibnamefont{et~al.} (\bibinfo{collaboration}{WMAP Collaboration}),
  \bibinfo{journal}{Astrophys. J. Suppl.} \textbf{\bibinfo{volume}{148}},
  \bibinfo{pages}{1} (\bibinfo{year}{2003}), \eprint{astro-ph/0302207}.

\bibitem[{\citenamefont{Bond et~al.}(1980)\citenamefont{Bond, Efstathiou, and
  Silk}}]{Bond:1980ha}
\bibinfo{author}{\bibfnamefont{J.~R.} \bibnamefont{Bond}},
  \bibinfo{author}{\bibfnamefont{G.}~\bibnamefont{Efstathiou}},
  \bibnamefont{and} \bibinfo{author}{\bibfnamefont{J.}~\bibnamefont{Silk}},
  \bibinfo{journal}{Phys. Rev. Lett.} \textbf{\bibinfo{volume}{45}},
  \bibinfo{pages}{1980} (\bibinfo{year}{1980}).

\bibitem[{\citenamefont{Tegmark}(2005)}]{Tegmark:2005cy}
\bibinfo{author}{\bibfnamefont{M.}~\bibnamefont{Tegmark}},
  \bibinfo{journal}{Phys. Scripta} \textbf{\bibinfo{volume}{T121}},
  \bibinfo{pages}{153} (\bibinfo{year}{2005}), \eprint{hep-ph/0503257}.

\bibitem[{\citenamefont{Barger et~al.}(2003)\citenamefont{Barger, Kneller, Lee,
  Marfatia, and Steigman}}]{Barger:2003zg}
\bibinfo{author}{\bibfnamefont{V.}~\bibnamefont{Barger}},
  \bibinfo{author}{\bibfnamefont{J.~P.} \bibnamefont{Kneller}},
  \bibinfo{author}{\bibfnamefont{H.-S.} \bibnamefont{Lee}},
  \bibinfo{author}{\bibfnamefont{D.}~\bibnamefont{Marfatia}}, \bibnamefont{and}
  \bibinfo{author}{\bibfnamefont{G.}~\bibnamefont{Steigman}},
  \bibinfo{journal}{Phys. Lett.} \textbf{\bibinfo{volume}{B566}},
  \bibinfo{pages}{8} (\bibinfo{year}{2003}), \eprint{hep-ph/0305075}.

\bibitem[{\citenamefont{Steigman}(2006{\natexlab{b}})}]{Steigman:2006mv}
\bibinfo{author}{\bibfnamefont{G.}~\bibnamefont{Steigman}}
  (\bibinfo{year}{2006}{\natexlab{b}}), \eprint{astro-ph/0610599}.

\bibitem[{\citenamefont{de~Bernardis et~al.}(2007)\citenamefont{de~Bernardis,
  Melchiorri, Verde, and Jimenez}}]{deBernardis:2007bu}
\bibinfo{author}{\bibfnamefont{F.}~\bibnamefont{de~Bernardis}},
  \bibinfo{author}{\bibfnamefont{A.}~\bibnamefont{Melchiorri}},
  \bibinfo{author}{\bibfnamefont{L.}~\bibnamefont{Verde}}, \bibnamefont{and}
  \bibinfo{author}{\bibfnamefont{R.}~\bibnamefont{Jimenez}}
  (\bibinfo{year}{2007}), \eprint{arXiv:0707.4170}.

\bibitem[{\citenamefont{Gershtein and Zeldovich}(1966)}]{Gershtein:1966gg}
\bibinfo{author}{\bibfnamefont{S.~S.} \bibnamefont{Gershtein}}
  \bibnamefont{and} \bibinfo{author}{\bibfnamefont{Y.~B.}
  \bibnamefont{Zeldovich}}, \bibinfo{journal}{JETP Lett.}
  \textbf{\bibinfo{volume}{4}}, \bibinfo{pages}{120} (\bibinfo{year}{1966}).

\bibitem[{\citenamefont{Cowsik and McClelland}(1972)}]{Cowsik:1972gh}
\bibinfo{author}{\bibfnamefont{R.}~\bibnamefont{Cowsik}} \bibnamefont{and}
  \bibinfo{author}{\bibfnamefont{J.}~\bibnamefont{McClelland}},
  \bibinfo{journal}{Phys. Rev. Lett.} \textbf{\bibinfo{volume}{29}},
  \bibinfo{pages}{669} (\bibinfo{year}{1972}).

\bibitem[{\citenamefont{Fogli et~al.}(2004)}]{Fogli:2004as}
\bibinfo{author}{\bibfnamefont{G.~L.} \bibnamefont{Fogli}}
  \bibnamefont{et~al.}, \bibinfo{journal}{Phys. Rev.}
  \textbf{\bibinfo{volume}{D70}}, \bibinfo{pages}{113003}
  (\bibinfo{year}{2004}), \eprint{hep-ph/0408045}.

\bibitem[{\citenamefont{Fukuda et~al.}(1998)}]{Fukuda:1998mi}
\bibinfo{author}{\bibfnamefont{Y.}~\bibnamefont{Fukuda}} \bibnamefont{et~al.}
  (\bibinfo{collaboration}{Super-Kamiokande Collaboration}),
  \bibinfo{journal}{Phys. Rev. Lett.} \textbf{\bibinfo{volume}{81}},
  \bibinfo{pages}{1562} (\bibinfo{year}{1998}), \eprint{hep-ex/9807003}.

\bibitem[{\citenamefont{Ahmad et~al.}(2002)}]{Ahmad:2002jz}
\bibinfo{author}{\bibfnamefont{Q.~R.} \bibnamefont{Ahmad}} \bibnamefont{et~al.}
  (\bibinfo{collaboration}{SNO Collaboration}), \bibinfo{journal}{Phys. Rev.
  Lett.} \textbf{\bibinfo{volume}{89}}, \bibinfo{pages}{011301}
  (\bibinfo{year}{2002}), \eprint{nucl-ex/0204008}.

\bibitem[{\citenamefont{Eguchi et~al.}(2003)}]{Eguchi:2002dm}
\bibinfo{author}{\bibfnamefont{K.}~\bibnamefont{Eguchi}} \bibnamefont{et~al.}
  (\bibinfo{collaboration}{KamLAND Collaboration}), \bibinfo{journal}{Phys.
  Rev. Lett.} \textbf{\bibinfo{volume}{90}}, \bibinfo{pages}{021802}
  (\bibinfo{year}{2003}), \eprint{hep-ex/0212021}.

\bibitem[{\citenamefont{Goldhaber et~al.}(1958)\citenamefont{Goldhaber,
  Grodzins, and Sunyar}}]{PhysRev.109.1015}
\bibinfo{author}{\bibfnamefont{M.}~\bibnamefont{Goldhaber}},
  \bibinfo{author}{\bibfnamefont{L.}~\bibnamefont{Grodzins}}, \bibnamefont{and}
  \bibinfo{author}{\bibfnamefont{A.~W.} \bibnamefont{Sunyar}},
  \bibinfo{journal}{Phys. Rev.} \textbf{\bibinfo{volume}{109}},
  \bibinfo{pages}{1015} (\bibinfo{year}{1958}).

\bibitem[{\citenamefont{Bardon et~al.}(1961)\citenamefont{Bardon, Franzini, and
  Lee}}]{PhysRevLett.7.23}
\bibinfo{author}{\bibfnamefont{M.}~\bibnamefont{Bardon}},
  \bibinfo{author}{\bibfnamefont{P.}~\bibnamefont{Franzini}}, \bibnamefont{and}
  \bibinfo{author}{\bibfnamefont{J.}~\bibnamefont{Lee}},
  \bibinfo{journal}{Phys. Rev. Lett.} \textbf{\bibinfo{volume}{7}},
  \bibinfo{pages}{23} (\bibinfo{year}{1961}).

\bibitem[{\citenamefont{Possoz et~al.}(1977)}]{Possoz:1977jf}
\bibinfo{author}{\bibfnamefont{A.}~\bibnamefont{Possoz}} \bibnamefont{et~al.},
  \bibinfo{journal}{Phys. Lett.} \textbf{\bibinfo{volume}{B70}},
  \bibinfo{pages}{265} (\bibinfo{year}{1977}), \bibinfo{note}{erratum-ibid.
  \textbf{B73,} 504 (1978)}.

\bibitem[{\citenamefont{Abe et~al.}(1997)}]{PhysRevLett.78.4691}
\bibinfo{author}{\bibfnamefont{K.}~\bibnamefont{Abe}} \bibnamefont{et~al.}
  (\bibinfo{collaboration}{SLD Collaboration}), \bibinfo{journal}{Phys. Rev.
  Lett.} \textbf{\bibinfo{volume}{78}}, \bibinfo{pages}{4691}
  (\bibinfo{year}{1997}).

\bibitem[{\citenamefont{Quigg}(2007)}]{CQNUSS}
\bibinfo{author}{\bibfnamefont{C.}~\bibnamefont{Quigg}} (\bibinfo{year}{2007}),
  \bibinfo{note}{\textit{Neutrinos in the Electroweak Theory,} Lectures at the
  2007 Fermilab/KEK Neutrino Physics Summer School},
  \urlprefix\url{lutece.fnal.gov/Talks/CQNuSchool.pdf}.

\bibitem[{\citenamefont{Parke}(2007)}]{Parke:2006mr}
\bibinfo{author}{\bibfnamefont{S.}~\bibnamefont{Parke}}, in
  \emph{\bibinfo{booktitle}{Advanced Summer School in Physics 2006: Frontiers
  in Contemporary Physics: EAV06}}, edited by
  \bibinfo{editor}{\bibfnamefont{O.}~\bibnamefont{Miranda}}
  \bibnamefont{et~al.} (\bibinfo{publisher}{American Institute of Physics},
  \bibinfo{address}{New York}, \bibinfo{year}{2007}), pp.
  \bibinfo{pages}{69--84}, \bibinfo{note}{{AIP} Conference Proceedings vol.
  885}, \urlprefix\url{link.aip.org/link/?APCPCS/885/69/1}.

\bibitem[{\citenamefont{Mohapatra}(2007)}]{RabiNUSS}
\bibinfo{author}{\bibfnamefont{R.~N.} \bibnamefont{Mohapatra}}
  (\bibinfo{year}{2007}), \bibinfo{note}{\textit{Physics of Neutrino Mass,}
  Lectures at the 2007 Fermilab/KEK Neutrino Physics Summer School},
  \urlprefix\url{projects.fnal.gov/nuss/lectures/RabiM_1.pdf}.

\bibitem[{\citenamefont{Drexlin}(2006)}]{Drexlin}
\bibinfo{author}{\bibfnamefont{G.}~\bibnamefont{Drexlin}}
  (\bibinfo{year}{2006}), \bibinfo{note}{\textit{The KATRIN Experiment,} talk
  at the GERDA Collaboration Meeting, MPIK Heidelberg},
  \urlprefix\url{www-ik.fzk.de/~katrin/publications/talks/mpik2006.pdf}.

\bibitem[{\citenamefont{Raffelt}(1985)}]{PhysRevD.31.3002}
\bibinfo{author}{\bibfnamefont{G.~G.} \bibnamefont{Raffelt}},
  \bibinfo{journal}{Phys. Rev. D} \textbf{\bibinfo{volume}{31}},
  \bibinfo{pages}{3002} (\bibinfo{year}{1985}).

\bibitem[{\citenamefont{Mirizzi et~al.}(2007)\citenamefont{Mirizzi, Montanino,
  and Serpico}}]{Mirizzi:2007jd}
\bibinfo{author}{\bibfnamefont{A.}~\bibnamefont{Mirizzi}},
  \bibinfo{author}{\bibfnamefont{D.}~\bibnamefont{Montanino}},
  \bibnamefont{and} \bibinfo{author}{\bibfnamefont{P.~D.}
  \bibnamefont{Serpico}}, \bibinfo{journal}{Phys. Rev.}
  \textbf{\bibinfo{volume}{D76}}, \bibinfo{pages}{053007}
  (\bibinfo{year}{2007}), \eprint{arXiv:0705.4667}.

\bibitem[{\citenamefont{Beacom and Bell}(2002)}]{Beacom:2002cb}
\bibinfo{author}{\bibfnamefont{J.~F.} \bibnamefont{Beacom}} \bibnamefont{and}
  \bibinfo{author}{\bibfnamefont{N.~F.} \bibnamefont{Bell}},
  \bibinfo{journal}{Phys. Rev.} \textbf{\bibinfo{volume}{D65}},
  \bibinfo{pages}{113009} (\bibinfo{year}{2002}), \eprint{hep-ph/0204111}.

\bibitem[{\citenamefont{Kayser}(2004)}]{Kayser:2005cd}
\bibinfo{author}{\bibfnamefont{B.}~\bibnamefont{Kayser}},
  \bibinfo{journal}{ECONF} \textbf{\bibinfo{volume}{C040802}},
  \bibinfo{pages}{L004} (\bibinfo{year}{2004}), \eprint{hep-ph/0506165}.

\bibitem[{\citenamefont{Davis}(2003)}]{Davis:2003kh}
\bibinfo{author}{\bibfnamefont{R.}~\bibnamefont{Davis}}, \bibinfo{journal}{Rev.
  Mod. Phys.} \textbf{\bibinfo{volume}{75}}, \bibinfo{pages}{985}
  (\bibinfo{year}{2003}),
  \urlprefix\url{nobelprize.org/nobel_prizes/physics/laureates/2002/davis-lect%
ure.html}.

\bibitem[{\citenamefont{Koshiba}(2003)}]{Koshiba:2003xy}
\bibinfo{author}{\bibfnamefont{M.}~\bibnamefont{Koshiba}},
  \bibinfo{journal}{Rev. Mod. Phys.} \textbf{\bibinfo{volume}{75}},
  \bibinfo{pages}{1011} (\bibinfo{year}{2003}),
  \urlprefix\url{nobelprize.org/nobel_prizes/physics/laureates/2002/koshiba-le%
cture.html}.

\bibitem[{\citenamefont{Strumia and Vissani}(2006--2008)}]{Strumia:2006db}
\bibinfo{author}{\bibfnamefont{A.}~\bibnamefont{Strumia}} \bibnamefont{and}
  \bibinfo{author}{\bibfnamefont{F.}~\bibnamefont{Vissani}}
  (\bibinfo{year}{2006--2008}), \bibinfo{note}{{``Neutrino masses and mixings
  and \ldots ''} an organic review, the most recent version is available at
  \url{www.pi.infn.it/~astrumia/review.html}}, \eprint{hep-ph/0606054}.

\bibitem[{\citenamefont{Hirata et~al.}(1992)}]{Hirata:1992ku}
\bibinfo{author}{\bibfnamefont{K.~S.} \bibnamefont{Hirata}}
  \bibnamefont{et~al.} (\bibinfo{collaboration}{Kamiokande-II Collaboration}),
  \bibinfo{journal}{Phys. Lett.} \textbf{\bibinfo{volume}{B280}},
  \bibinfo{pages}{146} (\bibinfo{year}{1992}).

\bibitem[{\citenamefont{Ashie et~al.}(2005)}]{Ashie:2005ik}
\bibinfo{author}{\bibfnamefont{Y.}~\bibnamefont{Ashie}} \bibnamefont{et~al.}
  (\bibinfo{collaboration}{Super-Kamiokande Collaboration}),
  \bibinfo{journal}{Phys. Rev.} \textbf{\bibinfo{volume}{D71}},
  \bibinfo{pages}{112005} (\bibinfo{year}{2005}), \eprint{hep-ex/0501064}.

\bibitem[{\citenamefont{Ahn et~al.}(2006)}]{Ahn:2006zz}
\bibinfo{author}{\bibfnamefont{M.~H.} \bibnamefont{Ahn}} \bibnamefont{et~al.}
  (\bibinfo{collaboration}{K2K Collaboration}), \bibinfo{journal}{Phys. Rev.}
  \textbf{\bibinfo{volume}{D74}}, \bibinfo{pages}{072003}
  (\bibinfo{year}{2006}), \eprint{hep-ex/0606032}.

\bibitem[{\citenamefont{Michael et~al.}(2006)}]{Michael:2006rx}
\bibinfo{author}{\bibfnamefont{D.~G.} \bibnamefont{Michael}}
  \bibnamefont{et~al.} (\bibinfo{collaboration}{MINOS Collaboration}),
  \bibinfo{journal}{Phys. Rev. Lett.} \textbf{\bibinfo{volume}{97}},
  \bibinfo{pages}{191801} (\bibinfo{year}{2006}), \eprint{hep-ex/0607088}.

\bibitem[{\citenamefont{{MINOS Collaboration}}(2007)}]{MINOS:2007zz}
\bibinfo{author}{\bibnamefont{{MINOS Collaboration}}} (\bibinfo{year}{2007}),
  \eprint{arXiv:0708.1495}.

\bibitem[{\citenamefont{Ashie et~al.}(2004)}]{Ashie:2004mr}
\bibinfo{author}{\bibfnamefont{Y.}~\bibnamefont{Ashie}} \bibnamefont{et~al.}
  (\bibinfo{collaboration}{Super-Kamiokande Collaboration}),
  \bibinfo{journal}{Phys. Rev. Lett.} \textbf{\bibinfo{volume}{93}},
  \bibinfo{pages}{101801} (\bibinfo{year}{2004}), \eprint{hep-ex/0404034}.

\bibitem[{\citenamefont{Araki et~al.}(2005{\natexlab{b}})}]{Araki:2004mb}
\bibinfo{author}{\bibfnamefont{T.}~\bibnamefont{Araki}} \bibnamefont{et~al.}
  (\bibinfo{collaboration}{KamLAND Collaboration}), \bibinfo{journal}{Phys.
  Rev. Lett.} \textbf{\bibinfo{volume}{94}}, \bibinfo{pages}{081801}
  (\bibinfo{year}{2005}{\natexlab{b}}), \eprint{hep-ex/0406035}.

\bibitem[{\citenamefont{Abe et~al.}(2008)}]{KamLAND:2008ee}
\bibinfo{author}{\bibfnamefont{S.}~\bibnamefont{Abe}} \bibnamefont{et~al.}
  (\bibinfo{collaboration}{KamLAND Collaboration}) (\bibinfo{year}{2008}),
  \eprint{arXiv:0801.4589}.

\bibitem[{\citenamefont{Arpesella et~al.}(2008)}]{Arpesella:2007xf}
\bibinfo{author}{\bibfnamefont{C.}~\bibnamefont{Arpesella}}
  \bibnamefont{et~al.} (\bibinfo{collaboration}{Borexino Collaboration}),
  \bibinfo{journal}{Phys. Lett.} \textbf{\bibinfo{volume}{B658}},
  \bibinfo{pages}{101} (\bibinfo{year}{2008}), \eprint{arXiv:0708.2251}.

\bibitem[{\citenamefont{Fukuda et~al.}(2001)}]{Fukuda:2001nj}
\bibinfo{author}{\bibfnamefont{S.}~\bibnamefont{Fukuda}} \bibnamefont{et~al.}
  (\bibinfo{collaboration}{Super-Kamiokande Collaboration}),
  \bibinfo{journal}{Phys. Rev. Lett.} \textbf{\bibinfo{volume}{86}},
  \bibinfo{pages}{5651} (\bibinfo{year}{2001}), \eprint{hep-ex/0103032}.

\bibitem[{\citenamefont{Ahmad et~al.}(2001)}]{Ahmad:2001an}
\bibinfo{author}{\bibfnamefont{Q.~R.} \bibnamefont{Ahmad}} \bibnamefont{et~al.}
  (\bibinfo{collaboration}{SNO Collaboration}), \bibinfo{journal}{Phys. Rev.
  Lett.} \textbf{\bibinfo{volume}{87}}, \bibinfo{pages}{071301}
  (\bibinfo{year}{2001}), \eprint{nucl-ex/0106015}.

\bibitem[{\citenamefont{Bahcall et~al.}(2005)\citenamefont{Bahcall, Serenelli,
  and Basu}}]{Bahcall:2004pz}
\bibinfo{author}{\bibfnamefont{J.~N.} \bibnamefont{Bahcall}},
  \bibinfo{author}{\bibfnamefont{A.~M.} \bibnamefont{Serenelli}},
  \bibnamefont{and} \bibinfo{author}{\bibfnamefont{S.}~\bibnamefont{Basu}},
  \bibinfo{journal}{Astrophys. J.} \textbf{\bibinfo{volume}{621}},
  \bibinfo{pages}{L85} (\bibinfo{year}{2005}), \eprint{astro-ph/0412440}.

\bibitem[{\citenamefont{Fukuda et~al.}(2002)}]{Fukuda:2002pe}
\bibinfo{author}{\bibfnamefont{S.}~\bibnamefont{Fukuda}} \bibnamefont{et~al.}
  (\bibinfo{collaboration}{Super-Kamiokande Collaboration}),
  \bibinfo{journal}{Phys. Lett.} \textbf{\bibinfo{volume}{B539}},
  \bibinfo{pages}{179} (\bibinfo{year}{2002}), \eprint{hep-ex/0205075}.

\bibitem[{\citenamefont{Aharmim et~al.}(2005)}]{Aharmim:2005gt}
\bibinfo{author}{\bibfnamefont{B.}~\bibnamefont{Aharmim}} \bibnamefont{et~al.}
  (\bibinfo{collaboration}{SNO}), \bibinfo{journal}{Phys. Rev.}
  \textbf{\bibinfo{volume}{C72}}, \bibinfo{pages}{055502}
  (\bibinfo{year}{2005}), \eprint{nucl-ex/0502021}.

\bibitem[{\citenamefont{Aguilar-Arevalo et~al.}(2007)}]{AguilarArevalo:2007it}
\bibinfo{author}{\bibfnamefont{A.~A.} \bibnamefont{Aguilar-Arevalo}}
  \bibnamefont{et~al.} (\bibinfo{collaboration}{MiniBooNE Collaboration}),
  \bibinfo{journal}{Phys. Rev. Lett.} \textbf{\bibinfo{volume}{98}},
  \bibinfo{pages}{231801} (\bibinfo{year}{2007}), \eprint{arXiv:0704.1500}.

\bibitem[{\citenamefont{Steigman}(1979)}]{Steigman:1979kw}
\bibinfo{author}{\bibfnamefont{G.}~\bibnamefont{Steigman}},
  \bibinfo{journal}{Ann. Rev. Nucl. Part. Sci.} \textbf{\bibinfo{volume}{29}},
  \bibinfo{pages}{313} (\bibinfo{year}{1979}).

\bibitem[{\citenamefont{Barenboim et~al.}(2005)\citenamefont{Barenboim,
  Mena~Requejo, and Quigg}}]{Barenboim:2004di}
\bibinfo{author}{\bibfnamefont{G.}~\bibnamefont{Barenboim}},
  \bibinfo{author}{\bibfnamefont{O.}~\bibnamefont{Mena~Requejo}},
  \bibnamefont{and} \bibinfo{author}{\bibfnamefont{C.}~\bibnamefont{Quigg}},
  \bibinfo{journal}{Phys. Rev.} \textbf{\bibinfo{volume}{D71}},
  \bibinfo{pages}{083002} (\bibinfo{year}{2005}), \eprint{hep-ph/0412122}.

\bibitem[{\citenamefont{Lopez et~al.}(1999)\citenamefont{Lopez, Dodelson,
  Heckler, and Turner}}]{Lopez:1998aq}
\bibinfo{author}{\bibfnamefont{R.~E.} \bibnamefont{Lopez}},
  \bibinfo{author}{\bibfnamefont{S.}~\bibnamefont{Dodelson}},
  \bibinfo{author}{\bibfnamefont{A.}~\bibnamefont{Heckler}}, \bibnamefont{and}
  \bibinfo{author}{\bibfnamefont{M.~S.} \bibnamefont{Turner}},
  \bibinfo{journal}{Phys. Rev. Lett.} \textbf{\bibinfo{volume}{82}},
  \bibinfo{pages}{3952} (\bibinfo{year}{1999}), \eprint{astro-ph/9803095}.

\bibitem[{\citenamefont{Cowsik and McClelland}(1973)}]{Cowsik:1973yj}
\bibinfo{author}{\bibfnamefont{R.}~\bibnamefont{Cowsik}} \bibnamefont{and}
  \bibinfo{author}{\bibfnamefont{J.}~\bibnamefont{McClelland}},
  \bibinfo{journal}{Astrophys. J.} \textbf{\bibinfo{volume}{180}},
  \bibinfo{pages}{7} (\bibinfo{year}{1973}).

\bibitem[{\citenamefont{Szalay and Marx}(1976)}]{Szalay:1976ef}
\bibinfo{author}{\bibfnamefont{A.~S.} \bibnamefont{Szalay}} \bibnamefont{and}
  \bibinfo{author}{\bibfnamefont{G.}~\bibnamefont{Marx}},
  \bibinfo{journal}{Astron. Astrophys.} \textbf{\bibinfo{volume}{49}},
  \bibinfo{pages}{437} (\bibinfo{year}{1976}).

\bibitem[{\citenamefont{Beacom et~al.}(2004)\citenamefont{Beacom, Bell, and
  Dodelson}}]{Beacom:2004yd}
\bibinfo{author}{\bibfnamefont{J.~F.} \bibnamefont{Beacom}},
  \bibinfo{author}{\bibfnamefont{N.~F.} \bibnamefont{Bell}}, \bibnamefont{and}
  \bibinfo{author}{\bibfnamefont{S.}~\bibnamefont{Dodelson}},
  \bibinfo{journal}{Phys. Rev. Lett.} \textbf{\bibinfo{volume}{93}},
  \bibinfo{pages}{121302} (\bibinfo{year}{2004}), \eprint{astro-ph/0404585}.

\bibitem[{\citenamefont{Gelmini}(2005)}]{Gelmini:2004hg}
\bibinfo{author}{\bibfnamefont{G.~B.} \bibnamefont{Gelmini}},
  \bibinfo{journal}{Phys. Scripta} \textbf{\bibinfo{volume}{T121}},
  \bibinfo{pages}{131} (\bibinfo{year}{2005}), \eprint{hep-ph/0412305}.

\bibitem[{\citenamefont{Fukugita}(2006)}]{Fukugita:2005sb}
\bibinfo{author}{\bibfnamefont{M.}~\bibnamefont{Fukugita}},
  \bibinfo{journal}{Nucl. Phys. Proc. Suppl.} \textbf{\bibinfo{volume}{155}},
  \bibinfo{pages}{10} (\bibinfo{year}{2006}), \eprint{hep-ph/0511068}.

\bibitem[{\citenamefont{Giunti and Laveder}(2007)}]{NuinC}
\bibinfo{author}{\bibfnamefont{C.}~\bibnamefont{Giunti}} \bibnamefont{and}
  \bibinfo{author}{\bibfnamefont{M.}~\bibnamefont{Laveder}}
  (\bibinfo{year}{2007}), \bibinfo{note}{\textit{Neutrino Mass Limits}},
  \urlprefix\url{www.nu.to.infn.it/Neutrino_Cosmology}.

\bibitem[{\citenamefont{Lesgourgues and Pastor}(2006)}]{Lesgourgues:2006nd}
\bibinfo{author}{\bibfnamefont{J.}~\bibnamefont{Lesgourgues}} \bibnamefont{and}
  \bibinfo{author}{\bibfnamefont{S.}~\bibnamefont{Pastor}},
  \bibinfo{journal}{Phys. Rept.} \textbf{\bibinfo{volume}{429}},
  \bibinfo{pages}{307} (\bibinfo{year}{2006}), \eprint{astro-ph/0603494}.

\bibitem[{\citenamefont{Dodelson}(2007{\natexlab{b}})}]{ScottNUSS}
\bibinfo{author}{\bibfnamefont{S.}~\bibnamefont{Dodelson}}
  (\bibinfo{year}{2007}{\natexlab{b}}), \bibinfo{note}{\textit{Neutrinos and
  the Universe---Cosmology,} Lecture at the 2007 Fermilab/KEK Neutrino Physics
  Summer School}, \urlprefix\url{theory.fnal.gov/jetp/talks/dodelson.pdf}.

\bibitem[{\citenamefont{Tegmark et~al.}(2004)}]{Tegmark:2003ud}
\bibinfo{author}{\bibfnamefont{M.}~\bibnamefont{Tegmark}} \bibnamefont{et~al.}
  (\bibinfo{collaboration}{SDSS Collaboration}), \bibinfo{journal}{Phys. Rev.}
  \textbf{\bibinfo{volume}{D69}}, \bibinfo{pages}{103501}
  (\bibinfo{year}{2004}), \eprint{astro-ph/0310723}.

\bibitem[{\citenamefont{Spergel et~al.}(2007)}]{Spergel:2006hy}
\bibinfo{author}{\bibfnamefont{D.~N.} \bibnamefont{Spergel}}
  \bibnamefont{et~al.} (\bibinfo{collaboration}{WMAP Collaboration}),
  \bibinfo{journal}{Astrophys. J. Suppl.} \textbf{\bibinfo{volume}{170}},
  \bibinfo{pages}{377} (\bibinfo{year}{2007}), \eprint{astro-ph/0603449}.

\bibitem[{\citenamefont{Goobar et~al.}(2006)\citenamefont{Goobar, Hannestad,
  Mortsell, and Tu}}]{Goobar:2006xz}
\bibinfo{author}{\bibfnamefont{A.}~\bibnamefont{Goobar}},
  \bibinfo{author}{\bibfnamefont{S.}~\bibnamefont{Hannestad}},
  \bibinfo{author}{\bibfnamefont{E.}~\bibnamefont{Mortsell}}, \bibnamefont{and}
  \bibinfo{author}{\bibfnamefont{H.}~\bibnamefont{Tu}}, \bibinfo{journal}{JCAP}
  \textbf{\bibinfo{volume}{0606}}, \bibinfo{pages}{019} (\bibinfo{year}{2006}),
  \eprint{astro-ph/0602155}.

\bibitem[{\citenamefont{Melchiorri et~al.}(2006)\citenamefont{Melchiorri,
  Serra, Dodelson, and Slosar}}]{Melchiorri:2006nj}
\bibinfo{author}{\bibfnamefont{A.}~\bibnamefont{Melchiorri}},
  \bibinfo{author}{\bibfnamefont{P.}~\bibnamefont{Serra}},
  \bibinfo{author}{\bibfnamefont{S.}~\bibnamefont{Dodelson}}, \bibnamefont{and}
  \bibinfo{author}{\bibfnamefont{A.}~\bibnamefont{Slosar}},
  \bibinfo{journal}{New Astron. Rev.} \textbf{\bibinfo{volume}{50}},
  \bibinfo{pages}{1020} (\bibinfo{year}{2006}).

\bibitem[{\citenamefont{Seljak et~al.}(2006)\citenamefont{Seljak, Slosar, and
  McDonald}}]{Seljak:2006bg}
\bibinfo{author}{\bibfnamefont{U.}~\bibnamefont{Seljak}},
  \bibinfo{author}{\bibfnamefont{A.}~\bibnamefont{Slosar}}, \bibnamefont{and}
  \bibinfo{author}{\bibfnamefont{P.}~\bibnamefont{McDonald}},
  \bibinfo{journal}{JCAP} \textbf{\bibinfo{volume}{0610}}, \bibinfo{pages}{014}
  (\bibinfo{year}{2006}), \eprint{astro-ph/0604335}.

\bibitem[{\citenamefont{Cirelli and Strumia}(2006)}]{Cirelli:2006kt}
\bibinfo{author}{\bibfnamefont{M.}~\bibnamefont{Cirelli}} \bibnamefont{and}
  \bibinfo{author}{\bibfnamefont{A.}~\bibnamefont{Strumia}},
  \bibinfo{journal}{JCAP} \textbf{\bibinfo{volume}{0612}}, \bibinfo{pages}{013}
  (\bibinfo{year}{2006}), \eprint{astro-ph/0607086}.

\bibitem[{\citenamefont{Serpico}(2007)}]{Serpico:2007pt}
\bibinfo{author}{\bibfnamefont{P.~D.} \bibnamefont{Serpico}},
  \bibinfo{journal}{Phys. Rev. Lett.} \textbf{\bibinfo{volume}{98}},
  \bibinfo{pages}{171301} (\bibinfo{year}{2007}), \eprint{astro-ph/0701699}.

\bibitem[{\citenamefont{Cooray}(1999)}]{Cooray:1999rv}
\bibinfo{author}{\bibfnamefont{A.~R.} \bibnamefont{Cooray}},
  \bibinfo{journal}{Astron. Astrophys.} \textbf{\bibinfo{volume}{348}},
  \bibinfo{pages}{31} (\bibinfo{year}{1999}), \eprint{astro-ph/9904246}.

\bibitem[{\citenamefont{Bilenky et~al.}(2003)\citenamefont{Bilenky, Giunti,
  Grifols, and Mass\'{o}}}]{Bilenky:2002aw}
\bibinfo{author}{\bibfnamefont{S.~M.} \bibnamefont{Bilenky}},
  \bibinfo{author}{\bibfnamefont{C.}~\bibnamefont{Giunti}},
  \bibinfo{author}{\bibfnamefont{J.~A.} \bibnamefont{Grifols}},
  \bibnamefont{and}
  \bibinfo{author}{\bibfnamefont{E.}~\bibnamefont{Mass\'{o}}},
  \bibinfo{journal}{Phys. Rept.} \textbf{\bibinfo{volume}{379}},
  \bibinfo{pages}{69} (\bibinfo{year}{2003}), \eprint{hep-ph/0211462}.

\bibitem[{\citenamefont{King}(2004)}]{King:2003jb}
\bibinfo{author}{\bibfnamefont{S.~F.} \bibnamefont{King}},
  \bibinfo{journal}{Rept. Prog. Phys.} \textbf{\bibinfo{volume}{67}},
  \bibinfo{pages}{107} (\bibinfo{year}{2004}), \eprint{hep-ph/0310204}.

\bibitem[{\citenamefont{Mohapatra}(2004)}]{Mohapatra:2004ht}
\bibinfo{author}{\bibfnamefont{R.~N.} \bibnamefont{Mohapatra}},
  \bibinfo{journal}{ECONF} \textbf{\bibinfo{volume}{C040802}},
  \bibinfo{pages}{L011} (\bibinfo{year}{2004}), \eprint{hep-ph/0411131}.

\bibitem[{\citenamefont{Mohapatra and Smirnov}(2006)}]{Mohapatra:2006gs}
\bibinfo{author}{\bibfnamefont{R.~N.} \bibnamefont{Mohapatra}}
  \bibnamefont{and} \bibinfo{author}{\bibfnamefont{A.~Y.}
  \bibnamefont{Smirnov}}, \bibinfo{journal}{Ann. Rev. Nucl. Part. Sci.}
  \textbf{\bibinfo{volume}{56}}, \bibinfo{pages}{569} (\bibinfo{year}{2006}),
  \eprint{hep-ph/0603118}.

\bibitem[{\citenamefont{King}(2007)}]{King:2007nw}
\bibinfo{author}{\bibfnamefont{S.~F.} \bibnamefont{King}},
  \bibinfo{journal}{Contemporary Physics} \textbf{\bibinfo{volume}{48}},
  \bibinfo{pages}{195 } (\bibinfo{year}{2007}), \eprint{arXiv:0712.1750}.

\bibitem[{\citenamefont{Elliott and Vogel}(2002)}]{Elliott:2002xe}
\bibinfo{author}{\bibfnamefont{S.~R.} \bibnamefont{Elliott}} \bibnamefont{and}
  \bibinfo{author}{\bibfnamefont{P.}~\bibnamefont{Vogel}},
  \bibinfo{journal}{Ann. Rev. Nucl. Part. Sci.} \textbf{\bibinfo{volume}{52}},
  \bibinfo{pages}{115} (\bibinfo{year}{2002}), \eprint{hep-ph/0202264}.

\bibitem[{\citenamefont{{Avignone, III} et~al.}(2007)\citenamefont{{Avignone,
  III}, Elliott, and Engel}}]{Avignone:2007fu}
\bibinfo{author}{\bibfnamefont{F.~T.} \bibnamefont{{Avignone, III}}},
  \bibinfo{author}{\bibfnamefont{S.~R.} \bibnamefont{Elliott}},
  \bibnamefont{and} \bibinfo{author}{\bibfnamefont{J.}~\bibnamefont{Engel}}
  (\bibinfo{year}{2007}), \eprint{arXiv:0708.1033}.

\bibitem[{\citenamefont{Piquemal}(2007)}]{Piquemal}
\bibinfo{author}{\bibfnamefont{F.}~\bibnamefont{Piquemal}}
  (\bibinfo{year}{2007}), \bibinfo{note}{\textit{Beta Decay Experiments,} talk
  at the XXIII International Symposium on Lepton and Photon Interactions at
  High Energies}, \urlprefix\url{chep.knu.ac.kr/lp07/htm/S12/S12_36.pdf}.

\bibitem[{\citenamefont{Vogel and Piepke}(2007)}]{VogelPiepkeBB}
\bibinfo{author}{\bibfnamefont{P.}~\bibnamefont{Vogel}} \bibnamefont{and}
  \bibinfo{author}{\bibfnamefont{A.}~\bibnamefont{Piepke}}
  (\bibinfo{year}{2007}), \bibinfo{note}{in~\cite{Yao:2006px}},
  \urlprefix\url{pdg.lbl.gov/2007/reviews/betabeta_s076.pdf}.

\bibitem[{\citenamefont{Gelmini and Roncadelli}(1981)}]{Gelmini:1980re}
\bibinfo{author}{\bibfnamefont{G.~B.} \bibnamefont{Gelmini}} \bibnamefont{and}
  \bibinfo{author}{\bibfnamefont{M.}~\bibnamefont{Roncadelli}},
  \bibinfo{journal}{Phys. Lett.} \textbf{\bibinfo{volume}{B99}},
  \bibinfo{pages}{411} (\bibinfo{year}{1981}).

\bibitem[{\citenamefont{Weinberg}(1979)}]{Weinberg:1979sa}
\bibinfo{author}{\bibfnamefont{S.}~\bibnamefont{Weinberg}},
  \bibinfo{journal}{Phys. Rev. Lett.} \textbf{\bibinfo{volume}{43}},
  \bibinfo{pages}{1566} (\bibinfo{year}{1979}).

\bibitem[{\citenamefont{Minkowski}(1977)}]{Minkowski:1977sc}
\bibinfo{author}{\bibfnamefont{P.}~\bibnamefont{Minkowski}},
  \bibinfo{journal}{Phys. Lett.} \textbf{\bibinfo{volume}{B67}},
  \bibinfo{pages}{421} (\bibinfo{year}{1977}).

\bibitem[{\citenamefont{Yanagida}(1979)}]{Yanagida:1979as}
\bibinfo{author}{\bibfnamefont{T.}~\bibnamefont{Yanagida}}, in
  \emph{\bibinfo{booktitle}{{Proceedings of the Workshop on the Baryon Number
  of the Universe and Unified Theories}}}, edited by
  \bibinfo{editor}{\bibfnamefont{O.}~\bibnamefont{Sawada}} \bibnamefont{and}
  \bibinfo{editor}{\bibfnamefont{A.}~\bibnamefont{Sugamoto}}
  (\bibinfo{publisher}{KEK}, \bibinfo{address}{Tsukuba}, \bibinfo{year}{1979}),
  pp. \bibinfo{pages}{95--98}, \bibinfo{note}{[reprinted in \textit{SEESAW 25:
  Proceedings of the International Conference on the Seesaw Mechanism} ed
  Orloff J, Lavignac S and Cribier M (Singapore: World Scientific, 2005) pp
  261-264]}.

\bibitem[{\citenamefont{Gell-Mann et~al.}(1979)\citenamefont{Gell-Mann, Ramond,
  and Slansky}}]{Gell-Mann:1980vs}
\bibinfo{author}{\bibfnamefont{M.}~\bibnamefont{Gell-Mann}},
  \bibinfo{author}{\bibfnamefont{P.}~\bibnamefont{Ramond}}, \bibnamefont{and}
  \bibinfo{author}{\bibfnamefont{R.}~\bibnamefont{Slansky}}, in
  \emph{\bibinfo{booktitle}{Supergravity}}, edited by
  \bibinfo{editor}{\bibnamefont{{van Nieuwenhuizen, P.}}} \bibnamefont{and}
  \bibinfo{editor}{\bibfnamefont{D.~Z.} \bibnamefont{Freedman}}
  (\bibinfo{publisher}{North-Holland}, \bibinfo{address}{Amsterdam},
  \bibinfo{year}{1979}), pp. \bibinfo{pages}{95--101}.

\bibitem[{\citenamefont{Mohapatra and Senjanovic}(1980)}]{Mohapatra:1979ia}
\bibinfo{author}{\bibfnamefont{R.~N.} \bibnamefont{Mohapatra}}
  \bibnamefont{and}
  \bibinfo{author}{\bibfnamefont{G.}~\bibnamefont{Senjanovic}},
  \bibinfo{journal}{Phys. Rev. Lett.} \textbf{\bibinfo{volume}{44}},
  \bibinfo{pages}{912} (\bibinfo{year}{1980}).

\bibitem[{\citenamefont{Schechter and Valle}(1980)}]{Schechter:1980gr}
\bibinfo{author}{\bibfnamefont{J.}~\bibnamefont{Schechter}} \bibnamefont{and}
  \bibinfo{author}{\bibfnamefont{J.~W.~F.} \bibnamefont{Valle}},
  \bibinfo{journal}{Phys. Rev.} \textbf{\bibinfo{volume}{D22}},
  \bibinfo{pages}{2227} (\bibinfo{year}{1980}).

\bibitem[{\citenamefont{Smirnov}(2005)}]{Smirnov:2004hs}
\bibinfo{author}{\bibfnamefont{A.~Y.} \bibnamefont{Smirnov}}, in
  \emph{\bibinfo{booktitle}{{SEESAW 25: Proceedings of the International
  Conference on the Seesaw Mechanism}}}, edited by
  \bibinfo{editor}{\bibfnamefont{J.}~\bibnamefont{Orloff}},
  \bibinfo{editor}{\bibfnamefont{S.}~\bibnamefont{Lavignac}}, \bibnamefont{and}
  \bibinfo{editor}{\bibfnamefont{M.}~\bibnamefont{Cribier}}
  (\bibinfo{publisher}{World Scientific}, \bibinfo{address}{Singapore},
  \bibinfo{year}{2005}), pp. \bibinfo{pages}{221--236},
  \eprint{hep-ph/0411194}.

\bibitem[{\citenamefont{Chen et~al.}(2007)\citenamefont{Chen, de~Gouv\^{e}a,
  and Dobrescu}}]{Chen:2006hn}
\bibinfo{author}{\bibfnamefont{M.-C.} \bibnamefont{Chen}},
  \bibinfo{author}{\bibfnamefont{A.}~\bibnamefont{de~Gouv\^{e}a}},
  \bibnamefont{and} \bibinfo{author}{\bibfnamefont{B.~A.}
  \bibnamefont{Dobrescu}}, \bibinfo{journal}{Phys. Rev.}
  \textbf{\bibinfo{volume}{D75}}, \bibinfo{pages}{055009}
  (\bibinfo{year}{2007}), \eprint{hep-ph/0612017}.

\bibitem[{\citenamefont{Lee and Shrock}(1977)}]{Lee:1977ti}
\bibinfo{author}{\bibfnamefont{B.~W.} \bibnamefont{Lee}} \bibnamefont{and}
  \bibinfo{author}{\bibfnamefont{R.~E.} \bibnamefont{Shrock}},
  \bibinfo{journal}{Phys. Rev.} \textbf{\bibinfo{volume}{D16}},
  \bibinfo{pages}{1444} (\bibinfo{year}{1977}).

\bibitem[{\citenamefont{Pontecorvo}(1957)}]{Pontecorvo:1957qd}
\bibinfo{author}{\bibfnamefont{B.}~\bibnamefont{Pontecorvo}},
  \bibinfo{journal}{Zh. Eksp. Teor. Fiz.} \textbf{\bibinfo{volume}{34}},
  \bibinfo{pages}{247} (\bibinfo{year}{1957}), \bibinfo{note}{{[English
  transl.: \textit{Sov. Phys. JETP} \textbf{7} 172-173 (1958)]}}.

\bibitem[{\citenamefont{Maki et~al.}(1962)\citenamefont{Maki, Nakagawa, and
  Sakata}}]{Maki:1962mu}
\bibinfo{author}{\bibfnamefont{Z.}~\bibnamefont{Maki}},
  \bibinfo{author}{\bibfnamefont{M.}~\bibnamefont{Nakagawa}}, \bibnamefont{and}
  \bibinfo{author}{\bibfnamefont{S.}~\bibnamefont{Sakata}},
  \bibinfo{journal}{Prog. Theor. Phys.} \textbf{\bibinfo{volume}{28}},
  \bibinfo{pages}{870} (\bibinfo{year}{1962}).

\bibitem[{\citenamefont{Gonzalez-Garcia}(2005)}]{Gonzalez-Garcia:2004jd}
\bibinfo{author}{\bibfnamefont{M.~C.} \bibnamefont{Gonzalez-Garcia}},
  \bibinfo{journal}{Phys. Scripta} \textbf{\bibinfo{volume}{T121}},
  \bibinfo{pages}{72} (\bibinfo{year}{2005}), \eprint{hep-ph/0410030}.

\bibitem[{\citenamefont{Gonzalez-Garcia and
  Maltoni}(2007)}]{GonzalezGarcia:2007ib}
\bibinfo{author}{\bibfnamefont{M.~C.} \bibnamefont{Gonzalez-Garcia}}
  \bibnamefont{and} \bibinfo{author}{\bibfnamefont{M.}~\bibnamefont{Maltoni}}
  (\bibinfo{year}{2007}), \eprint{arXiv:0704.1800}.

\bibitem[{\citenamefont{Maltoni et~al.}(2004)\citenamefont{Maltoni, Schwetz,
  Tortola, and Valle}}]{Maltoni:2004ei}
\bibinfo{author}{\bibfnamefont{M.}~\bibnamefont{Maltoni}},
  \bibinfo{author}{\bibfnamefont{T.}~\bibnamefont{Schwetz}},
  \bibinfo{author}{\bibfnamefont{M.~A.} \bibnamefont{Tortola}},
  \bibnamefont{and} \bibinfo{author}{\bibfnamefont{J.~W.~F.}
  \bibnamefont{Valle}}, \bibinfo{journal}{New J. Phys.}
  \textbf{\bibinfo{volume}{6}}, \bibinfo{pages}{122} (\bibinfo{year}{2004}),
  \eprint{hep-ph/0405172v6}.

\bibitem[{\citenamefont{Ackermann et~al.}(2007)}]{Ackermann:2007km}
\bibinfo{author}{\bibfnamefont{M.}~\bibnamefont{Ackermann}}
  \bibnamefont{et~al.} (\bibinfo{collaboration}{IceCube Collaboration})
  (\bibinfo{year}{2007}), \eprint{arXiv:0711.3022}.

\bibitem[{\citenamefont{Albuquerque}(2006)}]{Albuquerque:2006fd}
\bibinfo{author}{\bibfnamefont{I.~F.~M.} \bibnamefont{Albuquerque}}
  (\bibinfo{year}{2006}), \eprint{hep-ph/0612090}.

\bibitem[{\citenamefont{Hoffman}(2007)}]{KaraSSI}
\bibinfo{author}{\bibfnamefont{K.}~\bibnamefont{Hoffman}}
  (\bibinfo{year}{2007}), \bibinfo{note}{\textit{Recent Results from AMANDA and
  Prospects for IceCube,} Topical Conference Lecture at the 2007 SLAC Summer
  Institute}, \urlprefix\url{www-conf.slac.stanford.edu/ssi/2007/}.

\bibitem[{\citenamefont{Andreev et~al.}(1979)\citenamefont{Andreev, Berezinsky,
  and Smirnov}}]{Andreev:1979cp}
\bibinfo{author}{\bibfnamefont{Y.~M.} \bibnamefont{Andreev}},
  \bibinfo{author}{\bibfnamefont{V.~S.} \bibnamefont{Berezinsky}},
  \bibnamefont{and} \bibinfo{author}{\bibfnamefont{A.~Y.}
  \bibnamefont{Smirnov}}, \bibinfo{journal}{Phys. Lett.}
  \textbf{\bibinfo{volume}{B84}}, \bibinfo{pages}{247} (\bibinfo{year}{1979}).

\bibitem[{\citenamefont{Eichten et~al.}(1984)\citenamefont{Eichten, Hinchliffe,
  Lane, and Quigg}}]{Eichten:1984eu}
\bibinfo{author}{\bibfnamefont{E.}~\bibnamefont{Eichten}},
  \bibinfo{author}{\bibfnamefont{I.}~\bibnamefont{Hinchliffe}},
  \bibinfo{author}{\bibfnamefont{K.~D.} \bibnamefont{Lane}}, \bibnamefont{and}
  \bibinfo{author}{\bibfnamefont{C.}~\bibnamefont{Quigg}},
  \bibinfo{journal}{Rev. Mod. Phys.} \textbf{\bibinfo{volume}{56}},
  \bibinfo{pages}{579} (\bibinfo{year}{1984}).

\bibitem[{\citenamefont{Quigg et~al.}(1986)\citenamefont{Quigg, Reno, and
  Walker}}]{Quigg:1986mb}
\bibinfo{author}{\bibfnamefont{C.}~\bibnamefont{Quigg}},
  \bibinfo{author}{\bibfnamefont{M.~H.} \bibnamefont{Reno}}, \bibnamefont{and}
  \bibinfo{author}{\bibfnamefont{T.~P.} \bibnamefont{Walker}},
  \bibinfo{journal}{Phys. Rev. Lett.} \textbf{\bibinfo{volume}{57}},
  \bibinfo{pages}{774} (\bibinfo{year}{1986}).

\bibitem[{\citenamefont{Reno and Quigg}(1988)}]{Reno:1987zf}
\bibinfo{author}{\bibfnamefont{M.~H.} \bibnamefont{Reno}} \bibnamefont{and}
  \bibinfo{author}{\bibfnamefont{C.}~\bibnamefont{Quigg}},
  \bibinfo{journal}{Phys. Rev.} \textbf{\bibinfo{volume}{D37}},
  \bibinfo{pages}{657} (\bibinfo{year}{1988}).

\bibitem[{\citenamefont{Gandhi et~al.}(1996)\citenamefont{Gandhi, Quigg, Reno,
  and Sarcevic}}]{Gandhi:1995tf}
\bibinfo{author}{\bibfnamefont{R.}~\bibnamefont{Gandhi}},
  \bibinfo{author}{\bibfnamefont{C.}~\bibnamefont{Quigg}},
  \bibinfo{author}{\bibfnamefont{M.~H.} \bibnamefont{Reno}}, \bibnamefont{and}
  \bibinfo{author}{\bibfnamefont{I.}~\bibnamefont{Sarcevic}},
  \bibinfo{journal}{Astropart. Phys.} \textbf{\bibinfo{volume}{5}},
  \bibinfo{pages}{81} (\bibinfo{year}{1996}), \eprint{hep-ph/9512364}.

\bibitem[{\citenamefont{Gandhi et~al.}(1998)\citenamefont{Gandhi, Quigg, Reno,
  and Sarcevic}}]{Gandhi:1998ri}
\bibinfo{author}{\bibfnamefont{R.}~\bibnamefont{Gandhi}},
  \bibinfo{author}{\bibfnamefont{C.}~\bibnamefont{Quigg}},
  \bibinfo{author}{\bibfnamefont{M.~H.} \bibnamefont{Reno}}, \bibnamefont{and}
  \bibinfo{author}{\bibfnamefont{I.}~\bibnamefont{Sarcevic}},
  \bibinfo{journal}{Phys. Rev.} \textbf{\bibinfo{volume}{D58}},
  \bibinfo{pages}{093009} (\bibinfo{year}{1998}), \eprint{hep-ph/9807264}.

\bibitem[{\citenamefont{Reno}(2005)}]{Reno:2004cx}
\bibinfo{author}{\bibfnamefont{M.~H.} \bibnamefont{Reno}},
  \bibinfo{journal}{Nucl. Phys. Proc. Suppl.} \textbf{\bibinfo{volume}{143}},
  \bibinfo{pages}{407} (\bibinfo{year}{2005}), \eprint{hep-ph/0410109}.

\bibitem[{\citenamefont{Cooper-Sarkar and Sarkar}(2008)}]{CooperSarkar:2007cv}
\bibinfo{author}{\bibfnamefont{A.}~\bibnamefont{Cooper-Sarkar}}
  \bibnamefont{and} \bibinfo{author}{\bibfnamefont{S.}~\bibnamefont{Sarkar}},
  \bibinfo{journal}{JHEP} \textbf{\bibinfo{volume}{01}}, \bibinfo{pages}{075}
  (\bibinfo{year}{2008}), \eprint{arXiv:0710.5303}.

\bibitem[{\citenamefont{Pumplin et~al.}(2002)}]{Pumplin:2002vw}
\bibinfo{author}{\bibfnamefont{J.}~\bibnamefont{Pumplin}} \bibnamefont{et~al.}
  (\bibinfo{collaboration}{CTEQ Collaboration}), \bibinfo{journal}{JHEP}
  \textbf{\bibinfo{volume}{07}}, \bibinfo{pages}{012} (\bibinfo{year}{2002}),
  \eprint{hep-ph/0201195}.

\bibitem[{\citenamefont{Berger et~al.}(2007)\citenamefont{Berger, Block, McKay,
  and Tan}}]{Berger:2007ic}
\bibinfo{author}{\bibfnamefont{E.~L.} \bibnamefont{Berger}},
  \bibinfo{author}{\bibfnamefont{M.~M.} \bibnamefont{Block}},
  \bibinfo{author}{\bibfnamefont{D.~W.} \bibnamefont{McKay}}, \bibnamefont{and}
  \bibinfo{author}{\bibfnamefont{C.-I.} \bibnamefont{Tan}}
  (\bibinfo{year}{2007}), \eprint{arXiv:0708.1960}.

\bibitem[{\citenamefont{Adloff et~al.}(2003)}]{Adloff:2003uh}
\bibinfo{author}{\bibfnamefont{C.}~\bibnamefont{Adloff}} \bibnamefont{et~al.}
  (\bibinfo{collaboration}{H1 Collaboration}), \bibinfo{journal}{Eur. Phys. J.}
  \textbf{\bibinfo{volume}{C30}}, \bibinfo{pages}{1} (\bibinfo{year}{2003}),
  \eprint{hep-ex/0304003}.

\bibitem[{\citenamefont{Chekanov et~al.}(2003)}]{Chekanov:2002pv}
\bibinfo{author}{\bibfnamefont{S.}~\bibnamefont{Chekanov}} \bibnamefont{et~al.}
  (\bibinfo{collaboration}{ZEUS Collaboration}), \bibinfo{journal}{Phys. Rev.}
  \textbf{\bibinfo{volume}{D67}}, \bibinfo{pages}{012007}
  (\bibinfo{year}{2003}), \eprint{hep-ex/0208023}.

\bibitem[{\citenamefont{Chekanov et~al.}(2007)}]{Chekanov:2006ff}
\bibinfo{author}{\bibfnamefont{S.}~\bibnamefont{Chekanov}} \bibnamefont{et~al.}
  (\bibinfo{collaboration}{ZEUS Collaboration}), \bibinfo{journal}{Eur. Phys.
  J.} \textbf{\bibinfo{volume}{C49}}, \bibinfo{pages}{523}
  (\bibinfo{year}{2007}), \eprint{hep-ex/0608014}.

\bibitem[{\citenamefont{Anchordoqui et~al.}(2006)\citenamefont{Anchordoqui,
  Cooper-Sarkar, Hooper, and Sarkar}}]{Anchordoqui:2006ta}
\bibinfo{author}{\bibfnamefont{L.~A.} \bibnamefont{Anchordoqui}},
  \bibinfo{author}{\bibfnamefont{A.~M.} \bibnamefont{Cooper-Sarkar}},
  \bibinfo{author}{\bibfnamefont{D.}~\bibnamefont{Hooper}}, \bibnamefont{and}
  \bibinfo{author}{\bibfnamefont{S.}~\bibnamefont{Sarkar}},
  \bibinfo{journal}{Phys. Rev.} \textbf{\bibinfo{volume}{D74}},
  \bibinfo{pages}{043008} (\bibinfo{year}{2006}), \eprint{hep-ph/0605086}.

\bibitem[{\citenamefont{Learned and Pakvasa}(1995)}]{Learned:1994wg}
\bibinfo{author}{\bibfnamefont{J.~G.} \bibnamefont{Learned}} \bibnamefont{and}
  \bibinfo{author}{\bibfnamefont{S.}~\bibnamefont{Pakvasa}},
  \bibinfo{journal}{Astropart. Phys.} \textbf{\bibinfo{volume}{3}},
  \bibinfo{pages}{267} (\bibinfo{year}{1995}), \eprint{hep-ph/9405296}.

\bibitem[{\citenamefont{de~Holanda and Smirnov}(2003)}]{deHolanda:2002iv}
\bibinfo{author}{\bibfnamefont{P.~C.} \bibnamefont{de~Holanda}}
  \bibnamefont{and} \bibinfo{author}{\bibfnamefont{A.~Y.}
  \bibnamefont{Smirnov}}, \bibinfo{journal}{JCAP}
  \textbf{\bibinfo{volume}{0302}}, \bibinfo{pages}{001} (\bibinfo{year}{2003}),
  \eprint{hep-ph/0212270}.

\bibitem[{\citenamefont{Barenboim and Quigg}(2003)}]{Barenboim:2003jm}
\bibinfo{author}{\bibfnamefont{G.}~\bibnamefont{Barenboim}} \bibnamefont{and}
  \bibinfo{author}{\bibfnamefont{C.}~\bibnamefont{Quigg}},
  \bibinfo{journal}{Phys. Rev.} \textbf{\bibinfo{volume}{D67}},
  \bibinfo{pages}{073024} (\bibinfo{year}{2003}), \eprint{hep-ph/0301220}.

\bibitem[{\citenamefont{Kashti and Waxman}(2005)}]{Kashti:2005qa}
\bibinfo{author}{\bibfnamefont{T.}~\bibnamefont{Kashti}} \bibnamefont{and}
  \bibinfo{author}{\bibfnamefont{E.}~\bibnamefont{Waxman}},
  \bibinfo{journal}{Phys. Rev. Lett.} \textbf{\bibinfo{volume}{95}},
  \bibinfo{pages}{181101} (\bibinfo{year}{2005}), \eprint{astro-ph/0507599}.

\bibitem[{\citenamefont{Valle}(1983)}]{Valle:1983ua}
\bibinfo{author}{\bibfnamefont{J.~W.~F.} \bibnamefont{Valle}},
  \bibinfo{journal}{Phys. Lett.} \textbf{\bibinfo{volume}{B131}},
  \bibinfo{pages}{87} (\bibinfo{year}{1983}).

\bibitem[{\citenamefont{Gelmini and Valle}(1984)}]{Gelmini:1983ea}
\bibinfo{author}{\bibfnamefont{G.~B.} \bibnamefont{Gelmini}} \bibnamefont{and}
  \bibinfo{author}{\bibfnamefont{J.~W.~F.} \bibnamefont{Valle}},
  \bibinfo{journal}{Phys. Lett.} \textbf{\bibinfo{volume}{B142}},
  \bibinfo{pages}{181} (\bibinfo{year}{1984}).

\bibitem[{\citenamefont{Beacom et~al.}(2003)\citenamefont{Beacom, Bell, Hooper,
  Pakvasa, and Weiler}}]{Beacom:2002vi}
\bibinfo{author}{\bibfnamefont{J.~F.} \bibnamefont{Beacom}},
  \bibinfo{author}{\bibfnamefont{N.~F.} \bibnamefont{Bell}},
  \bibinfo{author}{\bibfnamefont{D.}~\bibnamefont{Hooper}},
  \bibinfo{author}{\bibfnamefont{S.}~\bibnamefont{Pakvasa}}, \bibnamefont{and}
  \bibinfo{author}{\bibfnamefont{T.~J.} \bibnamefont{Weiler}},
  \bibinfo{journal}{Phys. Rev. Lett.} \textbf{\bibinfo{volume}{90}},
  \bibinfo{pages}{181301} (\bibinfo{year}{2003}), \eprint{hep-ph/0211305}.

\bibitem[{\citenamefont{Chacko et~al.}(2004)\citenamefont{Chacko, Hall, Okui,
  and Oliver}}]{Chacko:2003dt}
\bibinfo{author}{\bibfnamefont{Z.}~\bibnamefont{Chacko}},
  \bibinfo{author}{\bibfnamefont{L.~J.} \bibnamefont{Hall}},
  \bibinfo{author}{\bibfnamefont{T.}~\bibnamefont{Okui}}, \bibnamefont{and}
  \bibinfo{author}{\bibfnamefont{S.~J.} \bibnamefont{Oliver}},
  \bibinfo{journal}{Phys. Rev.} \textbf{\bibinfo{volume}{D70}},
  \bibinfo{pages}{085008} (\bibinfo{year}{2004}), \eprint{hep-ph/0312267}.

\bibitem[{\citenamefont{Glashow}(1960)}]{PhysRev.118.316}
\bibinfo{author}{\bibfnamefont{S.~L.} \bibnamefont{Glashow}},
  \bibinfo{journal}{Phys. Rev.} \textbf{\bibinfo{volume}{118}},
  \bibinfo{pages}{316} (\bibinfo{year}{1960}).

\bibitem[{\citenamefont{Weiler}(1982)}]{Weiler:1982qy}
\bibinfo{author}{\bibfnamefont{T.~J.} \bibnamefont{Weiler}},
  \bibinfo{journal}{Phys. Rev. Lett.} \textbf{\bibinfo{volume}{49}},
  \bibinfo{pages}{234} (\bibinfo{year}{1982}).

\bibitem[{\citenamefont{Weiler}(1984)}]{Weiler:1983xx}
\bibinfo{author}{\bibfnamefont{T.~J.} \bibnamefont{Weiler}},
  \bibinfo{journal}{Astrophys. J.} \textbf{\bibinfo{volume}{285}},
  \bibinfo{pages}{495} (\bibinfo{year}{1984}).

\bibitem[{\citenamefont{Roulet}(1993)}]{Roulet:1992pz}
\bibinfo{author}{\bibfnamefont{E.}~\bibnamefont{Roulet}},
  \bibinfo{journal}{Phys. Rev.} \textbf{\bibinfo{volume}{D47}},
  \bibinfo{pages}{5247} (\bibinfo{year}{1993}).

\bibitem[{\citenamefont{Gondolo et~al.}(1993)\citenamefont{Gondolo, Gelmini,
  and Sarkar}}]{Gondolo:1991rn}
\bibinfo{author}{\bibfnamefont{P.}~\bibnamefont{Gondolo}},
  \bibinfo{author}{\bibfnamefont{G.}~\bibnamefont{Gelmini}}, \bibnamefont{and}
  \bibinfo{author}{\bibfnamefont{S.}~\bibnamefont{Sarkar}},
  \bibinfo{journal}{Nucl. Phys.} \textbf{\bibinfo{volume}{B392}},
  \bibinfo{pages}{111} (\bibinfo{year}{1993}), \eprint{hep-ph/9209236}.

\bibitem[{\citenamefont{Yoshida et~al.}(1997)\citenamefont{Yoshida, Dai, Jui,
  and Sommers}}]{Yoshida:1996ie}
\bibinfo{author}{\bibfnamefont{S.}~\bibnamefont{Yoshida}},
  \bibinfo{author}{\bibfnamefont{H.-y.} \bibnamefont{Dai}},
  \bibinfo{author}{\bibfnamefont{C.~C.~H.} \bibnamefont{Jui}},
  \bibnamefont{and} \bibinfo{author}{\bibfnamefont{P.}~\bibnamefont{Sommers}},
  \bibinfo{journal}{Astrophys. J.} \textbf{\bibinfo{volume}{479}},
  \bibinfo{pages}{547} (\bibinfo{year}{1997}), \eprint{astro-ph/9608186}.

\bibitem[{\citenamefont{Fargion et~al.}(1999)\citenamefont{Fargion, Mele, and
  Salis}}]{Fargion:1997ft}
\bibinfo{author}{\bibfnamefont{D.}~\bibnamefont{Fargion}},
  \bibinfo{author}{\bibfnamefont{B.}~\bibnamefont{Mele}}, \bibnamefont{and}
  \bibinfo{author}{\bibfnamefont{A.}~\bibnamefont{Salis}},
  \bibinfo{journal}{Astrophys. J.} \textbf{\bibinfo{volume}{517}},
  \bibinfo{pages}{725} (\bibinfo{year}{1999}), \eprint{astro-ph/9710029}.

\bibitem[{\citenamefont{Weiler}(1999)}]{Weiler:1997sh}
\bibinfo{author}{\bibfnamefont{T.~J.} \bibnamefont{Weiler}},
  \bibinfo{journal}{Astropart. Phys.} \textbf{\bibinfo{volume}{11}},
  \bibinfo{pages}{303} (\bibinfo{year}{1999}), \eprint{hep-ph/9710431}.

\bibitem[{\citenamefont{Eberle et~al.}(2004)\citenamefont{Eberle, Ringwald,
  Song, and Weiler}}]{Eberle:2004ua}
\bibinfo{author}{\bibfnamefont{B.}~\bibnamefont{Eberle}},
  \bibinfo{author}{\bibfnamefont{A.}~\bibnamefont{Ringwald}},
  \bibinfo{author}{\bibfnamefont{L.}~\bibnamefont{Song}}, \bibnamefont{and}
  \bibinfo{author}{\bibfnamefont{T.~J.} \bibnamefont{Weiler}},
  \bibinfo{journal}{Phys. Rev.} \textbf{\bibinfo{volume}{D70}},
  \bibinfo{pages}{023007} (\bibinfo{year}{2004}), \eprint{hep-ph/0401203}.

\bibitem[{\citenamefont{Tremaine and Gunn}(1979)}]{Tremaine:1979we}
\bibinfo{author}{\bibfnamefont{S.}~\bibnamefont{Tremaine}} \bibnamefont{and}
  \bibinfo{author}{\bibfnamefont{J.~E.} \bibnamefont{Gunn}},
  \bibinfo{journal}{Phys. Rev. Lett.} \textbf{\bibinfo{volume}{42}},
  \bibinfo{pages}{407} (\bibinfo{year}{1979}).

\bibitem[{\citenamefont{Schramm and Steigman}(1981)}]{Schramm:1980xw}
\bibinfo{author}{\bibfnamefont{D.~N.} \bibnamefont{Schramm}} \bibnamefont{and}
  \bibinfo{author}{\bibfnamefont{G.}~\bibnamefont{Steigman}},
  \bibinfo{journal}{Gen. Rel. Grav.} \textbf{\bibinfo{volume}{13}},
  \bibinfo{pages}{101} (\bibinfo{year}{1981}).

\bibitem[{\citenamefont{Primack}(2007)}]{JoelPSSI}
\bibinfo{author}{\bibfnamefont{J.}~\bibnamefont{Primack}}
  (\bibinfo{year}{2007}), \bibinfo{note}{\textit{A Brief History of Dark
  Matter,} Lecture at the 2007 SLAC Summer Institute},
  \urlprefix\url{www-conf.slac.stanford.edu/ssi/2007/}.

\bibitem[{\citenamefont{Bertone et~al.}(2005)\citenamefont{Bertone, Hooper, and
  Silk}}]{Bertone:2004pz}
\bibinfo{author}{\bibfnamefont{G.}~\bibnamefont{Bertone}},
  \bibinfo{author}{\bibfnamefont{D.}~\bibnamefont{Hooper}}, \bibnamefont{and}
  \bibinfo{author}{\bibfnamefont{J.}~\bibnamefont{Silk}},
  \bibinfo{journal}{Phys. Rept.} \textbf{\bibinfo{volume}{405}},
  \bibinfo{pages}{279} (\bibinfo{year}{2005}), \eprint{hep-ph/0404175}.

\bibitem[{\citenamefont{Gondolo et~al.}(2004)}]{Gondolo:2004sc}
\bibinfo{author}{\bibfnamefont{P.}~\bibnamefont{Gondolo}} \bibnamefont{et~al.},
  \bibinfo{journal}{JCAP} \textbf{\bibinfo{volume}{0407}}, \bibinfo{pages}{008}
  (\bibinfo{year}{2004}), \eprint{astro-ph/0406204}.

\bibitem[{\citenamefont{Cirelli et~al.}(2005)}]{Cirelli:2005gh}
\bibinfo{author}{\bibfnamefont{M.}~\bibnamefont{Cirelli}} \bibnamefont{et~al.},
  \bibinfo{journal}{Nucl. Phys.} \textbf{\bibinfo{volume}{B727}},
  \bibinfo{pages}{99} (\bibinfo{year}{2005}), \eprint{hep-ph/0506298v3}.

\bibitem[{\citenamefont{Lehnert and Weiler}(2007)}]{Lehnert:2007fv}
\bibinfo{author}{\bibfnamefont{R.}~\bibnamefont{Lehnert}} \bibnamefont{and}
  \bibinfo{author}{\bibfnamefont{T.~J.} \bibnamefont{Weiler}}
  (\bibinfo{year}{2007}), \eprint{arXiv:0708.1035}.

\bibitem[{\citenamefont{Barger et~al.}(2007)\citenamefont{Barger, Keung,
  Shaughnessy, and Tregre}}]{Barger:2007xf}
\bibinfo{author}{\bibfnamefont{V.}~\bibnamefont{Barger}},
  \bibinfo{author}{\bibfnamefont{W.-Y.} \bibnamefont{Keung}},
  \bibinfo{author}{\bibfnamefont{G.}~\bibnamefont{Shaughnessy}},
  \bibnamefont{and} \bibinfo{author}{\bibfnamefont{A.}~\bibnamefont{Tregre}}
  (\bibinfo{year}{2007}), \eprint{arXiv:0708.1325}.

\bibitem[{\citenamefont{Berezinsky et~al.}(2003)\citenamefont{Berezinsky,
  Dokuchaev, and Eroshenko}}]{Berezinsky:2003vn}
\bibinfo{author}{\bibfnamefont{V.}~\bibnamefont{Berezinsky}},
  \bibinfo{author}{\bibfnamefont{V.}~\bibnamefont{Dokuchaev}},
  \bibnamefont{and}
  \bibinfo{author}{\bibfnamefont{Y.}~\bibnamefont{Eroshenko}},
  \bibinfo{journal}{Phys. Rev.} \textbf{\bibinfo{volume}{D68}},
  \bibinfo{pages}{103003} (\bibinfo{year}{2003}), \eprint{astro-ph/0301551}.

\bibitem[{\citenamefont{Carena et~al.}(2007)\citenamefont{Carena, Hooper, and
  Vallinotto}}]{Carena:2006nv}
\bibinfo{author}{\bibfnamefont{M.~S.} \bibnamefont{Carena}},
  \bibinfo{author}{\bibfnamefont{D.}~\bibnamefont{Hooper}}, \bibnamefont{and}
  \bibinfo{author}{\bibfnamefont{A.}~\bibnamefont{Vallinotto}},
  \bibinfo{journal}{Phys. Rev.} \textbf{\bibinfo{volume}{D75}},
  \bibinfo{pages}{055010} (\bibinfo{year}{2007}), \eprint{hep-ph/0611065}.

\bibitem[{\citenamefont{Peskin}(2007)}]{Peskin:2007nk}
\bibinfo{author}{\bibfnamefont{M.~E.} \bibnamefont{Peskin}}
  (\bibinfo{year}{2007}), \eprint{arXiv:0707.1536}.

\bibitem[{\citenamefont{Barenboim et~al.}(2006)\citenamefont{Barenboim,
  Mena~Requejo, and Quigg}}]{Barenboim:2006dj}
\bibinfo{author}{\bibfnamefont{G.}~\bibnamefont{Barenboim}},
  \bibinfo{author}{\bibfnamefont{O.}~\bibnamefont{Mena~Requejo}},
  \bibnamefont{and} \bibinfo{author}{\bibfnamefont{C.}~\bibnamefont{Quigg}},
  \bibinfo{journal}{Phys. Rev.} \textbf{\bibinfo{volume}{D74}},
  \bibinfo{pages}{023006} (\bibinfo{year}{2006}), \eprint{astro-ph/0604215}.

\bibitem[{\citenamefont{Buchmuller et~al.}(2005)\citenamefont{Buchmuller,
  Peccei, and Yanagida}}]{Buchmuller:2005eh}
\bibinfo{author}{\bibfnamefont{W.}~\bibnamefont{Buchmuller}},
  \bibinfo{author}{\bibfnamefont{R.~D.} \bibnamefont{Peccei}},
  \bibnamefont{and} \bibinfo{author}{\bibfnamefont{T.}~\bibnamefont{Yanagida}},
  \bibinfo{journal}{Ann. Rev. Nucl. Part. Sci.} \textbf{\bibinfo{volume}{55}},
  \bibinfo{pages}{311} (\bibinfo{year}{2005}), \eprint{hep-ph/0502169}.

\bibitem[{\citenamefont{Chen}(2007)}]{Chen:2007fv}
\bibinfo{author}{\bibfnamefont{M.-C.} \bibnamefont{Chen}}
  (\bibinfo{year}{2007}), \bibinfo{note}{lectures at \textit{TASI 2006:
  Exploring New Frontiers Using Colliders and Neutrinos}},
  \eprint{hep-ph/0703087}.

\bibitem[{\citenamefont{Hung}(2000)}]{Hung:2000yg}
\bibinfo{author}{\bibfnamefont{P.~Q.} \bibnamefont{Hung}}
  (\bibinfo{year}{2000}), \eprint{hep-ph/0010126}.

\bibitem[{\citenamefont{Fardon et~al.}(2004)\citenamefont{Fardon, Nelson, and
  Weiner}}]{Fardon:2003eh}
\bibinfo{author}{\bibfnamefont{R.}~\bibnamefont{Fardon}},
  \bibinfo{author}{\bibfnamefont{A.~E.} \bibnamefont{Nelson}},
  \bibnamefont{and} \bibinfo{author}{\bibfnamefont{N.}~\bibnamefont{Weiner}},
  \bibinfo{journal}{JCAP} \textbf{\bibinfo{volume}{0410}}, \bibinfo{pages}{005}
  (\bibinfo{year}{2004}), \eprint{astro-ph/0309800}.

\bibitem[{\citenamefont{Kaplan et~al.}(2004)\citenamefont{Kaplan, Nelson, and
  Weiner}}]{Kaplan:2004dq}
\bibinfo{author}{\bibfnamefont{D.~B.} \bibnamefont{Kaplan}},
  \bibinfo{author}{\bibfnamefont{A.~E.} \bibnamefont{Nelson}},
  \bibnamefont{and} \bibinfo{author}{\bibfnamefont{N.}~\bibnamefont{Weiner}},
  \bibinfo{journal}{Phys. Rev. Lett.} \textbf{\bibinfo{volume}{93}},
  \bibinfo{pages}{091801} (\bibinfo{year}{2004}), \eprint{hep-ph/0401099}.

\end{thebibliography}
\end{document}